\documentclass[prb,floatfix,twocolumn,superscriptaddress,nofootinbib,longbibliography]{revtex4-2}
\usepackage{amsmath,graphicx,amssymb,braket,enumitem,mathtools,dsfont}
\usepackage[colorlinks=true,citecolor=blue,linkcolor=red]{hyperref}

\usepackage{bm}
\usepackage{hyperref}

\usepackage{clipboard}
\usepackage{pifont}
\usepackage[dvipsnames]{xcolor}
\usepackage{subfigure}

\newcommand{\br}[1]{\left[#1\right]}
\newcommand{\pr}[1]{\left(#1\right)}
\newcommand{\sr}[1]{\left\{#1\right\}}
\newcommand{\abs}[1]{\left|#1\right|}
\newcommand{\absq}[1]{\abs{#1}^2}
\newcommand{\expect}[1]{\left\langle#1\right\rangle}

\newcommand{\Tr}[1]{\ensuremath{\mathrm{Tr}\br{#1}}}
\DeclareMathOperator*{\argmax}{arg\,max}

\newcommand{\uV}{\mu\mathrm{V}}
\newcommand{\Hz}{\mathrm{Hz}}
\newcommand{\kHz}{\mathrm{kHz}}
\newcommand{\MHz}{\mathrm{MHz}}
\newcommand{\GHz}{\mathrm{GHz}}

\newcommand{\ns}{\mathrm{ns}}
\newcommand{\us}{\mu\mathrm{s}}

\newcommand{\s}{\mathrm{s}}

\newcommand{\V}{\mathrm{V}}
\newcommand{\mV}{\mathrm{mV}}
\newcommand{\identity}{\mathds 1}

\newcommand{\indicator}[2]{\mathbf1_{#1}\pr{#2}}

\newcommand{\beq}{\begin{equation}}
\newcommand{\eeq}{\end{equation}}
\newcommand{\beqnn}{\begin{equation*}}
\newcommand{\eeqnn}{\end{equation*}}
\newcommand{\bea}{\begin{eqnarray}}
\newcommand{\eea}{\end{eqnarray}}
\newcommand{\beann}{\begin{eqnarray*}}
\newcommand{\eeann}{\end{eqnarray*}}
\newcommand{\bes} {\begin{subequations}}
\newcommand{\ees} {\end{subequations}}

\begin{document}

\title{Model validation and error attribution for a drifting qubit}

\author{Malick A. Gaye}
\email{mgaye3@jhu.edu}
\thanks{Present affiliation: University of Rochester}
\affiliation{Center for Computing Research, Sandia National Laboratories, Albuquerque NM, 87185 USA}
\affiliation{Johns Hopkins University, Baltimore MD, 21218 USA}

\author{Dylan Albrecht}
\affiliation{Sandia National Laboratories, Albuquerque NM, 87185 USA}

\author{Steve Young}
\affiliation{Center for Computing Research, Sandia National Laboratories, Albuquerque NM, 87185 USA}

\author{Tameem Albash}
\affiliation{Center for Computing Research, Sandia National Laboratories, Albuquerque NM, 87185 USA}

\author{N. Tobias Jacobson}
\email{ntjacob@sandia.gov}
\affiliation{Center for Computing Research, Sandia National Laboratories, Albuquerque NM, 87185 USA}

\begin{abstract}
Qubit performance is often reported in terms of a variety of single-value metrics, each providing a facet of the underlying noise mechanism limiting performance.  However, the value of these metrics may drift over long timescales, and reporting a single number for qubit performance fails to account for the low-frequency noise processes that give rise to this drift. In this work, we demonstrate how we can use the distribution of these values to validate or invalidate candidate noise models.  We focus on the case of randomized benchmarking (RB), where typically a single error rate is reported but this error rate can drift over time when multiple passes of RB are performed. We show that using a statistical test as simple as the Kolmogorov-Smirnov statistic on the distribution of RB error rates can be used to rule out noise models, assuming the experiment is performed over a long enough time interval to capture relevant low frequency noise.  With confidence in a noise model, we show how care must be exercised when performing error attribution using the distribution of drifting RB error rate.
\end{abstract}

\maketitle

\section{Introduction}
Advances in materials, fabrication, and experimental control have fueled the development of a variety of quantum information processing platforms. While each platform enjoys its own set of advantages, an incomplete understanding of their respective noise sources and how to completely mitigate or suppress them means that decoherence continues to limit the performance of these platforms. While low enough error rates have been reached for early demonstrations of quantum error correction~\cite{Google2023,Postler2023,Bluvstein2024,daSilva2024,Mayer2024}, even when error rates have unequivocally crossed fault tolerance thresholds, further lowering the intrinsic physical noise will still be advantageous to minimize resource overheads of error correction. This motivates efforts to understand and reduce noise.

Spectral properties of the noise can be probed~\cite{Harris2008,Bylander2011,Chan2018} and are an important component of characterizing the noise, but they may not be able to cover the full frequency range needed. To quantify the effect of noise on device or qubit performance, a variety of ``single number" metrics are routinely reported, such as the $T_2^\ast$ dephasing time that quantifies the time for the loss of phase information in a qubit or the randomized benchmarking (RB) error rate~\cite{Knill2008,Magesan2011,Magesan2012b} that is interpreted~\cite{Proctor2017} as the average error per Clifford gate. While each number captures an important aspect of the noise, even the collection of them may not fully characterize the noise or sufficiently constrain noise models. For example, noise with significant low frequency components may mean that a property measured at one point in time may \emph{drift} to a different value when the experiment is repeated at a later point in time~\cite{Fogarty2015,Klimov2018,Proctor2020}. In this case, instead of a single $T_2^\ast$ number, experiments performed on a device over a long time period may instead give a distribution of $T_2^\ast$ numbers. Furthermore, compensating for drifting device parameters may require re-calibration, effectively changing the device and its performance over time. 

We consider the problem of validating a noise model with significant drift using a subset of characterization data. A variety of noise models may be identified using available decoherence rates and spectroscopy data, and we ask how we can validate or invalidate these models using additional data arising from another characterization protocol, in our case RB. For concreteness and physical relevance, we focus on a single qubit encoded in the zero-magnetization singlet and triplet states of two spin-1/2 systems~\cite{Levy2002}, such as those realized using two electrons in a pair of semiconductor quantum dots \cite{Petta2005,Borselli2012,Shulman2012,Wu2014,Nichol2017,Jock2018,Cerfontaine2020,Cerfontaine2020b,Fedele2021,Jirovec2021}. Such systems are susceptible to both magnetic and charge noise, both of which have been observed~\cite{Connors2019,Struck2020,Connors2022,Yoneda2023} to have $1/f$ character~\cite{Dutta1981,Paladino2014} and hence exhibit noise over both short and long timescales. Using full ``wall-clock'' simulations of RB that account for noise correlations across all relevant timescales of a realistic RB experiment, we show how the time-dependent behavior of the RB error rate can be used to further validate and invalidate noise models that were consistent with the previous characterization data. A primary motivation for our approach is to facilitate making use of data experimentalists are already likely to have to further refine error models for their devices.

The paper is organized as follows. In Sec.~\ref{sec:Background} we describe the assumed qubit model, how the parametrization of noise relates to dephasing time measurements, and the details of the benchmarking protocol we simulate. In Sec.~\ref{sec:statistical_methods}, we investigate how one may validate or invalidate noise models based on the distribution of RB error rates and per-circuit survival probabilities. In Sec.~\ref{sec:Error_attribution}, we explore how a validated model may be used to attribute errors on a per-mechanism and/or per-spectral weight basis. Finally, we conclude in Sec.~\ref{sec:Discussion}.

\section{Background}
\label{sec:Background}
\subsection{The singlet-triplet encoding}\label{sec:singlet_triplet_encoding}
Although the methods described in this work can be applied to arbitrary qubit encodings, we focus on the singlet-triplet encoding \cite{Levy2002}, whereby the computational basis states of a single qubit are encoded into the zero-magnetization singlet and triplet states of two spin-$1/2$ systems.
The Hamiltonian for the encoded singlet-triplet qubit can be expressed as
\begin{equation}
\begin{split}
    H_{ST_0}(t)&=\frac{J(t)}2\sigma^z+\frac{\Delta b_z(t)}2\sigma^x, 
\end{split}
    \label{eq:st0_hamiltonian}
\end{equation}
where $\Delta b_z(t)$ is the magnetic field gradient, and $\sigma^x$ and $\sigma^z$ are the Pauli operators acting on the computational states $\ket{0}$ and $\ket{1}$, i.e. $\sigma^z \ket{0} = \ket{0}, \sigma^z \ket{1} = -\ket{1}, \sigma^x \ket{0} = \ket{1}$. A nonzero $J(t)$ is used to generate rotations about the axis $\hat{n} \propto \pr{\Delta b_z(t), 0 , J(t)}^T$, while at $J(t) = 0$ the (always-on) magnetic field gradient generates rotations about the $x$-axis.  Together, these two Hamiltonian terms can be used to implement universal control of the single qubit.

In semiconductor quantum dots, the exchange interaction $J(t)$ is controlled electrically by varying the voltage of a gate electrode, and the mapping between voltage and exchange is often modeled as an exponential dependence~\cite{Burkard2023}, $J(t)=J_0 e^{V(t)/\mathcal{I}}$, where $J_0$ is the residual exchange with a typical value of $J_0 \sim 10^{-2} \ \MHz$, and $\mathcal{I}$ is the insensitivity with a typical value of $\mathcal{I} \sim10^{1} \ \mathrm{mV}$ assuming tunnel barrier control in order to make use of better performance through operating at the symmetric operating point \cite{Reed2016, Martins2016}. Due to charge noise on the electrode and fluctuations in the environment of the quantum dot, both $J(t)$ and $\Delta b_z(t)$ have time-dependent noise, which we model as
\bes
\begin{align}
V(t) &= V_0(t)+\delta V(t) \ , \\
\Delta b_z(t) &= \Delta b_z+\delta b_z(t) \ ,
\end{align}
\ees
where $V_0(t)$ is the ideal (programmed) control voltage pulses  and $\Delta b_z$ is the static magnetic field gradient with typical values of $\Delta b_z/h\sim10 \ \MHz$~\cite{Cerfontaine2020}. The fluctuation $\delta V(t)$ on the control voltage induces noise on the exchange interaction, and $\delta b_z(t)$ captures fluctuations in the magnetic field gradient. Our modeling of these noise terms is the subject of Sec.~\ref{sec:noise_model_and_calibration}, and we give further details of our parameter choices for our model in Appendix~\ref{app:SimParameters}.

Since the static field gradient $\Delta b_z$ is always on, our control axes are not orthogonal. As a consequence, we must compile the desired set of qubit operations in terms of non-orthogonal rotations. To do this we perform a numerical optimization of control voltage timeseries for given values of $\Delta b_z$, $\mathcal{I}$, and $J_0$ (defined in Appendix~\ref{app:SimParameters}) subject to control bandwidth and pulse shaping constraints. Since we will be interested in simulating RB, we perform these optimizations for $\pm \pi/2$ rotations about the $x$ and $z$ axis, which can be used as the generators of the single-qubit Clifford quotient group $\mathbf C_1$ to realize the remaining 20 unitary operators of the group.
The decomposition of the Clifford operators in terms of generators that we use are tabulated in Ref.~\cite{Xue2019}. We give more details of our numerical gate compilation in Appendix~\ref{sec:simulation_framework}.

\subsection{Noise model and calibration}\label{sec:noise_model_and_calibration}
In this work we model correlated noise as sums of independent Ornstein-Uhlenbeck (OU) processes $\eta_i(t)$, where each $\eta_{i}(t)$ is a Gaussian stochastic process having autocorrelation function 
$\langle \eta_{i}(t) \eta_{i}(t') \rangle = (p_{i}/2) e^{-2 \pi f_{i} \vert t- t'\vert}$, with $p_{i}/2$ the noise power and $f_{i}$ the characteristic frequency of each OU process. The sum process $\mathcal{N}_{j,k}(t)=\sum_{i=j}^k\eta_i(t)$ has a one-sided power spectral density of the form
\begin{equation}
S(f) = \sum_{i=j}^k\frac{p_{i} f_i}{\pi(f_i^2+f^2)} \ .
\label{eq:psd}
\end{equation}
In general, varying the relative OU noise powers $p_{i}$ and characteristic frequencies $f_{i}$ can be used to approximate a $1/f^{\alpha}$ power spectral density over a desired frequency band for $\alpha \in [0,2]$. Note that we report frequencies in true frequency rather than angular frequency. In this work we will consider the case where all OU components have the same noise powers $p_{i} = p$, so that we obtain approximately $1/f$ behavior between the low and high-frequency components $f_j$ and $f_k$ that we refer to as IR and UV cutoffs, respectively. In our analysis here, we typically include one OU process per decade.

A convenience of using OU processes is the ability to easily ``fast-forward'' the noise over long windows of time while preserving temporal correlations, such as during state preparation and measurement (SPAM) or idling periods, as we discuss in Appendix~\ref{sec:noise_propagation}. The assumption of Gaussian noise is not essential for our analysis, as it could be extended straightforwardly to include non-Gaussian noise components such as two-level fluctuators \cite{Kirton1989,Muller2015,Muller2019}.

\begin{figure}
    \centering
    \includegraphics[width=\linewidth]{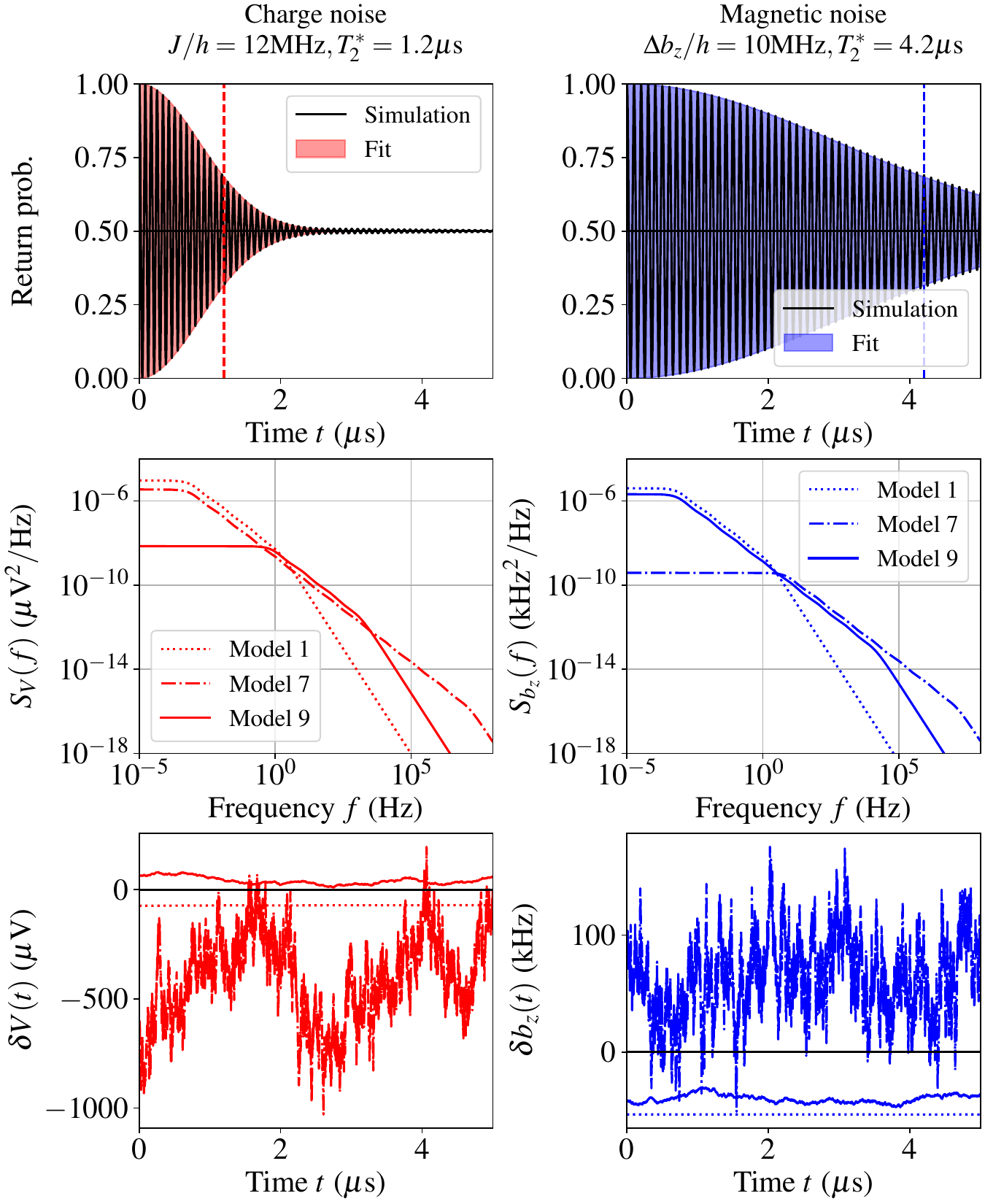}
    \caption{ \textbf{Models with equivalent charge and magnetic $T_2^{*}$ times.} Various charge (red/left column) and magnetic (blue/right column) noise models of the form $S(f)$~\eqref{eq:psd}, described in Table~\ref{Tab:model_data}. All three charge noise models have $T_2^*=1.2\us$ given exchange driven at $12\MHz$ with sensitivity parameter $\mathcal{I}=18m\V$, and all three magnetic noise models have $T_2^*=4.2\us$. First row: Ramsey oscillations averaged over $10^4$ noise realizations; a curve $\overline{P(...)}$ for each model is plotted but differences between them are negligible. We also show the Gaussian fit $\frac12\pr{1+e^{-(t/T_2^*)^2}}$ (shaded red/blue) that estimates the $T_2^*$ (dashed line). 
    Second row: Power spectral densities $S(f)$ for all noise models shown on log-log axes.
    Third row: Example time traces for each noise model.}
    \label{fig:sample_noise_plot}
\end{figure}

To mock up an ``experimentally informed'' noise model for our singlet-triplet qubit, we identify noise powers and frequency ranges $(f_j, f_k)$ that match free induction decay (FID) experiments.  We consider two types of FID experiments in order to characterize the magnetic and charge noise:
\begin{enumerate}
\item To characterize the charge noise, the global magnetic field is turned off while keeping the exchange interaction on, and the qubit is prepared in the superposition state $\vert + \rangle = \frac{1}{\sqrt{2}} \left( \ket{0} + \ket{1} \right)$.
The qubit is then allowed to precess freely along the $z$-axis, with a noisy frequency $J(t)/h=J_0e^{V_{\rm FID}/\mathcal{I}+\delta V/I}=(J_{\rm FID}+\delta J(t))/h$ and $\Delta b_z(t)=0$, where for simplicity we have turned off the magnetic noise.
In the absence of charge noise ($\delta J(t)=0$), the probability of returning to the initial state is given by $P_{\rm FID}({+|+,(V_{\rm FID},0),t})=\cos^2 \left( J_{\rm FID}t/2\hbar \right)$.
In the presence of charge noise, there will be a decay envelope around the oscillation, the shape and scale of which is dependent on the noise. For the noise model given in Eq.~\eqref{eq:psd}, the probability of reading the input state at time $t$ averaged over all noise realizations can be expressed as
\begin{eqnarray}
\langle P_{\rm FID}\pr{+|+,(V(t),0),t} \rangle &=& \nonumber \\
&& \hspace{-2cm} \frac{1}{2} \left(1+e^{-\sigma^2_{\delta\theta_V}(t)/2} \cos(J_{\rm FID} t / \hbar) \right) \ ,
\label{eq:fid_analytic_1}
\end{eqnarray}
where
\begin{eqnarray}
\sigma^2_{\delta\theta_V}(t) &=& \nonumber \\
&& \hspace{-1cm} \left(\frac{J_{\rm FID}}{\mathcal{I}} \right)^2\sum_{i}\frac{p_V\pr{e^{-2\pi f_it}+2\pi f_it-1}}{(hf_i)^2}  \ ,
\end{eqnarray}
and where we have made the approximation $e^{\delta V(t)/\mathcal{I}}\approx1+\delta V(t) / \mathcal{I}$.  Further details of this derivation are given in Appendix~\ref{sec:analytical_results_for_free_induction_decay}.
The decay term in Eq.~\eqref{eq:fid_analytic_1} is approximately Gaussian in $t$ for $f_i t \ll 1$, and consequently $T_2^*$ may be estimated by fitting the envelope of the finite-sample average of the return probability $\expect{P_{\rm FID}\pr{+|+,(V(t),0),t}}$ to $\frac12\pr{1\pm e^{-(t/T_2^*)^2}}$. The charge noise power $p_V$ is then chosen such that this estimated $T_2^*$ is approximately equal to its experimental analogue. We find that a charge $T_2^*$ time of $1.2\us$ corresponds to charge noise powers of roughly $p_V\sim (150\uV)^2$. Exact values for this noise power depend on the chosen set of constituent process frequencies.

\item To characterize the magnetic noise, the exchange interaction is turned off while the global magnetic field is applied. The qubit is prepared in the $\ket{0}$ state and allowed to precess freely around the $x$-axis. The resulting Ramsey oscillations are then described by
\begin{eqnarray}
\expect{P_{\rm FID}\pr{0|0,(0,\Delta b_z(t)),t}} &=& \nonumber \\
&& \hspace{-2cm} \frac12 \left(1+e^{-\sigma^2_{\delta\theta_{b_z}}(t)/2}\cos(\Delta b_z t/\hbar) \right) \ ,
\label{eq:fid_analytic_2}
\end{eqnarray}
with
\beq
\sigma^2_{\delta\theta_{b_z}}(t)=\sum_{i}\frac{p_{b_z}\pr{e^{-2\pi f_it}+2\pi f_it-1}}{(hf_i)^2} \ .
\eeq
Typical magnetic noise powers are then estimated in a fashion similar to the charge noise powers. For a magnetic $T_2^\ast$ time of about $4.2 \mu s$, we find $p_{b_z}/h^2\sim(30\kHz)^2$.
\end{enumerate}
In Table~\ref{Tab:model_data}, we give the parameters for ten noise models that are consistent with these constraints. In Fig.~\ref{fig:sample_noise_plot}, we show three charge noise models each with $T_2^*=1.2\us$ and three magnetic noise models each with $T_2^*=4.2\us$. While we only use magnetic and charge $T_2^\ast$ times to constrain our noise models for illustration purposes,  additional noise spectroscopy data, when available, could be used to further constrain the candidate noise models.

\begin{table}[h]
\scalebox{0.8}{\begin{tabular}{|c|c|c|c|c|c|}
\hline
\textrm{\textbf{Model}}& $\sqrt{p_V}$ \phantom i($\uV$)&$f_{V,\mathrm{IR}},f_{V,\mathrm{UV}}$ \phantom i($\Hz$)&$\sqrt{p_{b_z}}/h$ \phantom i($\kHz$)&$f_{b_z,\mathrm{IR}},f_{b_z,\mathrm{UV}}$ \phantom i($\Hz$)\\
\hline
 1 & {$201$}& $10^{-3},10^0$ & {$37.8$} & $10^{-3},10^0$\\
 2 & {$201$}& $10^{-3},10^0$ & $26.9$  & $10^{-3},10^4$\\
 3 & {$142$}& $10^{-3},10^4$ & {$37.8$}& $10^{-3},10^0$\\
 4 & {$142$}& $10^{-3},10^4$ & $26.9$ & $10^{-3},10^4$\\
 5 & {$201$}& $10^{-3},10^0$ & {$37.9$}& $10^0,10^3$\\
 6 & {$142$}& $10^{-3},10^4$ & {$37.9$}& $10^0,10^3$\\
 7 & {$134$}& $10^{-3},10^7$ & {$35.8$}& $10^1,10^7$\\
 8 & {$201$}& $10^0,10^3$    & {$37.8$}& $10^{-3},10^0$\\
 9 & {$201$}& $10^0,10^3$    & $26.9$ & $10^{-3},10^4$\\
 10 &{$201$} & $10^0,10^3$ & {$37.9$} & $10^{0},10^3$\\
\hline
\end{tabular}}
\caption{\textbf{Noise models with equivalent $T_{2}^{*}$.} The noise models associated with charge noise $T_2^*=1.2\us$ and magnetic noise $T_2^*=4.2\us$, used in Sec.~\ref{sec:model_validation_results}. Both charge and magnetic noise are turned on, with each OU process having power $p_V$ (charge noise) or $p_{b_z}$ (magnetic noise). $f_{\cdot/\mathrm{IR}},f_{\cdot/\mathrm{UV}}$ refer to the chosen IR and UV frequencies respectively, with one process per decade between these frequencies ($f_{i+1}=10f_i$).}
\label{Tab:model_data}
\end{table}

\subsection{Randomized benchmarking}\label{sec:rb_outline}
\begin{figure*}[ht]
    \centering
    \includegraphics[width=\linewidth]{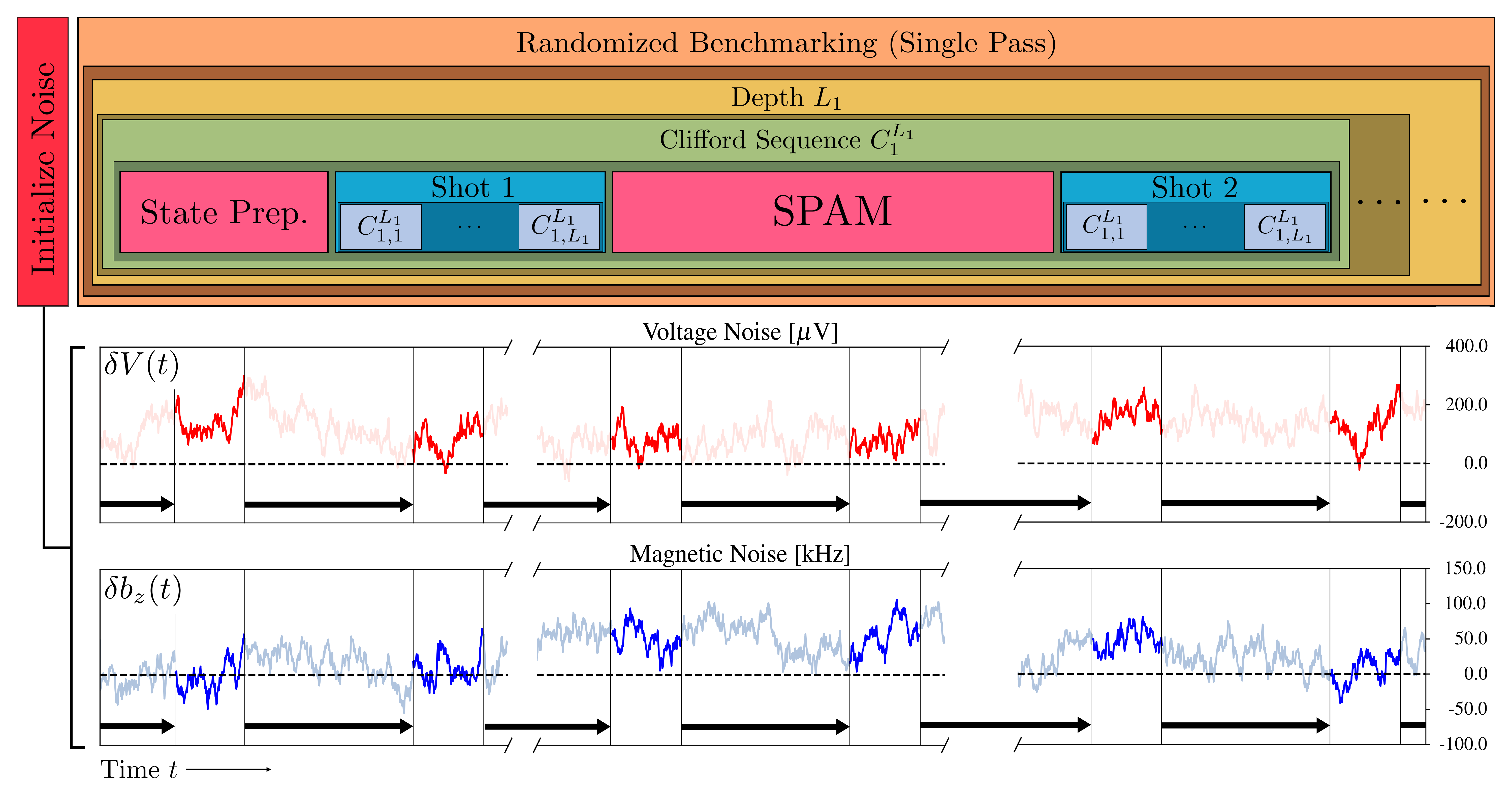}
    \caption{\textbf{Wall clock simulation schedule.} A visual outline of the randomized benchmarking procedure described in Sec.~\ref{sec:rb_outline}. A single noise trajectory $\Delta=(V(t),\Delta b_z(t))$ is initialized at time $t=0$, and persists across several successive RB passes. Each circuit is separated by SPAM, which we fast-forward over in our noise processes. This is indicated in the visual for the noise trajectory -- the noise acting on the qubit during our circuit run times is shown as dark red/blue, and we fast forward (see: arrow) through the noise during SPAM windows (light red/blue). Multiple RB passes may be done in sequence while maintaining this single noise trajectory, separated by arbitrarily long idling periods. Depicted SPAM times are not to scale.}
    \label{fig:rb_schematic}
\end{figure*}
RB~\cite{Knill2008,Magesan2011,Magesan2012b} is a popular experimental protocol for estimating average Clifford error rates. RB consists of applying multiple depth-$L$ Clifford circuits that correspond to an identity operation, taking the system back to its initial state in the absence of error.  We focus on the case of single-qubit RB.  The qubit is initialized in the state $\ket{0}$, and for each depth-$L$ circuit the first $L-1$ gates $\left\{ C_{i,k_j} \right\}_{j=1}^{L-1}$ are picked randomly from the Clifford quotient group $\mathbf{C}_1$.  The composition of these unitaries, $\tilde{U}_i$, is another element of the group:
\beq
\tilde{U}_i=\prod_{j=1}^{L-1}C_{i,k_j}, \quad C_{i,k_j} \in \mathbf{C}_1 \ ,
\eeq
and the $L$-th gate is taken to be $\tilde{U}_i^\dagger$.

In the ideal case, the probability of return to the initial state, or the survival probability, is one. 
In the presence of noise, we instead have $P_{\tilde{U}_i}(0|0,\Delta,t_i)\leq1$ with $\Delta\equiv(V(t), \Delta b_z(t))$ denoting a specific noise trajectory. By averaging over multiple circuits $\left\{ \tilde{U}_i \right\}_i$, we can calculate the average survival probability $P_{\mathrm{avg}}$ as a function of sequence depth $L$, which is expected to decay exponentially for Markovian noise:
\beq \label{eqt:RBfit}
P_{\mathrm{avg}} \sim \frac12+\frac12(1-2r)^L \ , 
\eeq
where we have neglected SPAM error.
The value $r \in [0, 1/2]$ is the RB error rate and is often interpreted as an average per-Clifford gate error~\cite{Magesan2012b,Proctor2017}. For temporally correlated noise, we should expect some deviation from the purely exponential fit in Eq.~\eqref{eqt:RBfit}~\cite{Ball2016,Fogarty2015,Qi2021}.  We show examples of our fits in Appendix~\ref{sec:rb_fit_quality}.

For simplicity, we have assumed no SPAM error as part of our noise models. Including a constant SPAM error should only change the visibility of the circuit survival probabilities and not affect the RB number, which is one of the key advantages of using RB. In principle, we could consider a drifting SPAM error as part of our noise model, which could be extracted from the RB protocol in addition to the RB number.

To mimic an experimental realization of RB, we perform ``wall-clock'' simulations of RB, where we maintain our noise trajectory $\Delta \equiv (V(t), \Delta b_z(t))$ over the entire duration of the simulation, ``fast-forwarding'' the noise over long SPAM/idling windows (see Appendix~\ref{sec:noise_propagation}). This means that we are maintaining the correlations in our noise over the entire RB protocol. We perform a single measurement for each of $10$ circuits per  depth $L$, and we take $L \in \left\{2, 4, 8, 16, ..., 512 \right\}$. We perform a total of 100 successive passes. We choose state preparation and measurement times to be $50\us$ in our simulations. We show a visual outline of the wall-clock RB procedure in Fig.~\ref{fig:rb_schematic}. The simulated laboratory time after 100 passes is approximately $0.99\s$. For each noise model, we perform 100 independent wall-clock simulations with different noise trajectories, corresponding to different random number seeds. 

While we have focused on single qubit RB in our study, the extension to multiqubit RB~\cite{Magesan2011} is straightforward. However, implementing a random $n$-qubit Clifford operation in terms of a universal gate set of 1- and 2-qubit gates becomes significantly more involved~\cite{Bravyi2021}, resulting in much longer circuits for a single Clifford operation. This is likely why experimental demonstrations of multiqubit RB have been limited to $n \leq 3$ \cite{McKay2019} so far.  For this reason, alternative methods, motivated by RB, have been proposed, such as in Refs.~\cite{Proctor2019,Hines2023}. We expect drift to be relevant for performance in these other benchmarks as well, and we expect our methodology to be equally applicable to them.

\section{Model validation} \label{sec:statistical_methods}
\subsection{Motivation and description}
We now address the question of how we can validate different noise models in the presence of drift. Given some set of of spectroscopic data, a set of noise models is identified that is consistent with this data. In our case, we treat one of the 10 noise models identified in Table~\ref{Tab:model_data} as our ``experiment'' or ``reference,'' and the remaining 9 models as noise models that reproduce consistent magnetic and charge $T_2^\ast$ values. We generate wall-clock RB results for our experiment and noise models, and our goal is to identify which noise models are consistent with the experimental data.

In what follows, we use different aspects of the RB data. 
If multiple passes of RB are performed, each pass provides an $r$ value, in contrast with averaging over all passes and reporting a single $r$ value. The $r$ value drifts from pass to pass due to the slow frequency components of the noise, giving a distribution of $r$ values (see Appendix~\ref{sec:rb_fit_quality} for an example). Our aim is to use statistical tests on the distribution of $r$ values in order to validate or invalidate our noise models.

A more fine-grained comparison is to consider the drift of the per-circuit survival probability. In our RB protocol, each pass runs the same set of circuits, so the per-circuit survival probability for a given circuit also drifts from pass to pass.  We can therefore also use the distribution of circuit survival probabilities of a given circuit to probe consistency between model and experiment.

\subsection{Statistical methods}

To compare two datasets, we use the two-sample Kolmogorov-Smirnov (K-S) statistic, which considers the largest difference in the cumulative distributions of two datasets. Denoting $X_{js}$ as the dataset of metric $X$ generated by model $j$ using a random number seed $s$, the K-S statistic between two datasets is given by $D_{X, jj', ss'}$. We give details for calculating this quantity in Appendix~\ref{app:KS}. 

The expectation is that $D_{X, jj', ss'}$ should be small when $j\sim j'$ ($\sim$ in this context means that the corresponding datasets are generated under the same noise model). We explicitly choose $s\neq s'$ when $j=j'$ such that we do not encounter $D_{X, jj', ss'}=0$ when $j=j'$ since this result provides no useful information, as we would be comparing identical data.

Conversely, large $D_{X, jj', ss'}$ indicates that the two datasets associated with $j, j'$ are unlikely to have been generated under the same noise model, and consequently we reject $j\sim j'$. The choice of the rejection threshold $d_{X,j}$ is a significant consideration in K-S testing, as an ideal test should have $D_{X, jj, ss'}\leq d_{X,j}$ for all $j$ and $D_{X, jj', ss'}>d_{X,j}$ for all $j\not\sim j'$; however, we anticipate some degree of test error in most settings and consequently must consider how to quantify them.

We primarily consider type I and II error rates in quantifying test accuracy. A type I (false negative) error occurs when the test rejects the hypothesis that two datasets are generated by the same noise model when they actually are, i.e., when $D_{X, jj, ss'}>d_{X,j}$. We label type I error rates as
\beq
\alpha_{X,j}\equiv P(D_{X, jj', ss'}>d_{X,j}|j\sim j') \ .
\eeq
A large type I error rate $\alpha_{X,j}$ indicates that data generated under model $j$ lacks in self-consistency, and that more noise samples may need to be taken and/or the scheduling of the experiment should change to more comprehensively capture the effects of the noise processes on the data (for more details see Appendix~\ref{sec:test_error_tradeoffs}). We enforce that all our models have the same level of self-consistency by choosing each rejection threshold $d_{X,j}$ such that $\alpha_{X,j}=\alpha_{X}$ is constant over $j$. This may be done by choosing $d_{X,j}$ to be the $p_X$-th percentile of $\sr{D_{X, jj, ss'}}_{s<s'}$. We then have constant type I error rates $\alpha_{X,j}=1-p_X/100$ for all models $j$. We use these choices of $d_{X,j}$ in our computations of type II error rates next.

Type II (false positive) errors occur when the test fails to reject the hypothesis (that models $j, j'$ are generated under the same noise model) when they are \textit{not} generated under the same noise model. That is, we have type II errors when $D_{X,jj',ss'}\leq d_{X,j}$ given $j\not\sim j'$. Here we specifically have an asymmetry in the rejection criterion in $j,j'$. This is because one of the models must be chosen as our reference model (the experiment), and the other as the model tested against. Here we label $j$ as the reference model ($d_{X,j}$ being the rejection threshold corresponding to this model), and $j'$ as the tested model. We label type II error rates as $\beta_{X,jj'}\equiv P(D_{X,jj',ss'}\leq d_{X,j}|j\not\sim j')$. Reference models with low type II error rates are more powerful discriminators in that they are able to more reliably discriminate models.

By choosing several models $j \in [n]$ and a metric $X$, we may concisely show type I/II errors in an $n\times n$ matrix with type I error rates $\alpha_{X,j}$ along the diagonal and type II error rates $\beta_{X,jj'}$ as off-diagonal elements. We call this our  empirical error rate grid. We demonstrate this visualization with our ten models from Table~\ref{Tab:model_data} in the next section.

\subsection{Results}\label{sec:model_validation_results}
%
\subsubsection{RB number test}\label{sec:rb_number_test}
\begin{figure}
    \centering
    \includegraphics[width=0.7\linewidth]{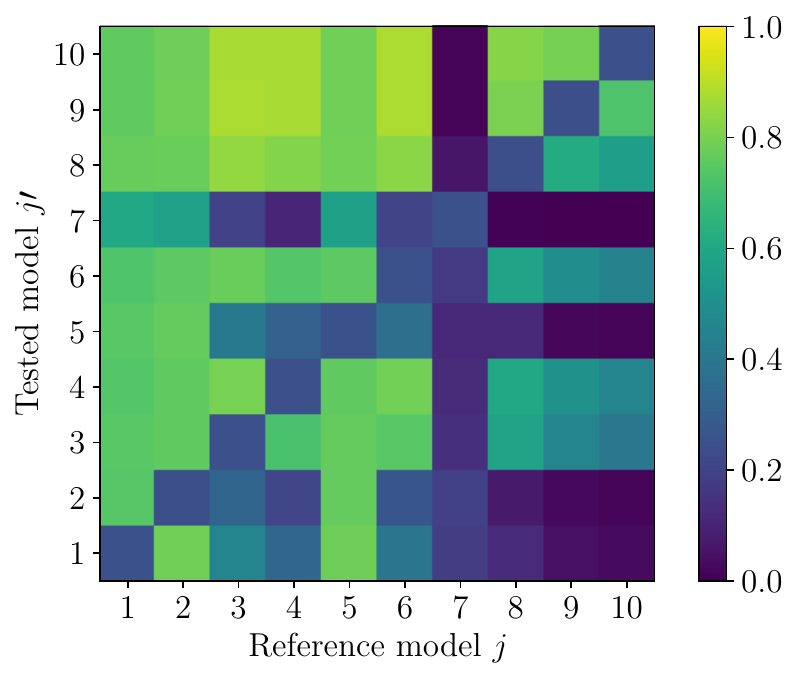}
    \caption{\textbf{Model validation test error rates.} The model comparison test error rate grid (with the figure of merit $X=r$ as the RB number). Along the diagonal are empirical type I error rates ($\alpha_X)$, and along the off-diagonal are the type II error rates ($\beta_{X,jj'}$); here we have 100 seeds per noise model. Columns with darker boxes than the diagonal boxes correspond to more powerful model discriminators.}
    \label{fig:ks_test_error_rates}
\end{figure}

The first metric we use in our model validation scheme outlined in Sec.~\ref{sec:statistical_methods} is the RB error rate $r$.
We define our rejection threshold for our K-S test to be $p_X=75$. We show the empirical error rate grid for the $X=r$ test in Fig.~\ref{fig:ks_test_error_rates}. Since these error rates are computed using a finite number of seeds, they are estimates of the true values. For example, models for which we perform 100 self-tests have an empirical type I error rate \emph{around} 0.25 instead of our chosen value $0.25$. 

A significant feature of the results in Fig.~\ref{fig:ks_test_error_rates} is that the type II error rates are particularly sensitive to the choice of IR and UV frequencies, especially for the charge noise. The results organize themselves according to grouping the models based on their charge noise frequencies: (a) Models 1, 2, and 5 with frequency range $\left[10^{-3}, 10^{0} \right]$Hz, (b) Models 3, 4, and 6 with frequency range $\left[10^{-3},10^{4} \right]$Hz, (c) Models 8, 9, and 10 with frequency range $[10^0, 10^3]$Hz, and (d) Model 7 with frequency range $[10^{-3}, 10^{7}]$Hz. We make the following observations based on this grouping:
\begin{itemize}
    \item Models 8, 9, and 10 can rule out the other models (Models 1-7) that are dominated by much lower frequency components.
    \item Models 3,4, and 6 can rule out Models 1,2,5, which have only low frequency components.
    \item Model 7 is a powerful discriminator because it has an error rate much higher than the other noise models. This is because it has noise components with high-frequency components $[10^4, 10^7]$, which is on the timescale of multiple gates and gives rise to higher coherent error rates.
    \item Models 1,2, and 5 have the largest threshold values, and this seems to give them little discriminating power.  These models only have low frequency components, which together suggests that even with 100 independent simulations (noise realizations), there is a lot of variability.  
    \item Models with similar high-frequency components are not able to discriminate between each other.  For example, Models 3, 4, and 6 have a harder time discriminating Models 8,9, and 10, and vice versa.
\end{itemize}  

We have also considered the pass-to-pass change in the RB number $X=\Delta r \equiv r_{i+1}-r_i$ as our metric, but we have largely seen that using the raw RB number results in a more powerful test.

\subsubsection{Per-circuit error test}\label{sec:per_circuit_error_test}
\begin{figure*}[h!t!]
    \centering
    \subfigure[Best]{\includegraphics[height=0.3\linewidth]{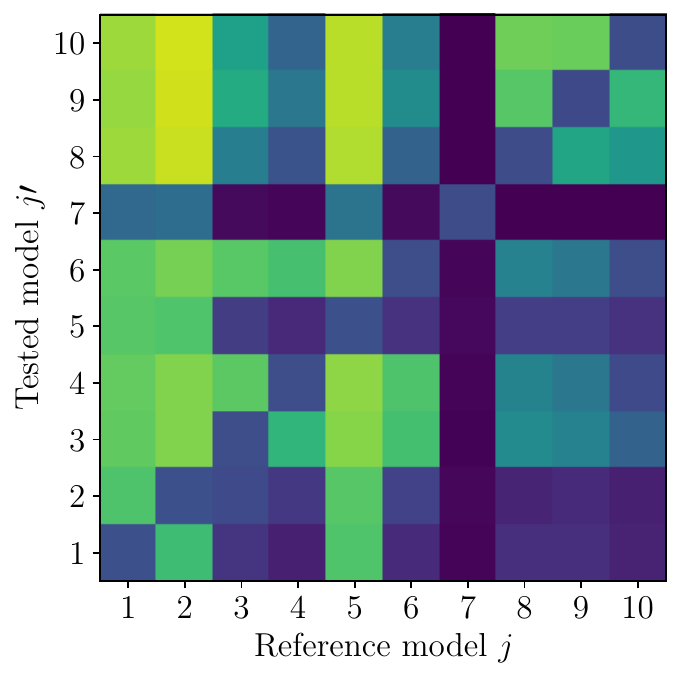}}
    \subfigure[Median]{\includegraphics[height=0.3\linewidth]{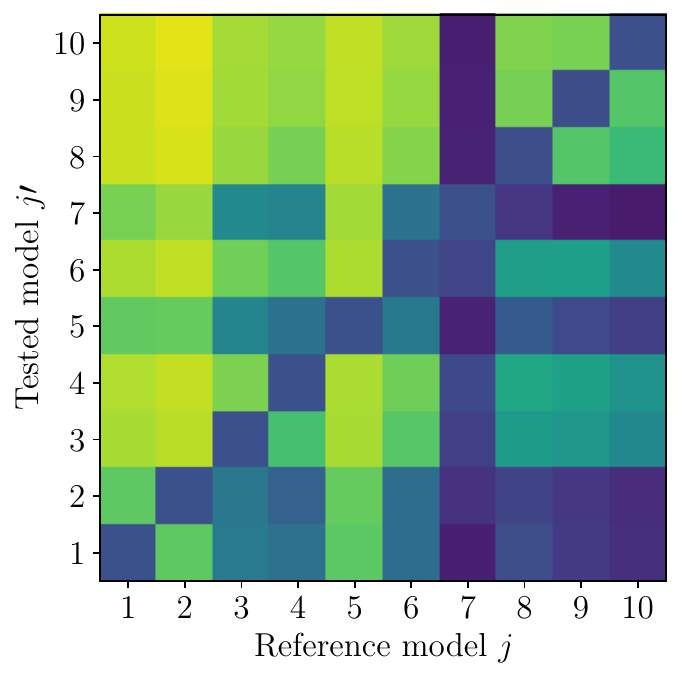}}
    \subfigure[Worst]{\includegraphics[height=0.3\linewidth]{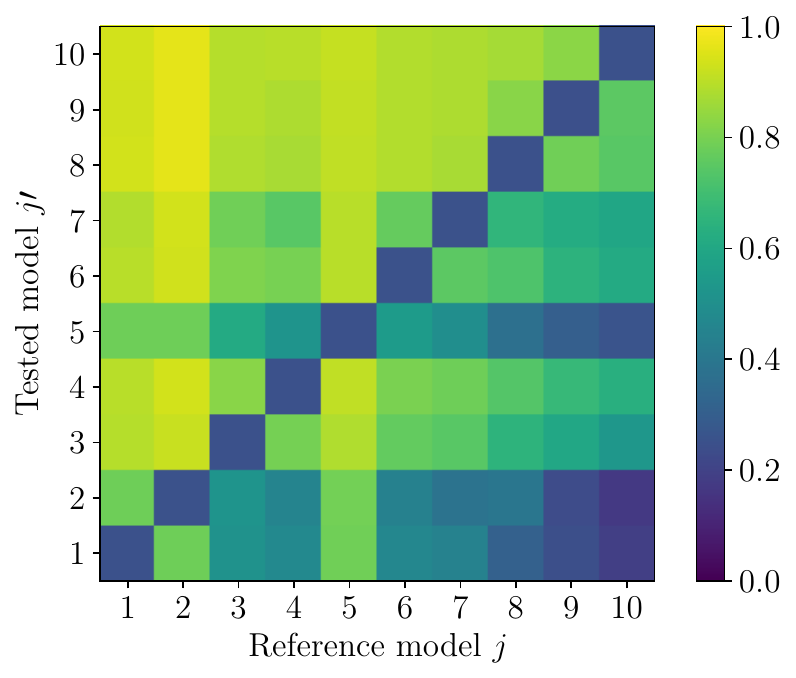}}
    \caption{\textbf{Per-circuit model validation test error rates.} The best-case, median-case, and worst-case error rates for the per-circuit model validation test outlined in Sec.~\ref{sec:per_circuit_error_test}. More specifically, each off-diagonal cell shown above is defined as the 0th, 50th, or 100th percentile of the set $\sr{\beta_{P_{U_i}(1|0,\Delta, t_i),jj'}|U_i\textrm{ with depth }L=256}$ respectively, and similarly for diagonals $\pr{\sr{\alpha_{P_{U_i}(1|0,\Delta, t_i),j}|U_i\textrm{ with depth }L=256}}$.}
    \label{fig:per_circuit_model_validation_256}
\end{figure*}
In addition to performing model validation using the RB number, we also consider the per-circuit bitflip probabilities $X=P_{U_i}(1|0,\Delta, t_i)$ for a circuit $U_i$. Using the same data generated by the 10 noise models from the previous section, we consider the K-S statistic for the per-circuit bitflip probability. We denote this by $D_{P_{U_i}(1|0,\Delta, t_i),jj',ss'}$ for models $j,j'$ with seeds $s,s'$, and we use a threshold $d_{P_{U_i}(1|0,\Delta, r_i),j}$ computed using $p_{X}=75$. 
We showcase the worst-case, median-case, and best-case type I and II error rates in Fig.~\ref{fig:per_circuit_model_validation_256} using the 10 circuits of depth $256$. In these results, we see many of the same trends as were in the RB number test results, especially for the best and median cases. 

\section{Error attribution}
\label{sec:Error_attribution}
In general, multiple sources of noise contribute to qubit error simultaneously. For example, in the context of the singlet-triplet qubit, classical Hamiltonian noise enters via the charge and magnetic noise terms, $\delta V(t)$ and $\delta b_z(t)$ respectively. After identifying a noise model that is consistent with the reference data, the noise model can be used to identify the contribution of different error sources to the total error. 
We consider two types of noise partitioning for error attribution: by separating contributions from charge ($Z$) or magnetic ($X$) noise and by separating contributions from independent OU processes.

\subsection{Breakdown of average gate infidelities into single-axis contributions}\label{sec:error_attribution_by_axis}

\begin{figure}[h!]
    \centering
    \subfigure[]{\includegraphics[width=\linewidth]{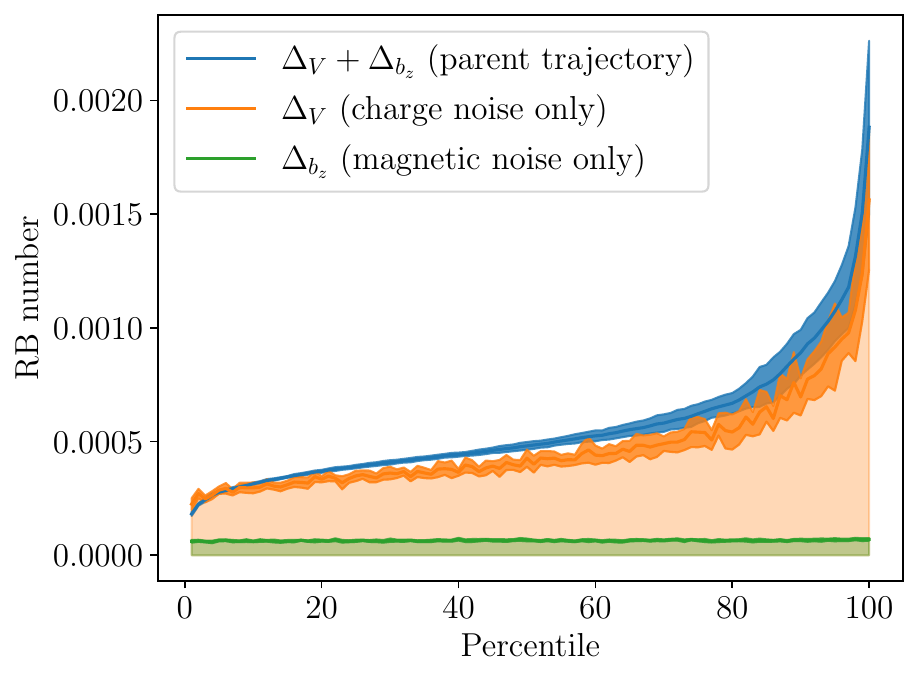}\label{subfig:ChargeMag}
    }
    \subfigure[]{\includegraphics[width=\linewidth]{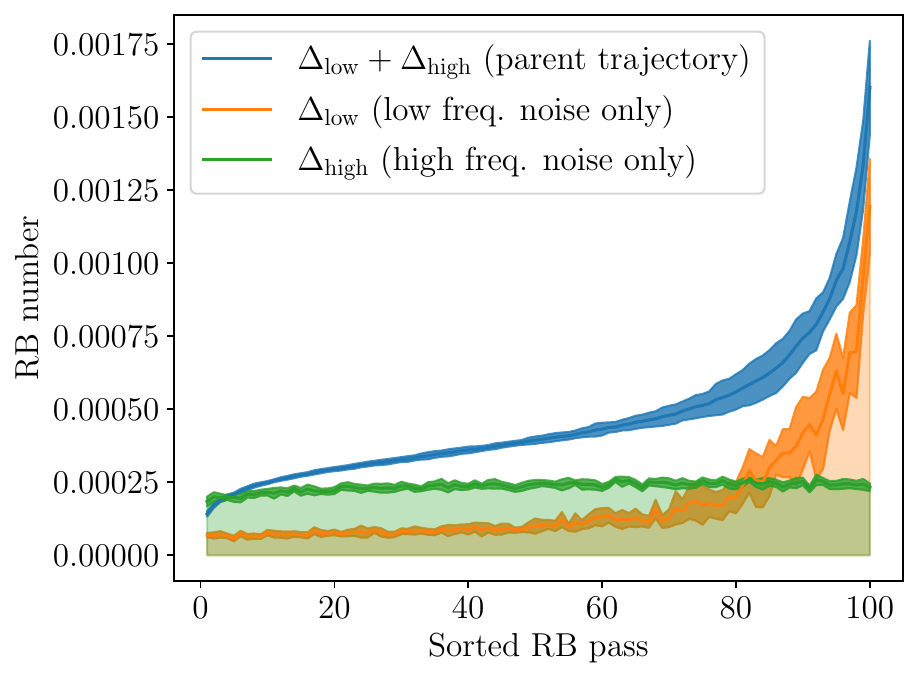}\label{subfig:LowHigh}}
    \caption{\textbf{RB error attribution for Model 7.} For 100 independent noise realizations, 100 RB passes are performed.  For each noise realization, the RB numbers from the 100 passes are ordered according to the parent trajectory. This same ordering is used for the split trajectories, where in (a) we split the noise into its different axes and in (b) we consider only charge noise and partition it into low and high-frequency components according to Eq.~\eqref{eq:partition}. The solid lines are the mean of the median of these sorted RB numbers, and the dark shaded regions are the 95\% confidence intervals.  Both are computed using a bootstrap over the 100 noise realizations.} \label{fig:rb_number_error_attribution}
    
\end{figure}

To identify the contributions of charge and magnetic noise to the infidelity, we perform wall-clock RB simulations under the following noise trajectories:
\begin{equation}
\begin{split}
    \Delta\in\{\overbrace{(\delta V(t),\delta b_z(t))}^{\Delta_V+\Delta_{b_z}},\overbrace{(\delta V(t),0)}^{\Delta_V},\overbrace{(0,\delta b_z(t))}^{\Delta_{b_z}}\} \ ,
\end{split}
\end{equation}
corresponding to having both sources of noise on, only charge noise, and only magnetic noise respectively. We emphasize that the OU processes in each case are identical, meaning the magnetic noise in $\Delta_{b_z}$ is identical to that in $\Delta_V + \Delta_{b_z}$. We show in Fig.~\ref{subfig:ChargeMag}  the error attribution for the median behavior over 100 independent noise realizations. It is clear that charge noise is the dominant source of noise and gives rise to the variation in the RB pass numbers, whereas the effect of the magnetic noise is constant across the passes. Additivity of the noise seems to hold for this median behavior, meaning that $r(t,\Delta_V+\Delta_{b_z})\approx r(t,\Delta_V)+r(t,\Delta_{b_z})$, even though single noise realizations do not necessarily exhibit this behavior, as we show in Appendix~\ref{sec:single_noise_error_attribution}.  In contrast, at very low percentiles, the sum of the individual contributions gives a larger RB number. Because we treat the noise coherently, it is possible that in these rare cases one of the noise components is effectively implementing a form of dynamical decoupling~\cite{Viola1998,Viola1999} resulting in a lower RB number.

\subsection{Attributing infidelities to independent noise components}\label{sec:error_attribution_by_OU_component}
We can perform a similar analysis for error contributions from independent noise processes that define a given noise trajectory. The magnetic and charge components of the ``parent'' noise trajectory are each a sum of independent OU processes $\left\{\eta_i(t) \right\}$, with each OU process described by a power and frequency $p_i, f_i$, such that $\Delta_\mathrm{all}\equiv\sum_i\eta_i(t)$.
We create partitions $\mathcal{P}_j$ of the set of trajectories $\sr{\eta_i(t)}_i$, such that $\bigcup_{{\mathcal{P}_j}\in \mathcal{P}}{\mathcal{P}_j}=\sr{\eta_i(t)}_i$ and the different $\mathcal{P}_j$'s are pairwise disjoint. 
We then define component noise trajectories $\Delta_{\mathcal{P}_j}$ as a sum over the part $\mathcal{P}_j$, $\Delta_{\mathcal{P}_j}\equiv\sum_{k\in {\mathcal{P}_j}}\eta_k(t)$. By definition these component noise trajectories sum to the parent, as $\sum_{{\mathcal{P}_j}\in \mathcal{P}}\Delta_{\mathcal{P}_j}=\sum_{{\mathcal{P}_j}\in \mathcal{P}}\sum_{k\in {\mathcal{P}_j}}\eta_k(t)=\sum_i\eta_i(t)=\Delta_\mathrm{all}$.

For simplicity, we turn the magnetic noise off ($p_{b_z}=0$) since charge noise is the dominant source of error. We want to analyze the contribution of low and high-frequency components of the noise to the error, but we find that results can vary based on how we choose to partition the OU components. We show this sensitivity in Appendix~\ref{app:differentpartition}. Here, we choose the partitioning of our noise based on the $1/e$ point of the survival probability as a function of circuit depth for the RB protocol.  For our noise parameters, this occurs around depth $512$, and running the 10 circuits of depth 512 takes approximately $1\mathrm{ms}$.  Therefore we choose to partition our noise into a low-frequency and high-frequency part:
\begin{eqnarray}  \label{eq:partition}
    \mathcal{P}_{\mathrm{low}}&=&\sr{\eta_{10^{-3}\Hz}(t),\eta_{10^{-2}\Hz}(t),...,\eta_{10^3\Hz}(t)} \nonumber \\
    \mathcal{P}_{\mathrm{high}}&=&\sr{\eta_{10^4\Hz}(t),\eta_{10^5\Hz}(t),...,\eta_{10^7\Hz}(t)} \ .
\end{eqnarray}
In Eq.~\eqref{eq:partition}, $\eta_{f_i}(t)$ is a process with power $p_i=p_V$ and frequency $f_i$.

We show in Fig.~\ref{subfig:LowHigh} the error attribution into low- and high-frequency components.  High-frequency noise contributes to an approximately constant error level across all percentiles, while the low-frequency noise components are responsible for a large fraction of the worst-case error rates. Unlike the error attribution by type as in the previous section, here we see that the sum of the individual components is \emph{higher} at lower percentiles. 
At lowest percentiles, because the OU components may add with different signs, their sum results in cancellations resulting in a lower error for the parent trajectory.  At higher percentiles, the additivity is restored, consistent with our previous observations.  

While we have focused on performing error attribution on the RB error rate here, one may perform a similar attribution analysis of the individual circuit sensitivities to various noise frequencies. We do so in Appendix~\ref{sec:per_circuit_error_attribution}.

\section{Discussion}
\label{sec:Discussion}
In this work, we have shown how one may use the drift in measurable performance parameters, such as the RB error rate, to validate or invalidate noise models.  Our starting point is that a certain amount of noise spectroscopy has been performed, but properties of the noise model have not been uniquely constrained.  In our case, we assume that magnetic and charge $T_2^\ast$ times for a singlet-triplet qubit have been measured, with the frequency and power distribution of independent OU processes in the model tuned to match for a set of candidate noise spectra. Because of time correlations in the noise, wall-clock simulations of the RB protocol exhibit drift in the RB error rate from pass to pass, as is expected and experimentally observed. We demonstrate that comparing different distributions of the RB error rates using the K-S test can be used to rule out noise models through comparison with a reference experimental or simulated set of RB error rates.

Unsurprisingly, we observe that the power of this test depends strongly on sampling all relevant frequencies of the noise. The timescale of the RB protocol must be at least as long as the inverse of the smallest noise frequency component or, equivalently, enough independent RB passes must be performed to sample the quasistatic component of the noise. Extending the duration of the RB measurement by introducing delays between RB passes is one way to better account for low-frequency components of the noise.

In our analysis using the K-S test, we have not exploited temporal correlations such as the power spectral density of the RB error rate.  However, more sophisticated statistical tests between time series \cite{Grenander1950,Fleischer1958,Diggle1991,Kim2006,MarajZygmat2023} could likely be adopted that may provide even more stringent model validation tests. We leave exploring these possibilities for future work. %

Once confidence in a noise model has been established by passing the tests described, the noise model can then be used to perform error attribution. This should be done with caution, since additivity of the different error components does not necessarily hold, even approximately. Noise components along orthogonal axis can partially reduce the total error, and interference between different noise sources along the same axis can reduce the total error.

While we have focused on the drift of the RB error rate or of the survival probability of a fixed depth circuit within the RB protocol, we expect that our model-centric approach should apply to other metrics of qubit performance that drift. One type of drift that we have not accounted for in this work is drift in state preparation and measurement. An example of SPAM variability in the context of semiconductor qubits is drift in the voltage offset of a Coulomb blockade peak \cite{Connors2019}, which may influence the quality of readout. To incorporate this, a model for SPAM with candidate associated noise processes would be needed.

Finally, we anticipate that developing more realistic device models and validating them using more rigorous statistical tests against experimental data will facilitate understanding and, ultimately, mitigating error mechanisms affecting quantum information processors.

\begin{acknowledgments}
This work was performed, in part, at the Center for Integrated Nanotechnologies, an Office of Science User Facility operated for the U.S. Department of Energy (DOE) Office of Science.


Sandia National Laboratories is a multi-mission laboratory managed and operated by National Technology \& Engineering Solutions of Sandia, LLC (NTESS), a wholly owned subsidiary of Honeywell International Inc., for the U.S. Department of Energy’s National Nuclear Security Administration (DOE/NNSA) under contract DE-NA0003525. This written work is authored by an employee of NTESS. The employee, not NTESS, owns the right, title and interest in and to the written work and is responsible for its contents. Any subjective views or opinions that might be expressed in the written work do not necessarily represent the views of the U.S. Government. The publisher acknowledges that the U.S. Government retains a non-exclusive, paid-up, irrevocable, world-wide license to publish or reproduce the published form of this written work or allow others to do so, for U.S. Government purposes. The DOE will provide public access to results of federally sponsored research in accordance with the DOE Public Access Plan\url{https://www.energy.gov/downloads/doe-public-access-plan.}.
\end{acknowledgments}

\begin{thebibliography}{58}%
\makeatletter
\providecommand \@ifxundefined [1]{%
 \@ifx{#1\undefined}
}%
\providecommand \@ifnum [1]{%
 \ifnum #1\expandafter \@firstoftwo
 \else \expandafter \@secondoftwo
 \fi
}%
\providecommand \@ifx [1]{%
 \ifx #1\expandafter \@firstoftwo
 \else \expandafter \@secondoftwo
 \fi
}%
\providecommand \natexlab [1]{#1}%
\providecommand \enquote  [1]{``#1''}%
\providecommand \bibnamefont  [1]{#1}%
\providecommand \bibfnamefont [1]{#1}%
\providecommand \citenamefont [1]{#1}%
\providecommand \href@noop [0]{\@secondoftwo}%
\providecommand \href [0]{\begingroup \@sanitize@url \@href}%
\providecommand \@href[1]{\@@startlink{#1}\@@href}%
\providecommand \@@href[1]{\endgroup#1\@@endlink}%
\providecommand \@sanitize@url [0]{\catcode `\\12\catcode `\$12\catcode
  `\&12\catcode `\#12\catcode `\^12\catcode `\_12\catcode `\%12\relax}%
\providecommand \@@startlink[1]{}%
\providecommand \@@endlink[0]{}%
\providecommand \url  [0]{\begingroup\@sanitize@url \@url }%
\providecommand \@url [1]{\endgroup\@href {#1}{\urlprefix }}%
\providecommand \urlprefix  [0]{URL }%
\providecommand \Eprint [0]{\href }%
\providecommand \doibase [0]{https://doi.org/}%
\providecommand \selectlanguage [0]{\@gobble}%
\providecommand \bibinfo  [0]{\@secondoftwo}%
\providecommand \bibfield  [0]{\@secondoftwo}%
\providecommand \translation [1]{[#1]}%
\providecommand \BibitemOpen [0]{}%
\providecommand \bibitemStop [0]{}%
\providecommand \bibitemNoStop [0]{.\EOS\space}%
\providecommand \EOS [0]{\spacefactor3000\relax}%
\providecommand \BibitemShut  [1]{\csname bibitem#1\endcsname}%
\let\auto@bib@innerbib\@empty
\bibitem [{\citenamefont {Acharya}\ \emph {et~al.}(2023)\citenamefont
  {Acharya}, \citenamefont {Aleiner}, \citenamefont {Allen}, \citenamefont
  {Andersen}, \citenamefont {Ansmann}, \citenamefont {Arute}, \citenamefont
  {Arya}, \citenamefont {Asfaw}, \citenamefont {Atalaya}, \citenamefont
  {Babbush}, \citenamefont {Bacon}, \citenamefont {Bardin}, \citenamefont
  {Basso}, \citenamefont {Bengtsson}, \citenamefont {Boixo}, \citenamefont
  {Bortoli}, \citenamefont {Bourassa}, \citenamefont {Bovaird}, \citenamefont
  {Brill}, \citenamefont {Broughton}, \citenamefont {Buckley}, \citenamefont
  {Buell}, \citenamefont {Burger}, \citenamefont {Burkett}, \citenamefont
  {Bushnell}, \citenamefont {Chen}, \citenamefont {Chen}, \citenamefont
  {Chiaro}, \citenamefont {Cogan}, \citenamefont {Collins}, \citenamefont
  {Conner}, \citenamefont {Courtney}, \citenamefont {Crook}, \citenamefont
  {Curtin}, \citenamefont {Debroy}, \citenamefont {Del Toro~Barba},
  \citenamefont {Demura}, \citenamefont {Dunsworth}, \citenamefont {Eppens},
  \citenamefont {Erickson}, \citenamefont {Faoro}, \citenamefont {Farhi},
  \citenamefont {Fatemi}, \citenamefont {Flores~Burgos}, \citenamefont
  {Forati}, \citenamefont {Fowler}, \citenamefont {Foxen}, \citenamefont
  {Giang}, \citenamefont {Gidney}, \citenamefont {Gilboa}, \citenamefont
  {Giustina}, \citenamefont {Grajales~Dau}, \citenamefont {Gross},
  \citenamefont {Habegger}, \citenamefont {Hamilton}, \citenamefont {Harrigan},
  \citenamefont {Harrington}, \citenamefont {Higgott}, \citenamefont {Hilton},
  \citenamefont {Hoffmann}, \citenamefont {Hong}, \citenamefont {Huang},
  \citenamefont {Huff}, \citenamefont {Huggins}, \citenamefont {Ioffe},
  \citenamefont {Isakov}, \citenamefont {Iveland}, \citenamefont {Jeffrey},
  \citenamefont {Jiang}, \citenamefont {Jones}, \citenamefont {Juhas},
  \citenamefont {Kafri}, \citenamefont {Kechedzhi}, \citenamefont {Kelly},
  \citenamefont {Khattar}, \citenamefont {Khezri}, \citenamefont
  {Kieferov{\'a}}, \citenamefont {Kim}, \citenamefont {Kitaev}, \citenamefont
  {Klimov}, \citenamefont {Klots}, \citenamefont {Korotkov}, \citenamefont
  {Kostritsa}, \citenamefont {Kreikebaum}, \citenamefont {Landhuis},
  \citenamefont {Laptev}, \citenamefont {Lau}, \citenamefont {Laws},
  \citenamefont {Lee}, \citenamefont {Lee}, \citenamefont {Lester},
  \citenamefont {Lill}, \citenamefont {Liu}, \citenamefont {Locharla},
  \citenamefont {Lucero}, \citenamefont {Malone}, \citenamefont {Marshall},
  \citenamefont {Martin}, \citenamefont {McClean}, \citenamefont {McCourt},
  \citenamefont {McEwen}, \citenamefont {Megrant}, \citenamefont
  {Meurer~Costa}, \citenamefont {Mi}, \citenamefont {Miao}, \citenamefont
  {Mohseni}, \citenamefont {Montazeri}, \citenamefont {Morvan}, \citenamefont
  {Mount}, \citenamefont {Mruczkiewicz}, \citenamefont {Naaman}, \citenamefont
  {Neeley}, \citenamefont {Neill}, \citenamefont {Nersisyan}, \citenamefont
  {Neven}, \citenamefont {Newman}, \citenamefont {Ng}, \citenamefont {Nguyen},
  \citenamefont {Nguyen}, \citenamefont {Niu}, \citenamefont {O'Brien},
  \citenamefont {Opremcak}, \citenamefont {Platt}, \citenamefont {Petukhov},
  \citenamefont {Potter}, \citenamefont {Pryadko}, \citenamefont {Quintana},
  \citenamefont {Roushan}, \citenamefont {Rubin}, \citenamefont {Saei},
  \citenamefont {Sank}, \citenamefont {Sankaragomathi}, \citenamefont
  {Satzinger}, \citenamefont {Schurkus}, \citenamefont {Schuster},
  \citenamefont {Shearn}, \citenamefont {Shorter}, \citenamefont {Shvarts},
  \citenamefont {Skruzny}, \citenamefont {Smelyanskiy}, \citenamefont {Smith},
  \citenamefont {Sterling}, \citenamefont {Strain}, \citenamefont {Szalay},
  \citenamefont {Torres}, \citenamefont {Vidal}, \citenamefont {Villalonga},
  \citenamefont {Vollgraff~Heidweiller}, \citenamefont {White}, \citenamefont
  {Xing}, \citenamefont {Yao}, \citenamefont {Yeh}, \citenamefont {Yoo},
  \citenamefont {Young}, \citenamefont {Zalcman}, \citenamefont {Zhang},
  \citenamefont {Zhu},\ and\ \citenamefont {AI}}]{Google2023}%
  \BibitemOpen
  \bibfield  {author} {\bibinfo {author} {\bibfnamefont {R.}~\bibnamefont
  {Acharya}}, \bibinfo {author} {\bibfnamefont {I.}~\bibnamefont {Aleiner}},
  \bibinfo {author} {\bibfnamefont {R.}~\bibnamefont {Allen}}, \bibinfo
  {author} {\bibfnamefont {T.~I.}\ \bibnamefont {Andersen}}, \bibinfo {author}
  {\bibfnamefont {M.}~\bibnamefont {Ansmann}}, \bibinfo {author} {\bibfnamefont
  {F.}~\bibnamefont {Arute}}, \bibinfo {author} {\bibfnamefont
  {K.}~\bibnamefont {Arya}}, \bibinfo {author} {\bibfnamefont {A.}~\bibnamefont
  {Asfaw}}, \bibinfo {author} {\bibfnamefont {J.}~\bibnamefont {Atalaya}},
  \bibinfo {author} {\bibfnamefont {R.}~\bibnamefont {Babbush}}, \bibinfo
  {author} {\bibfnamefont {D.}~\bibnamefont {Bacon}}, \bibinfo {author}
  {\bibfnamefont {J.~C.}\ \bibnamefont {Bardin}}, \bibinfo {author}
  {\bibfnamefont {J.}~\bibnamefont {Basso}}, \bibinfo {author} {\bibfnamefont
  {A.}~\bibnamefont {Bengtsson}}, \bibinfo {author} {\bibfnamefont
  {S.}~\bibnamefont {Boixo}}, \bibinfo {author} {\bibfnamefont
  {G.}~\bibnamefont {Bortoli}}, \bibinfo {author} {\bibfnamefont
  {A.}~\bibnamefont {Bourassa}}, \bibinfo {author} {\bibfnamefont
  {J.}~\bibnamefont {Bovaird}}, \bibinfo {author} {\bibfnamefont
  {L.}~\bibnamefont {Brill}}, \bibinfo {author} {\bibfnamefont
  {M.}~\bibnamefont {Broughton}}, \bibinfo {author} {\bibfnamefont {B.~B.}\
  \bibnamefont {Buckley}}, \bibinfo {author} {\bibfnamefont {D.~A.}\
  \bibnamefont {Buell}}, \bibinfo {author} {\bibfnamefont {T.}~\bibnamefont
  {Burger}}, \bibinfo {author} {\bibfnamefont {B.}~\bibnamefont {Burkett}},
  \bibinfo {author} {\bibfnamefont {N.}~\bibnamefont {Bushnell}}, \bibinfo
  {author} {\bibfnamefont {Y.}~\bibnamefont {Chen}}, \bibinfo {author}
  {\bibfnamefont {Z.}~\bibnamefont {Chen}}, \bibinfo {author} {\bibfnamefont
  {B.}~\bibnamefont {Chiaro}}, \bibinfo {author} {\bibfnamefont
  {J.}~\bibnamefont {Cogan}}, \bibinfo {author} {\bibfnamefont
  {R.}~\bibnamefont {Collins}}, \bibinfo {author} {\bibfnamefont
  {P.}~\bibnamefont {Conner}}, \bibinfo {author} {\bibfnamefont
  {W.}~\bibnamefont {Courtney}}, \bibinfo {author} {\bibfnamefont {A.~L.}\
  \bibnamefont {Crook}}, \bibinfo {author} {\bibfnamefont {B.}~\bibnamefont
  {Curtin}}, \bibinfo {author} {\bibfnamefont {D.~M.}\ \bibnamefont {Debroy}},
  \bibinfo {author} {\bibfnamefont {A.}~\bibnamefont {Del Toro~Barba}},
  \bibinfo {author} {\bibfnamefont {S.}~\bibnamefont {Demura}}, \bibinfo
  {author} {\bibfnamefont {A.}~\bibnamefont {Dunsworth}}, \bibinfo {author}
  {\bibfnamefont {D.}~\bibnamefont {Eppens}}, \bibinfo {author} {\bibfnamefont
  {C.}~\bibnamefont {Erickson}}, \bibinfo {author} {\bibfnamefont
  {L.}~\bibnamefont {Faoro}}, \bibinfo {author} {\bibfnamefont
  {E.}~\bibnamefont {Farhi}}, \bibinfo {author} {\bibfnamefont
  {R.}~\bibnamefont {Fatemi}}, \bibinfo {author} {\bibfnamefont
  {L.}~\bibnamefont {Flores~Burgos}}, \bibinfo {author} {\bibfnamefont
  {E.}~\bibnamefont {Forati}}, \bibinfo {author} {\bibfnamefont {A.~G.}\
  \bibnamefont {Fowler}}, \bibinfo {author} {\bibfnamefont {B.}~\bibnamefont
  {Foxen}}, \bibinfo {author} {\bibfnamefont {W.}~\bibnamefont {Giang}},
  \bibinfo {author} {\bibfnamefont {C.}~\bibnamefont {Gidney}}, \bibinfo
  {author} {\bibfnamefont {D.}~\bibnamefont {Gilboa}}, \bibinfo {author}
  {\bibfnamefont {M.}~\bibnamefont {Giustina}}, \bibinfo {author}
  {\bibfnamefont {A.}~\bibnamefont {Grajales~Dau}}, \bibinfo {author}
  {\bibfnamefont {J.~A.}\ \bibnamefont {Gross}}, \bibinfo {author}
  {\bibfnamefont {S.}~\bibnamefont {Habegger}}, \bibinfo {author}
  {\bibfnamefont {M.~C.}\ \bibnamefont {Hamilton}}, \bibinfo {author}
  {\bibfnamefont {M.~P.}\ \bibnamefont {Harrigan}}, \bibinfo {author}
  {\bibfnamefont {S.~D.}\ \bibnamefont {Harrington}}, \bibinfo {author}
  {\bibfnamefont {O.}~\bibnamefont {Higgott}}, \bibinfo {author} {\bibfnamefont
  {J.}~\bibnamefont {Hilton}}, \bibinfo {author} {\bibfnamefont
  {M.}~\bibnamefont {Hoffmann}}, \bibinfo {author} {\bibfnamefont
  {S.}~\bibnamefont {Hong}}, \bibinfo {author} {\bibfnamefont {T.}~\bibnamefont
  {Huang}}, \bibinfo {author} {\bibfnamefont {A.}~\bibnamefont {Huff}},
  \bibinfo {author} {\bibfnamefont {W.~J.}\ \bibnamefont {Huggins}}, \bibinfo
  {author} {\bibfnamefont {L.~B.}\ \bibnamefont {Ioffe}}, \bibinfo {author}
  {\bibfnamefont {S.~V.}\ \bibnamefont {Isakov}}, \bibinfo {author}
  {\bibfnamefont {J.}~\bibnamefont {Iveland}}, \bibinfo {author} {\bibfnamefont
  {E.}~\bibnamefont {Jeffrey}}, \bibinfo {author} {\bibfnamefont
  {Z.}~\bibnamefont {Jiang}}, \bibinfo {author} {\bibfnamefont
  {C.}~\bibnamefont {Jones}}, \bibinfo {author} {\bibfnamefont
  {P.}~\bibnamefont {Juhas}}, \bibinfo {author} {\bibfnamefont
  {D.}~\bibnamefont {Kafri}}, \bibinfo {author} {\bibfnamefont
  {K.}~\bibnamefont {Kechedzhi}}, \bibinfo {author} {\bibfnamefont
  {J.}~\bibnamefont {Kelly}}, \bibinfo {author} {\bibfnamefont
  {T.}~\bibnamefont {Khattar}}, \bibinfo {author} {\bibfnamefont
  {M.}~\bibnamefont {Khezri}}, \bibinfo {author} {\bibfnamefont
  {M.}~\bibnamefont {Kieferov{\'a}}}, \bibinfo {author} {\bibfnamefont
  {S.}~\bibnamefont {Kim}}, \bibinfo {author} {\bibfnamefont {A.}~\bibnamefont
  {Kitaev}}, \bibinfo {author} {\bibfnamefont {P.~V.}\ \bibnamefont {Klimov}},
  \bibinfo {author} {\bibfnamefont {A.~R.}\ \bibnamefont {Klots}}, \bibinfo
  {author} {\bibfnamefont {A.~N.}\ \bibnamefont {Korotkov}}, \bibinfo {author}
  {\bibfnamefont {F.}~\bibnamefont {Kostritsa}}, \bibinfo {author}
  {\bibfnamefont {J.~M.}\ \bibnamefont {Kreikebaum}}, \bibinfo {author}
  {\bibfnamefont {D.}~\bibnamefont {Landhuis}}, \bibinfo {author}
  {\bibfnamefont {P.}~\bibnamefont {Laptev}}, \bibinfo {author} {\bibfnamefont
  {K.-M.}\ \bibnamefont {Lau}}, \bibinfo {author} {\bibfnamefont
  {L.}~\bibnamefont {Laws}}, \bibinfo {author} {\bibfnamefont {J.}~\bibnamefont
  {Lee}}, \bibinfo {author} {\bibfnamefont {K.}~\bibnamefont {Lee}}, \bibinfo
  {author} {\bibfnamefont {B.~J.}\ \bibnamefont {Lester}}, \bibinfo {author}
  {\bibfnamefont {A.}~\bibnamefont {Lill}}, \bibinfo {author} {\bibfnamefont
  {W.}~\bibnamefont {Liu}}, \bibinfo {author} {\bibfnamefont {A.}~\bibnamefont
  {Locharla}}, \bibinfo {author} {\bibfnamefont {E.}~\bibnamefont {Lucero}},
  \bibinfo {author} {\bibfnamefont {F.~D.}\ \bibnamefont {Malone}}, \bibinfo
  {author} {\bibfnamefont {J.}~\bibnamefont {Marshall}}, \bibinfo {author}
  {\bibfnamefont {O.}~\bibnamefont {Martin}}, \bibinfo {author} {\bibfnamefont
  {J.~R.}\ \bibnamefont {McClean}}, \bibinfo {author} {\bibfnamefont
  {T.}~\bibnamefont {McCourt}}, \bibinfo {author} {\bibfnamefont
  {M.}~\bibnamefont {McEwen}}, \bibinfo {author} {\bibfnamefont
  {A.}~\bibnamefont {Megrant}}, \bibinfo {author} {\bibfnamefont
  {B.}~\bibnamefont {Meurer~Costa}}, \bibinfo {author} {\bibfnamefont
  {X.}~\bibnamefont {Mi}}, \bibinfo {author} {\bibfnamefont {K.~C.}\
  \bibnamefont {Miao}}, \bibinfo {author} {\bibfnamefont {M.}~\bibnamefont
  {Mohseni}}, \bibinfo {author} {\bibfnamefont {S.}~\bibnamefont {Montazeri}},
  \bibinfo {author} {\bibfnamefont {A.}~\bibnamefont {Morvan}}, \bibinfo
  {author} {\bibfnamefont {E.}~\bibnamefont {Mount}}, \bibinfo {author}
  {\bibfnamefont {W.}~\bibnamefont {Mruczkiewicz}}, \bibinfo {author}
  {\bibfnamefont {O.}~\bibnamefont {Naaman}}, \bibinfo {author} {\bibfnamefont
  {M.}~\bibnamefont {Neeley}}, \bibinfo {author} {\bibfnamefont
  {C.}~\bibnamefont {Neill}}, \bibinfo {author} {\bibfnamefont
  {A.}~\bibnamefont {Nersisyan}}, \bibinfo {author} {\bibfnamefont
  {H.}~\bibnamefont {Neven}}, \bibinfo {author} {\bibfnamefont
  {M.}~\bibnamefont {Newman}}, \bibinfo {author} {\bibfnamefont {J.~H.}\
  \bibnamefont {Ng}}, \bibinfo {author} {\bibfnamefont {A.}~\bibnamefont
  {Nguyen}}, \bibinfo {author} {\bibfnamefont {M.}~\bibnamefont {Nguyen}},
  \bibinfo {author} {\bibfnamefont {M.~Y.}\ \bibnamefont {Niu}}, \bibinfo
  {author} {\bibfnamefont {T.~E.}\ \bibnamefont {O'Brien}}, \bibinfo {author}
  {\bibfnamefont {A.}~\bibnamefont {Opremcak}}, \bibinfo {author}
  {\bibfnamefont {J.}~\bibnamefont {Platt}}, \bibinfo {author} {\bibfnamefont
  {A.}~\bibnamefont {Petukhov}}, \bibinfo {author} {\bibfnamefont
  {R.}~\bibnamefont {Potter}}, \bibinfo {author} {\bibfnamefont {L.~P.}\
  \bibnamefont {Pryadko}}, \bibinfo {author} {\bibfnamefont {C.}~\bibnamefont
  {Quintana}}, \bibinfo {author} {\bibfnamefont {P.}~\bibnamefont {Roushan}},
  \bibinfo {author} {\bibfnamefont {N.~C.}\ \bibnamefont {Rubin}}, \bibinfo
  {author} {\bibfnamefont {N.}~\bibnamefont {Saei}}, \bibinfo {author}
  {\bibfnamefont {D.}~\bibnamefont {Sank}}, \bibinfo {author} {\bibfnamefont
  {K.}~\bibnamefont {Sankaragomathi}}, \bibinfo {author} {\bibfnamefont
  {K.~J.}\ \bibnamefont {Satzinger}}, \bibinfo {author} {\bibfnamefont {H.~F.}\
  \bibnamefont {Schurkus}}, \bibinfo {author} {\bibfnamefont {C.}~\bibnamefont
  {Schuster}}, \bibinfo {author} {\bibfnamefont {M.~J.}\ \bibnamefont
  {Shearn}}, \bibinfo {author} {\bibfnamefont {A.}~\bibnamefont {Shorter}},
  \bibinfo {author} {\bibfnamefont {V.}~\bibnamefont {Shvarts}}, \bibinfo
  {author} {\bibfnamefont {J.}~\bibnamefont {Skruzny}}, \bibinfo {author}
  {\bibfnamefont {V.}~\bibnamefont {Smelyanskiy}}, \bibinfo {author}
  {\bibfnamefont {W.~C.}\ \bibnamefont {Smith}}, \bibinfo {author}
  {\bibfnamefont {G.}~\bibnamefont {Sterling}}, \bibinfo {author}
  {\bibfnamefont {D.}~\bibnamefont {Strain}}, \bibinfo {author} {\bibfnamefont
  {M.}~\bibnamefont {Szalay}}, \bibinfo {author} {\bibfnamefont
  {A.}~\bibnamefont {Torres}}, \bibinfo {author} {\bibfnamefont
  {G.}~\bibnamefont {Vidal}}, \bibinfo {author} {\bibfnamefont
  {B.}~\bibnamefont {Villalonga}}, \bibinfo {author} {\bibfnamefont
  {C.}~\bibnamefont {Vollgraff~Heidweiller}}, \bibinfo {author} {\bibfnamefont
  {T.}~\bibnamefont {White}}, \bibinfo {author} {\bibfnamefont
  {C.}~\bibnamefont {Xing}}, \bibinfo {author} {\bibfnamefont {Z.~J.}\
  \bibnamefont {Yao}}, \bibinfo {author} {\bibfnamefont {P.}~\bibnamefont
  {Yeh}}, \bibinfo {author} {\bibfnamefont {J.}~\bibnamefont {Yoo}}, \bibinfo
  {author} {\bibfnamefont {G.}~\bibnamefont {Young}}, \bibinfo {author}
  {\bibfnamefont {A.}~\bibnamefont {Zalcman}}, \bibinfo {author} {\bibfnamefont
  {Y.}~\bibnamefont {Zhang}}, \bibinfo {author} {\bibfnamefont
  {N.}~\bibnamefont {Zhu}},\ and\ \bibinfo {author} {\bibfnamefont {G.~Q.}\
  \bibnamefont {AI}},\ }\bibfield  {title} {\bibinfo {title} {Suppressing
  quantum errors by scaling a surface code logical qubit},\ }\href
  {https://doi.org/10.1038/s41586-022-05434-1} {\bibfield  {journal} {\bibinfo
  {journal} {Nature}\ }\textbf {\bibinfo {volume} {614}},\ \bibinfo {pages}
  {676} (\bibinfo {year} {2023})}\BibitemShut {NoStop}%
\bibitem [{\citenamefont {Postler}\ \emph {et~al.}(2024)\citenamefont
  {Postler}, \citenamefont {Butt}, \citenamefont {Pogorelov}, \citenamefont
  {Marciniak}, \citenamefont {Heu\ss{}en}, \citenamefont {Blatt}, \citenamefont
  {Schindler}, \citenamefont {Rispler}, \citenamefont {M\"uller},\ and\
  \citenamefont {Monz}}]{Postler2023}%
  \BibitemOpen
  \bibfield  {author} {\bibinfo {author} {\bibfnamefont {L.}~\bibnamefont
  {Postler}}, \bibinfo {author} {\bibfnamefont {F.}~\bibnamefont {Butt}},
  \bibinfo {author} {\bibfnamefont {I.}~\bibnamefont {Pogorelov}}, \bibinfo
  {author} {\bibfnamefont {C.~D.}\ \bibnamefont {Marciniak}}, \bibinfo {author}
  {\bibfnamefont {S.}~\bibnamefont {Heu\ss{}en}}, \bibinfo {author}
  {\bibfnamefont {R.}~\bibnamefont {Blatt}}, \bibinfo {author} {\bibfnamefont
  {P.}~\bibnamefont {Schindler}}, \bibinfo {author} {\bibfnamefont
  {M.}~\bibnamefont {Rispler}}, \bibinfo {author} {\bibfnamefont
  {M.}~\bibnamefont {M\"uller}},\ and\ \bibinfo {author} {\bibfnamefont
  {T.}~\bibnamefont {Monz}},\ }\bibfield  {title} {\bibinfo {title}
  {Demonstration of fault-tolerant steane quantum error correction},\ }\href
  {https://doi.org/10.1103/PRXQuantum.5.030326} {\bibfield  {journal} {\bibinfo
   {journal} {PRX Quantum}\ }\textbf {\bibinfo {volume} {5}},\ \bibinfo {pages}
  {030326} (\bibinfo {year} {2024})}\BibitemShut {NoStop}%
\bibitem [{\citenamefont {Bluvstein}\ \emph {et~al.}(2024)\citenamefont
  {Bluvstein}, \citenamefont {Evered}, \citenamefont {Geim}, \citenamefont
  {Li}, \citenamefont {Zhou}, \citenamefont {Manovitz}, \citenamefont {Ebadi},
  \citenamefont {Cain}, \citenamefont {Kalinowski}, \citenamefont {Hangleiter},
  \citenamefont {Bonilla~Ataides}, \citenamefont {Maskara}, \citenamefont
  {Cong}, \citenamefont {Gao}, \citenamefont {Sales~Rodriguez}, \citenamefont
  {Karolyshyn}, \citenamefont {Semeghini}, \citenamefont {Gullans},
  \citenamefont {Greiner}, \citenamefont {Vuleti{\'c}},\ and\ \citenamefont
  {Lukin}}]{Bluvstein2024}%
  \BibitemOpen
  \bibfield  {author} {\bibinfo {author} {\bibfnamefont {D.}~\bibnamefont
  {Bluvstein}}, \bibinfo {author} {\bibfnamefont {S.~J.}\ \bibnamefont
  {Evered}}, \bibinfo {author} {\bibfnamefont {A.~A.}\ \bibnamefont {Geim}},
  \bibinfo {author} {\bibfnamefont {S.~H.}\ \bibnamefont {Li}}, \bibinfo
  {author} {\bibfnamefont {H.}~\bibnamefont {Zhou}}, \bibinfo {author}
  {\bibfnamefont {T.}~\bibnamefont {Manovitz}}, \bibinfo {author}
  {\bibfnamefont {S.}~\bibnamefont {Ebadi}}, \bibinfo {author} {\bibfnamefont
  {M.}~\bibnamefont {Cain}}, \bibinfo {author} {\bibfnamefont {M.}~\bibnamefont
  {Kalinowski}}, \bibinfo {author} {\bibfnamefont {D.}~\bibnamefont
  {Hangleiter}}, \bibinfo {author} {\bibfnamefont {J.~P.}\ \bibnamefont
  {Bonilla~Ataides}}, \bibinfo {author} {\bibfnamefont {N.}~\bibnamefont
  {Maskara}}, \bibinfo {author} {\bibfnamefont {I.}~\bibnamefont {Cong}},
  \bibinfo {author} {\bibfnamefont {X.}~\bibnamefont {Gao}}, \bibinfo {author}
  {\bibfnamefont {P.}~\bibnamefont {Sales~Rodriguez}}, \bibinfo {author}
  {\bibfnamefont {T.}~\bibnamefont {Karolyshyn}}, \bibinfo {author}
  {\bibfnamefont {G.}~\bibnamefont {Semeghini}}, \bibinfo {author}
  {\bibfnamefont {M.~J.}\ \bibnamefont {Gullans}}, \bibinfo {author}
  {\bibfnamefont {M.}~\bibnamefont {Greiner}}, \bibinfo {author} {\bibfnamefont
  {V.}~\bibnamefont {Vuleti{\'c}}},\ and\ \bibinfo {author} {\bibfnamefont
  {M.~D.}\ \bibnamefont {Lukin}},\ }\bibfield  {title} {\bibinfo {title}
  {Logical quantum processor based on reconfigurable atom arrays},\ }\href
  {https://doi.org/10.1038/s41586-023-06927-3} {\bibfield  {journal} {\bibinfo
  {journal} {Nature}\ }\textbf {\bibinfo {volume} {626}},\ \bibinfo {pages}
  {58} (\bibinfo {year} {2024})}\BibitemShut {NoStop}%
\bibitem [{\citenamefont {{da Silva}}\ \emph {et~al.}(2024)\citenamefont {{da
  Silva}}, \citenamefont {{Ryan-Anderson}}, \citenamefont {{Bello-Rivas}},
  \citenamefont {{Chernoguzov}}, \citenamefont {{Dreiling}}, \citenamefont
  {{Foltz}}, \citenamefont {{Frachon}}, \citenamefont {{Gaebler}},
  \citenamefont {{Gatterman}}, \citenamefont {{Grans-Samuelsson}},
  \citenamefont {{Hayes}}, \citenamefont {{Hewitt}}, \citenamefont
  {{Johansen}}, \citenamefont {{Lucchetti}}, \citenamefont {{Mills}},
  \citenamefont {{Moses}}, \citenamefont {{Neyenhuis}}, \citenamefont {{Paz}},
  \citenamefont {{Pino}}, \citenamefont {{Siegfried}}, \citenamefont
  {{Strabley}}, \citenamefont {{Sundaram}}, \citenamefont {{Tom}},
  \citenamefont {{Wernli}}, \citenamefont {{Zanner}}, \citenamefont {{Stutz}},\
  and\ \citenamefont {{Svore}}}]{daSilva2024}%
  \BibitemOpen
  \bibfield  {author} {\bibinfo {author} {\bibfnamefont {M.~P.}\ \bibnamefont
  {{da Silva}}}, \bibinfo {author} {\bibfnamefont {C.}~\bibnamefont
  {{Ryan-Anderson}}}, \bibinfo {author} {\bibfnamefont {J.~M.}\ \bibnamefont
  {{Bello-Rivas}}}, \bibinfo {author} {\bibfnamefont {A.}~\bibnamefont
  {{Chernoguzov}}}, \bibinfo {author} {\bibfnamefont {J.~M.}\ \bibnamefont
  {{Dreiling}}}, \bibinfo {author} {\bibfnamefont {C.}~\bibnamefont {{Foltz}}},
  \bibinfo {author} {\bibfnamefont {F.}~\bibnamefont {{Frachon}}}, \bibinfo
  {author} {\bibfnamefont {J.~P.}\ \bibnamefont {{Gaebler}}}, \bibinfo {author}
  {\bibfnamefont {T.~M.}\ \bibnamefont {{Gatterman}}}, \bibinfo {author}
  {\bibfnamefont {L.}~\bibnamefont {{Grans-Samuelsson}}}, \bibinfo {author}
  {\bibfnamefont {D.}~\bibnamefont {{Hayes}}}, \bibinfo {author} {\bibfnamefont
  {N.}~\bibnamefont {{Hewitt}}}, \bibinfo {author} {\bibfnamefont
  {J.}~\bibnamefont {{Johansen}}}, \bibinfo {author} {\bibfnamefont
  {D.}~\bibnamefont {{Lucchetti}}}, \bibinfo {author} {\bibfnamefont
  {M.}~\bibnamefont {{Mills}}}, \bibinfo {author} {\bibfnamefont {S.~A.}\
  \bibnamefont {{Moses}}}, \bibinfo {author} {\bibfnamefont {B.}~\bibnamefont
  {{Neyenhuis}}}, \bibinfo {author} {\bibfnamefont {A.}~\bibnamefont {{Paz}}},
  \bibinfo {author} {\bibfnamefont {J.}~\bibnamefont {{Pino}}}, \bibinfo
  {author} {\bibfnamefont {P.}~\bibnamefont {{Siegfried}}}, \bibinfo {author}
  {\bibfnamefont {J.}~\bibnamefont {{Strabley}}}, \bibinfo {author}
  {\bibfnamefont {A.}~\bibnamefont {{Sundaram}}}, \bibinfo {author}
  {\bibfnamefont {D.}~\bibnamefont {{Tom}}}, \bibinfo {author} {\bibfnamefont
  {S.~J.}\ \bibnamefont {{Wernli}}}, \bibinfo {author} {\bibfnamefont
  {M.}~\bibnamefont {{Zanner}}}, \bibinfo {author} {\bibfnamefont {R.~P.}\
  \bibnamefont {{Stutz}}},\ and\ \bibinfo {author} {\bibfnamefont {K.~M.}\
  \bibnamefont {{Svore}}},\ }\bibfield  {title} {\bibinfo {title}
  {{Demonstration of logical qubits and repeated error correction with
  better-than-physical error rates}},\ }\href
  {https://doi.org/10.48550/arXiv.2404.02280} {\bibfield  {journal} {\bibinfo
  {journal} {arXiv e-prints}\ ,\ \bibinfo {eid} {arXiv:2404.02280}} (\bibinfo
  {year} {2024})}\BibitemShut {NoStop}%
\bibitem [{\citenamefont {{Mayer}}\ \emph {et~al.}(2024)\citenamefont
  {{Mayer}}, \citenamefont {{Ryan-Anderson}}, \citenamefont {{Brown}},
  \citenamefont {{Durso-Sabina}}, \citenamefont {{Baldwin}}, \citenamefont
  {{Hayes}}, \citenamefont {{Dreiling}}, \citenamefont {{Foltz}}, \citenamefont
  {{Gaebler}}, \citenamefont {{Gatterman}}, \citenamefont {{Gerber}},
  \citenamefont {{Gilmore}}, \citenamefont {{Gresh}}, \citenamefont {{Hewitt}},
  \citenamefont {{Horst}}, \citenamefont {{Johansen}}, \citenamefont
  {{Mengle}}, \citenamefont {{Mills}}, \citenamefont {{Moses}}, \citenamefont
  {{Siegfried}}, \citenamefont {{Neyenhuis}}, \citenamefont {{Pino}},\ and\
  \citenamefont {{Stutz}}}]{Mayer2024}%
  \BibitemOpen
  \bibfield  {author} {\bibinfo {author} {\bibfnamefont {K.}~\bibnamefont
  {{Mayer}}}, \bibinfo {author} {\bibfnamefont {C.}~\bibnamefont
  {{Ryan-Anderson}}}, \bibinfo {author} {\bibfnamefont {N.}~\bibnamefont
  {{Brown}}}, \bibinfo {author} {\bibfnamefont {E.}~\bibnamefont
  {{Durso-Sabina}}}, \bibinfo {author} {\bibfnamefont {C.~H.}\ \bibnamefont
  {{Baldwin}}}, \bibinfo {author} {\bibfnamefont {D.}~\bibnamefont {{Hayes}}},
  \bibinfo {author} {\bibfnamefont {J.~M.}\ \bibnamefont {{Dreiling}}},
  \bibinfo {author} {\bibfnamefont {C.}~\bibnamefont {{Foltz}}}, \bibinfo
  {author} {\bibfnamefont {J.~P.}\ \bibnamefont {{Gaebler}}}, \bibinfo {author}
  {\bibfnamefont {T.~M.}\ \bibnamefont {{Gatterman}}}, \bibinfo {author}
  {\bibfnamefont {J.~A.}\ \bibnamefont {{Gerber}}}, \bibinfo {author}
  {\bibfnamefont {K.}~\bibnamefont {{Gilmore}}}, \bibinfo {author}
  {\bibfnamefont {D.}~\bibnamefont {{Gresh}}}, \bibinfo {author} {\bibfnamefont
  {N.}~\bibnamefont {{Hewitt}}}, \bibinfo {author} {\bibfnamefont {C.~V.}\
  \bibnamefont {{Horst}}}, \bibinfo {author} {\bibfnamefont {J.}~\bibnamefont
  {{Johansen}}}, \bibinfo {author} {\bibfnamefont {T.}~\bibnamefont
  {{Mengle}}}, \bibinfo {author} {\bibfnamefont {M.}~\bibnamefont {{Mills}}},
  \bibinfo {author} {\bibfnamefont {S.~A.}\ \bibnamefont {{Moses}}}, \bibinfo
  {author} {\bibfnamefont {P.~E.}\ \bibnamefont {{Siegfried}}}, \bibinfo
  {author} {\bibfnamefont {B.}~\bibnamefont {{Neyenhuis}}}, \bibinfo {author}
  {\bibfnamefont {J.}~\bibnamefont {{Pino}}},\ and\ \bibinfo {author}
  {\bibfnamefont {R.}~\bibnamefont {{Stutz}}},\ }\bibfield  {title} {\bibinfo
  {title} {{Benchmarking logical three-qubit quantum Fourier transform encoded
  in the Steane code on a trapped-ion quantum computer}},\ }\href
  {https://doi.org/10.48550/arXiv.2404.08616} {\bibfield  {journal} {\bibinfo
  {journal} {arXiv e-prints}\ ,\ \bibinfo {eid} {arXiv:2404.08616}} (\bibinfo
  {year} {2024})}\BibitemShut {NoStop}%
\bibitem [{\citenamefont {Harris}\ \emph {et~al.}(2008)\citenamefont {Harris},
  \citenamefont {Johnson}, \citenamefont {Han}, \citenamefont {Berkley},
  \citenamefont {Johansson}, \citenamefont {Bunyk}, \citenamefont {Ladizinsky},
  \citenamefont {Govorkov}, \citenamefont {Thom}, \citenamefont {Uchaikin},
  \citenamefont {Bumble}, \citenamefont {Fung}, \citenamefont {Kaul},
  \citenamefont {Kleinsasser}, \citenamefont {Amin},\ and\ \citenamefont
  {Averin}}]{Harris2008}%
  \BibitemOpen
  \bibfield  {author} {\bibinfo {author} {\bibfnamefont {R.}~\bibnamefont
  {Harris}}, \bibinfo {author} {\bibfnamefont {M.~W.}\ \bibnamefont {Johnson}},
  \bibinfo {author} {\bibfnamefont {S.}~\bibnamefont {Han}}, \bibinfo {author}
  {\bibfnamefont {A.~J.}\ \bibnamefont {Berkley}}, \bibinfo {author}
  {\bibfnamefont {J.}~\bibnamefont {Johansson}}, \bibinfo {author}
  {\bibfnamefont {P.}~\bibnamefont {Bunyk}}, \bibinfo {author} {\bibfnamefont
  {E.}~\bibnamefont {Ladizinsky}}, \bibinfo {author} {\bibfnamefont
  {S.}~\bibnamefont {Govorkov}}, \bibinfo {author} {\bibfnamefont {M.~C.}\
  \bibnamefont {Thom}}, \bibinfo {author} {\bibfnamefont {S.}~\bibnamefont
  {Uchaikin}}, \bibinfo {author} {\bibfnamefont {B.}~\bibnamefont {Bumble}},
  \bibinfo {author} {\bibfnamefont {A.}~\bibnamefont {Fung}}, \bibinfo {author}
  {\bibfnamefont {A.}~\bibnamefont {Kaul}}, \bibinfo {author} {\bibfnamefont
  {A.}~\bibnamefont {Kleinsasser}}, \bibinfo {author} {\bibfnamefont
  {M.~H.~S.}\ \bibnamefont {Amin}},\ and\ \bibinfo {author} {\bibfnamefont
  {D.~V.}\ \bibnamefont {Averin}},\ }\bibfield  {title} {\bibinfo {title}
  {Probing noise in flux qubits via macroscopic resonant tunneling},\ }\href
  {https://doi.org/10.1103/PhysRevLett.101.117003} {\bibfield  {journal}
  {\bibinfo  {journal} {Phys. Rev. Lett.}\ }\textbf {\bibinfo {volume} {101}},\
  \bibinfo {pages} {117003} (\bibinfo {year} {2008})}\BibitemShut {NoStop}%
\bibitem [{\citenamefont {Bylander}\ \emph {et~al.}(2011)\citenamefont
  {Bylander}, \citenamefont {Gustavsson}, \citenamefont {Yan}, \citenamefont
  {Yoshihara}, \citenamefont {Harrabi}, \citenamefont {Fitch}, \citenamefont
  {Cory}, \citenamefont {Nakamura}, \citenamefont {Tsai},\ and\ \citenamefont
  {Oliver}}]{Bylander2011}%
  \BibitemOpen
  \bibfield  {author} {\bibinfo {author} {\bibfnamefont {J.}~\bibnamefont
  {Bylander}}, \bibinfo {author} {\bibfnamefont {S.}~\bibnamefont
  {Gustavsson}}, \bibinfo {author} {\bibfnamefont {F.}~\bibnamefont {Yan}},
  \bibinfo {author} {\bibfnamefont {F.}~\bibnamefont {Yoshihara}}, \bibinfo
  {author} {\bibfnamefont {K.}~\bibnamefont {Harrabi}}, \bibinfo {author}
  {\bibfnamefont {G.}~\bibnamefont {Fitch}}, \bibinfo {author} {\bibfnamefont
  {D.~G.}\ \bibnamefont {Cory}}, \bibinfo {author} {\bibfnamefont
  {Y.}~\bibnamefont {Nakamura}}, \bibinfo {author} {\bibfnamefont {J.-S.}\
  \bibnamefont {Tsai}},\ and\ \bibinfo {author} {\bibfnamefont {W.~D.}\
  \bibnamefont {Oliver}},\ }\bibfield  {title} {\bibinfo {title} {Noise
  spectroscopy through dynamical decoupling with a superconducting flux
  qubit},\ }\href {https://doi.org/10.1038/nphys1994} {\bibfield  {journal}
  {\bibinfo  {journal} {Nature Physics}\ }\textbf {\bibinfo {volume} {7}},\
  \bibinfo {pages} {565} (\bibinfo {year} {2011})}\BibitemShut {NoStop}%
\bibitem [{\citenamefont {Chan}\ \emph {et~al.}(2018)\citenamefont {Chan},
  \citenamefont {Huang}, \citenamefont {Yang}, \citenamefont {Hwang},
  \citenamefont {Hensen}, \citenamefont {Tanttu}, \citenamefont {Hudson},
  \citenamefont {Itoh}, \citenamefont {Laucht}, \citenamefont {Morello},\ and\
  \citenamefont {Dzurak}}]{Chan2018}%
  \BibitemOpen
  \bibfield  {author} {\bibinfo {author} {\bibfnamefont {K.~W.}\ \bibnamefont
  {Chan}}, \bibinfo {author} {\bibfnamefont {W.}~\bibnamefont {Huang}},
  \bibinfo {author} {\bibfnamefont {C.~H.}\ \bibnamefont {Yang}}, \bibinfo
  {author} {\bibfnamefont {J.~C.~C.}\ \bibnamefont {Hwang}}, \bibinfo {author}
  {\bibfnamefont {B.}~\bibnamefont {Hensen}}, \bibinfo {author} {\bibfnamefont
  {T.}~\bibnamefont {Tanttu}}, \bibinfo {author} {\bibfnamefont {F.~E.}\
  \bibnamefont {Hudson}}, \bibinfo {author} {\bibfnamefont {K.~M.}\
  \bibnamefont {Itoh}}, \bibinfo {author} {\bibfnamefont {A.}~\bibnamefont
  {Laucht}}, \bibinfo {author} {\bibfnamefont {A.}~\bibnamefont {Morello}},\
  and\ \bibinfo {author} {\bibfnamefont {A.~S.}\ \bibnamefont {Dzurak}},\
  }\bibfield  {title} {\bibinfo {title} {Assessment of a silicon quantum dot
  spin qubit environment via noise spectroscopy},\ }\href
  {https://doi.org/10.1103/PhysRevApplied.10.044017} {\bibfield  {journal}
  {\bibinfo  {journal} {Phys. Rev. Appl.}\ }\textbf {\bibinfo {volume} {10}},\
  \bibinfo {pages} {044017} (\bibinfo {year} {2018})}\BibitemShut {NoStop}%
\bibitem [{\citenamefont {Knill}\ \emph {et~al.}(2008)\citenamefont {Knill},
  \citenamefont {Leibfried}, \citenamefont {Reichle}, \citenamefont {Britton},
  \citenamefont {Blakestad}, \citenamefont {Jost}, \citenamefont {Langer},
  \citenamefont {Ozeri}, \citenamefont {Seidelin},\ and\ \citenamefont
  {Wineland}}]{Knill2008}%
  \BibitemOpen
  \bibfield  {author} {\bibinfo {author} {\bibfnamefont {E.}~\bibnamefont
  {Knill}}, \bibinfo {author} {\bibfnamefont {D.}~\bibnamefont {Leibfried}},
  \bibinfo {author} {\bibfnamefont {R.}~\bibnamefont {Reichle}}, \bibinfo
  {author} {\bibfnamefont {J.}~\bibnamefont {Britton}}, \bibinfo {author}
  {\bibfnamefont {R.~B.}\ \bibnamefont {Blakestad}}, \bibinfo {author}
  {\bibfnamefont {J.~D.}\ \bibnamefont {Jost}}, \bibinfo {author}
  {\bibfnamefont {C.}~\bibnamefont {Langer}}, \bibinfo {author} {\bibfnamefont
  {R.}~\bibnamefont {Ozeri}}, \bibinfo {author} {\bibfnamefont
  {S.}~\bibnamefont {Seidelin}},\ and\ \bibinfo {author} {\bibfnamefont
  {D.~J.}\ \bibnamefont {Wineland}},\ }\bibfield  {title} {\bibinfo {title}
  {Randomized benchmarking of quantum gates},\ }\href
  {https://doi.org/10.1103/PhysRevA.77.012307} {\bibfield  {journal} {\bibinfo
  {journal} {Phys. Rev. A}\ }\textbf {\bibinfo {volume} {77}},\ \bibinfo
  {pages} {012307} (\bibinfo {year} {2008})}\BibitemShut {NoStop}%
\bibitem [{\citenamefont {Magesan}\ \emph {et~al.}(2011)\citenamefont
  {Magesan}, \citenamefont {Gambetta},\ and\ \citenamefont
  {Emerson}}]{Magesan2011}%
  \BibitemOpen
  \bibfield  {author} {\bibinfo {author} {\bibfnamefont {E.}~\bibnamefont
  {Magesan}}, \bibinfo {author} {\bibfnamefont {J.~M.}\ \bibnamefont
  {Gambetta}},\ and\ \bibinfo {author} {\bibfnamefont {J.}~\bibnamefont
  {Emerson}},\ }\bibfield  {title} {\bibinfo {title} {Scalable and robust
  randomized benchmarking of quantum processes},\ }\href
  {https://doi.org/10.1103/PhysRevLett.106.180504} {\bibfield  {journal}
  {\bibinfo  {journal} {Phys. Rev. Lett.}\ }\textbf {\bibinfo {volume} {106}},\
  \bibinfo {pages} {180504} (\bibinfo {year} {2011})}\BibitemShut {NoStop}%
\bibitem [{\citenamefont {Magesan}\ \emph {et~al.}(2012)\citenamefont
  {Magesan}, \citenamefont {Gambetta},\ and\ \citenamefont
  {Emerson}}]{Magesan2012b}%
  \BibitemOpen
  \bibfield  {author} {\bibinfo {author} {\bibfnamefont {E.}~\bibnamefont
  {Magesan}}, \bibinfo {author} {\bibfnamefont {J.~M.}\ \bibnamefont
  {Gambetta}},\ and\ \bibinfo {author} {\bibfnamefont {J.}~\bibnamefont
  {Emerson}},\ }\bibfield  {title} {\bibinfo {title} {Characterizing quantum
  gates via randomized benchmarking},\ }\href@noop {} {\bibfield  {journal}
  {\bibinfo  {journal} {Phys. Rev. A}\ }\textbf {\bibinfo {volume} {85}},\
  \bibinfo {pages} {042311} (\bibinfo {year} {2012})}\BibitemShut {NoStop}%
\bibitem [{\citenamefont {Proctor}\ \emph {et~al.}(2017)\citenamefont
  {Proctor}, \citenamefont {Rudinger}, \citenamefont {Young}, \citenamefont
  {Sarovar},\ and\ \citenamefont {Blume-Kohout}}]{Proctor2017}%
  \BibitemOpen
  \bibfield  {author} {\bibinfo {author} {\bibfnamefont {T.}~\bibnamefont
  {Proctor}}, \bibinfo {author} {\bibfnamefont {K.}~\bibnamefont {Rudinger}},
  \bibinfo {author} {\bibfnamefont {K.}~\bibnamefont {Young}}, \bibinfo
  {author} {\bibfnamefont {M.}~\bibnamefont {Sarovar}},\ and\ \bibinfo {author}
  {\bibfnamefont {R.}~\bibnamefont {Blume-Kohout}},\ }\bibfield  {title}
  {\bibinfo {title} {What randomized benchmarking actually measures},\ }\href
  {https://doi.org/10.1103/PhysRevLett.119.130502} {\bibfield  {journal}
  {\bibinfo  {journal} {Phys. Rev. Lett.}\ }\textbf {\bibinfo {volume} {119}},\
  \bibinfo {pages} {130502} (\bibinfo {year} {2017})}\BibitemShut {NoStop}%
\bibitem [{\citenamefont {Fogarty}\ \emph {et~al.}(2015)\citenamefont
  {Fogarty}, \citenamefont {Veldhorst}, \citenamefont {Harper}, \citenamefont
  {Yang}, \citenamefont {Bartlett}, \citenamefont {Flammia},\ and\
  \citenamefont {Dzurak}}]{Fogarty2015}%
  \BibitemOpen
  \bibfield  {author} {\bibinfo {author} {\bibfnamefont {M.~A.}\ \bibnamefont
  {Fogarty}}, \bibinfo {author} {\bibfnamefont {M.}~\bibnamefont {Veldhorst}},
  \bibinfo {author} {\bibfnamefont {R.}~\bibnamefont {Harper}}, \bibinfo
  {author} {\bibfnamefont {C.~H.}\ \bibnamefont {Yang}}, \bibinfo {author}
  {\bibfnamefont {S.~D.}\ \bibnamefont {Bartlett}}, \bibinfo {author}
  {\bibfnamefont {S.~T.}\ \bibnamefont {Flammia}},\ and\ \bibinfo {author}
  {\bibfnamefont {A.~S.}\ \bibnamefont {Dzurak}},\ }\bibfield  {title}
  {\bibinfo {title} {Nonexponential fidelity decay in randomized benchmarking
  with low-frequency noise},\ }\href
  {https://doi.org/10.1103/PhysRevA.92.022326} {\bibfield  {journal} {\bibinfo
  {journal} {Phys. Rev. A}\ }\textbf {\bibinfo {volume} {92}},\ \bibinfo
  {pages} {022326} (\bibinfo {year} {2015})}\BibitemShut {NoStop}%
\bibitem [{\citenamefont {Klimov}\ \emph {et~al.}(2018)\citenamefont {Klimov},
  \citenamefont {Kelly}, \citenamefont {Chen}, \citenamefont {Neeley},
  \citenamefont {Megrant}, \citenamefont {Burkett}, \citenamefont {Barends},
  \citenamefont {Arya}, \citenamefont {Chiaro}, \citenamefont {Chen},
  \citenamefont {Dunsworth}, \citenamefont {Fowler}, \citenamefont {Foxen},
  \citenamefont {Gidney}, \citenamefont {Giustina}, \citenamefont {Graff},
  \citenamefont {Huang}, \citenamefont {Jeffrey}, \citenamefont {Lucero},
  \citenamefont {Mutus}, \citenamefont {Naaman}, \citenamefont {Neill},
  \citenamefont {Quintana}, \citenamefont {Roushan}, \citenamefont {Sank},
  \citenamefont {Vainsencher}, \citenamefont {Wenner}, \citenamefont {White},
  \citenamefont {Boixo}, \citenamefont {Babbush}, \citenamefont {Smelyanskiy},
  \citenamefont {Neven},\ and\ \citenamefont {Martinis}}]{Klimov2018}%
  \BibitemOpen
  \bibfield  {author} {\bibinfo {author} {\bibfnamefont {P.~V.}\ \bibnamefont
  {Klimov}}, \bibinfo {author} {\bibfnamefont {J.}~\bibnamefont {Kelly}},
  \bibinfo {author} {\bibfnamefont {Z.}~\bibnamefont {Chen}}, \bibinfo {author}
  {\bibfnamefont {M.}~\bibnamefont {Neeley}}, \bibinfo {author} {\bibfnamefont
  {A.}~\bibnamefont {Megrant}}, \bibinfo {author} {\bibfnamefont
  {B.}~\bibnamefont {Burkett}}, \bibinfo {author} {\bibfnamefont
  {R.}~\bibnamefont {Barends}}, \bibinfo {author} {\bibfnamefont
  {K.}~\bibnamefont {Arya}}, \bibinfo {author} {\bibfnamefont {B.}~\bibnamefont
  {Chiaro}}, \bibinfo {author} {\bibfnamefont {Y.}~\bibnamefont {Chen}},
  \bibinfo {author} {\bibfnamefont {A.}~\bibnamefont {Dunsworth}}, \bibinfo
  {author} {\bibfnamefont {A.}~\bibnamefont {Fowler}}, \bibinfo {author}
  {\bibfnamefont {B.}~\bibnamefont {Foxen}}, \bibinfo {author} {\bibfnamefont
  {C.}~\bibnamefont {Gidney}}, \bibinfo {author} {\bibfnamefont
  {M.}~\bibnamefont {Giustina}}, \bibinfo {author} {\bibfnamefont
  {R.}~\bibnamefont {Graff}}, \bibinfo {author} {\bibfnamefont
  {T.}~\bibnamefont {Huang}}, \bibinfo {author} {\bibfnamefont
  {E.}~\bibnamefont {Jeffrey}}, \bibinfo {author} {\bibfnamefont
  {E.}~\bibnamefont {Lucero}}, \bibinfo {author} {\bibfnamefont {J.~Y.}\
  \bibnamefont {Mutus}}, \bibinfo {author} {\bibfnamefont {O.}~\bibnamefont
  {Naaman}}, \bibinfo {author} {\bibfnamefont {C.}~\bibnamefont {Neill}},
  \bibinfo {author} {\bibfnamefont {C.}~\bibnamefont {Quintana}}, \bibinfo
  {author} {\bibfnamefont {P.}~\bibnamefont {Roushan}}, \bibinfo {author}
  {\bibfnamefont {D.}~\bibnamefont {Sank}}, \bibinfo {author} {\bibfnamefont
  {A.}~\bibnamefont {Vainsencher}}, \bibinfo {author} {\bibfnamefont
  {J.}~\bibnamefont {Wenner}}, \bibinfo {author} {\bibfnamefont {T.~C.}\
  \bibnamefont {White}}, \bibinfo {author} {\bibfnamefont {S.}~\bibnamefont
  {Boixo}}, \bibinfo {author} {\bibfnamefont {R.}~\bibnamefont {Babbush}},
  \bibinfo {author} {\bibfnamefont {V.~N.}\ \bibnamefont {Smelyanskiy}},
  \bibinfo {author} {\bibfnamefont {H.}~\bibnamefont {Neven}},\ and\ \bibinfo
  {author} {\bibfnamefont {J.~M.}\ \bibnamefont {Martinis}},\ }\bibfield
  {title} {\bibinfo {title} {Fluctuations of energy-relaxation times in
  superconducting qubits},\ }\href
  {https://doi.org/10.1103/PhysRevLett.121.090502} {\bibfield  {journal}
  {\bibinfo  {journal} {Phys. Rev. Lett.}\ }\textbf {\bibinfo {volume} {121}},\
  \bibinfo {pages} {090502} (\bibinfo {year} {2018})}\BibitemShut {NoStop}%
\bibitem [{\citenamefont {Proctor}\ \emph {et~al.}(2020)\citenamefont
  {Proctor}, \citenamefont {Revelle}, \citenamefont {Nielsen}, \citenamefont
  {Rudinger}, \citenamefont {Lobser}, \citenamefont {Maunz}, \citenamefont
  {Blume-Kohout},\ and\ \citenamefont {Young}}]{Proctor2020}%
  \BibitemOpen
  \bibfield  {author} {\bibinfo {author} {\bibfnamefont {T.}~\bibnamefont
  {Proctor}}, \bibinfo {author} {\bibfnamefont {M.}~\bibnamefont {Revelle}},
  \bibinfo {author} {\bibfnamefont {E.}~\bibnamefont {Nielsen}}, \bibinfo
  {author} {\bibfnamefont {K.}~\bibnamefont {Rudinger}}, \bibinfo {author}
  {\bibfnamefont {D.}~\bibnamefont {Lobser}}, \bibinfo {author} {\bibfnamefont
  {P.}~\bibnamefont {Maunz}}, \bibinfo {author} {\bibfnamefont
  {R.}~\bibnamefont {Blume-Kohout}},\ and\ \bibinfo {author} {\bibfnamefont
  {K.}~\bibnamefont {Young}},\ }\bibfield  {title} {\bibinfo {title} {Detecting
  and tracking drift in quantum information processors},\ }\href
  {https://doi.org/10.1038/s41467-020-19074-4} {\bibfield  {journal} {\bibinfo
  {journal} {Nature Communications}\ }\textbf {\bibinfo {volume} {11}},\
  \bibinfo {pages} {5396} (\bibinfo {year} {2020})}\BibitemShut {NoStop}%
\bibitem [{\citenamefont {Levy}(2002)}]{Levy2002}%
  \BibitemOpen
  \bibfield  {author} {\bibinfo {author} {\bibfnamefont {J.}~\bibnamefont
  {Levy}},\ }\bibfield  {title} {\bibinfo {title} {Universal quantum
  computation with spin-$1/2$ pairs and heisenberg exchange},\ }\href
  {https://doi.org/10.1103/PhysRevLett.89.147902} {\bibfield  {journal}
  {\bibinfo  {journal} {Phys. Rev. Lett.}\ }\textbf {\bibinfo {volume} {89}},\
  \bibinfo {pages} {147902} (\bibinfo {year} {2002})}\BibitemShut {NoStop}%
\bibitem [{\citenamefont {Petta}\ \emph {et~al.}(2005)\citenamefont {Petta},
  \citenamefont {Johnson}, \citenamefont {Taylor}, \citenamefont {Laird},
  \citenamefont {Yacoby}, \citenamefont {Lukin}, \citenamefont {Marcus},
  \citenamefont {Hanson},\ and\ \citenamefont {Gossard}}]{Petta2005}%
  \BibitemOpen
  \bibfield  {author} {\bibinfo {author} {\bibfnamefont {J.~R.}\ \bibnamefont
  {Petta}}, \bibinfo {author} {\bibfnamefont {A.~C.}\ \bibnamefont {Johnson}},
  \bibinfo {author} {\bibfnamefont {J.~M.}\ \bibnamefont {Taylor}}, \bibinfo
  {author} {\bibfnamefont {E.~A.}\ \bibnamefont {Laird}}, \bibinfo {author}
  {\bibfnamefont {A.}~\bibnamefont {Yacoby}}, \bibinfo {author} {\bibfnamefont
  {M.~D.}\ \bibnamefont {Lukin}}, \bibinfo {author} {\bibfnamefont {C.~M.}\
  \bibnamefont {Marcus}}, \bibinfo {author} {\bibfnamefont {M.~P.}\
  \bibnamefont {Hanson}},\ and\ \bibinfo {author} {\bibfnamefont {A.~C.}\
  \bibnamefont {Gossard}},\ }\bibfield  {title} {\bibinfo {title} {Coherent
  manipulation of coupled electron spins in semiconductor quantum dots},\
  }\href {https://doi.org/10.1126/science.1116955} {\bibfield  {journal}
  {\bibinfo  {journal} {Science}\ }\textbf {\bibinfo {volume} {309}},\ \bibinfo
  {pages} {2180} (\bibinfo {year} {2005})}\BibitemShut {NoStop}%
\bibitem [{\citenamefont {Maune}\ \emph {et~al.}(2012)\citenamefont {Maune},
  \citenamefont {Borselli}, \citenamefont {Huang}, \citenamefont {Ladd},
  \citenamefont {Deelman}, \citenamefont {Holabird}, \citenamefont {Kiselev},
  \citenamefont {Alvarado-Rodriguez}, \citenamefont {Ross}, \citenamefont
  {Schmitz}, \citenamefont {Sokolich}, \citenamefont {Watson}, \citenamefont
  {Gyure},\ and\ \citenamefont {Hunter}}]{Borselli2012}%
  \BibitemOpen
  \bibfield  {author} {\bibinfo {author} {\bibfnamefont {B.~M.}\ \bibnamefont
  {Maune}}, \bibinfo {author} {\bibfnamefont {M.~G.}\ \bibnamefont {Borselli}},
  \bibinfo {author} {\bibfnamefont {B.}~\bibnamefont {Huang}}, \bibinfo
  {author} {\bibfnamefont {T.~D.}\ \bibnamefont {Ladd}}, \bibinfo {author}
  {\bibfnamefont {P.~W.}\ \bibnamefont {Deelman}}, \bibinfo {author}
  {\bibfnamefont {K.~S.}\ \bibnamefont {Holabird}}, \bibinfo {author}
  {\bibfnamefont {A.~A.}\ \bibnamefont {Kiselev}}, \bibinfo {author}
  {\bibfnamefont {I.}~\bibnamefont {Alvarado-Rodriguez}}, \bibinfo {author}
  {\bibfnamefont {R.~S.}\ \bibnamefont {Ross}}, \bibinfo {author}
  {\bibfnamefont {A.~E.}\ \bibnamefont {Schmitz}}, \bibinfo {author}
  {\bibfnamefont {M.}~\bibnamefont {Sokolich}}, \bibinfo {author}
  {\bibfnamefont {C.~A.}\ \bibnamefont {Watson}}, \bibinfo {author}
  {\bibfnamefont {M.~F.}\ \bibnamefont {Gyure}},\ and\ \bibinfo {author}
  {\bibfnamefont {A.~T.}\ \bibnamefont {Hunter}},\ }\bibfield  {title}
  {\bibinfo {title} {Coherent singlet-triplet oscillations in a silicon-based
  double quantum dot},\ }\href {https://doi.org/10.1038/nature10707} {\bibfield
   {journal} {\bibinfo  {journal} {Nature}\ }\textbf {\bibinfo {volume}
  {481}},\ \bibinfo {pages} {344} (\bibinfo {year} {2012})}\BibitemShut
  {NoStop}%
\bibitem [{\citenamefont {Shulman}\ \emph {et~al.}(2012)\citenamefont
  {Shulman}, \citenamefont {Dial}, \citenamefont {Harvey}, \citenamefont
  {Bluhm}, \citenamefont {Umansky},\ and\ \citenamefont
  {Yacoby}}]{Shulman2012}%
  \BibitemOpen
  \bibfield  {author} {\bibinfo {author} {\bibfnamefont {M.~D.}\ \bibnamefont
  {Shulman}}, \bibinfo {author} {\bibfnamefont {O.~E.}\ \bibnamefont {Dial}},
  \bibinfo {author} {\bibfnamefont {S.~P.}\ \bibnamefont {Harvey}}, \bibinfo
  {author} {\bibfnamefont {H.}~\bibnamefont {Bluhm}}, \bibinfo {author}
  {\bibfnamefont {V.}~\bibnamefont {Umansky}},\ and\ \bibinfo {author}
  {\bibfnamefont {A.}~\bibnamefont {Yacoby}},\ }\bibfield  {title} {\bibinfo
  {title} {Demonstration of entanglement of electrostatically coupled
  singlet-triplet qubits},\ }\href {https://doi.org/10.1126/science.1217692}
  {\bibfield  {journal} {\bibinfo  {journal} {Science}\ }\textbf {\bibinfo
  {volume} {336}},\ \bibinfo {pages} {202} (\bibinfo {year}
  {2012})}\BibitemShut {NoStop}%
\bibitem [{\citenamefont {Wu}\ \emph {et~al.}(2014)\citenamefont {Wu},
  \citenamefont {Ward}, \citenamefont {Prance}, \citenamefont {Kim},
  \citenamefont {Gamble}, \citenamefont {Mohr}, \citenamefont {Shi},
  \citenamefont {Savage}, \citenamefont {Lagally}, \citenamefont {Friesen},
  \citenamefont {Coppersmith},\ and\ \citenamefont {Eriksson}}]{Wu2014}%
  \BibitemOpen
  \bibfield  {author} {\bibinfo {author} {\bibfnamefont {X.}~\bibnamefont
  {Wu}}, \bibinfo {author} {\bibfnamefont {D.~R.}\ \bibnamefont {Ward}},
  \bibinfo {author} {\bibfnamefont {J.~R.}\ \bibnamefont {Prance}}, \bibinfo
  {author} {\bibfnamefont {D.}~\bibnamefont {Kim}}, \bibinfo {author}
  {\bibfnamefont {J.~K.}\ \bibnamefont {Gamble}}, \bibinfo {author}
  {\bibfnamefont {R.~T.}\ \bibnamefont {Mohr}}, \bibinfo {author}
  {\bibfnamefont {Z.}~\bibnamefont {Shi}}, \bibinfo {author} {\bibfnamefont
  {D.~E.}\ \bibnamefont {Savage}}, \bibinfo {author} {\bibfnamefont {M.~G.}\
  \bibnamefont {Lagally}}, \bibinfo {author} {\bibfnamefont {M.}~\bibnamefont
  {Friesen}}, \bibinfo {author} {\bibfnamefont {S.~N.}\ \bibnamefont
  {Coppersmith}},\ and\ \bibinfo {author} {\bibfnamefont {M.~A.}\ \bibnamefont
  {Eriksson}},\ }\bibfield  {title} {\bibinfo {title} {Two-axis control of a
  singlet–triplet qubit with an integrated micromagnet},\ }\href
  {https://doi.org/10.1073/pnas.1412230111} {\bibfield  {journal} {\bibinfo
  {journal} {Proceedings of the National Academy of Sciences}\ }\textbf
  {\bibinfo {volume} {111}},\ \bibinfo {pages} {11938} (\bibinfo {year}
  {2014})}\BibitemShut {NoStop}%
\bibitem [{\citenamefont {Nichol}\ \emph {et~al.}(2017)\citenamefont {Nichol},
  \citenamefont {Orona}, \citenamefont {Harvey}, \citenamefont {Fallahi},
  \citenamefont {Gardner}, \citenamefont {Manfra},\ and\ \citenamefont
  {Yacoby}}]{Nichol2017}%
  \BibitemOpen
  \bibfield  {author} {\bibinfo {author} {\bibfnamefont {J.~M.}\ \bibnamefont
  {Nichol}}, \bibinfo {author} {\bibfnamefont {L.~A.}\ \bibnamefont {Orona}},
  \bibinfo {author} {\bibfnamefont {S.~P.}\ \bibnamefont {Harvey}}, \bibinfo
  {author} {\bibfnamefont {S.}~\bibnamefont {Fallahi}}, \bibinfo {author}
  {\bibfnamefont {G.~C.}\ \bibnamefont {Gardner}}, \bibinfo {author}
  {\bibfnamefont {M.~J.}\ \bibnamefont {Manfra}},\ and\ \bibinfo {author}
  {\bibfnamefont {A.}~\bibnamefont {Yacoby}},\ }\bibfield  {title} {\bibinfo
  {title} {High-fidelity entangling gate for double-quantum-dot spin qubits},\
  }\href {https://doi.org/10.1038/s41534-016-0003-1} {\bibfield  {journal}
  {\bibinfo  {journal} {npj Quantum Information}\ }\textbf {\bibinfo {volume}
  {3}},\ \bibinfo {pages} {3} (\bibinfo {year} {2017})}\BibitemShut {NoStop}%
\bibitem [{\citenamefont {Jock}\ \emph {et~al.}(2018)\citenamefont {Jock},
  \citenamefont {Jacobson}, \citenamefont {Harvey-Collard}, \citenamefont
  {Mounce}, \citenamefont {Srinivasa}, \citenamefont {Ward}, \citenamefont
  {Anderson}, \citenamefont {Manginell}, \citenamefont {Wendt}, \citenamefont
  {Rudolph}, \citenamefont {Pluym}, \citenamefont {Gamble}, \citenamefont
  {Baczewski}, \citenamefont {Witzel},\ and\ \citenamefont
  {Carroll}}]{Jock2018}%
  \BibitemOpen
  \bibfield  {author} {\bibinfo {author} {\bibfnamefont {R.~M.}\ \bibnamefont
  {Jock}}, \bibinfo {author} {\bibfnamefont {N.~T.}\ \bibnamefont {Jacobson}},
  \bibinfo {author} {\bibfnamefont {P.}~\bibnamefont {Harvey-Collard}},
  \bibinfo {author} {\bibfnamefont {A.~M.}\ \bibnamefont {Mounce}}, \bibinfo
  {author} {\bibfnamefont {V.}~\bibnamefont {Srinivasa}}, \bibinfo {author}
  {\bibfnamefont {D.~R.}\ \bibnamefont {Ward}}, \bibinfo {author}
  {\bibfnamefont {J.}~\bibnamefont {Anderson}}, \bibinfo {author}
  {\bibfnamefont {R.}~\bibnamefont {Manginell}}, \bibinfo {author}
  {\bibfnamefont {J.~R.}\ \bibnamefont {Wendt}}, \bibinfo {author}
  {\bibfnamefont {M.}~\bibnamefont {Rudolph}}, \bibinfo {author} {\bibfnamefont
  {T.}~\bibnamefont {Pluym}}, \bibinfo {author} {\bibfnamefont {J.~K.}\
  \bibnamefont {Gamble}}, \bibinfo {author} {\bibfnamefont {A.~D.}\
  \bibnamefont {Baczewski}}, \bibinfo {author} {\bibfnamefont {W.~M.}\
  \bibnamefont {Witzel}},\ and\ \bibinfo {author} {\bibfnamefont {M.~S.}\
  \bibnamefont {Carroll}},\ }\bibfield  {title} {\bibinfo {title} {A silicon
  metal-oxide-semiconductor electron spin-orbit qubit},\ }\href
  {https://doi.org/10.1038/s41467-018-04200-0} {\bibfield  {journal} {\bibinfo
  {journal} {Nature Communications}\ }\textbf {\bibinfo {volume} {9}},\
  \bibinfo {pages} {1768} (\bibinfo {year} {2018})}\BibitemShut {NoStop}%
\bibitem [{\citenamefont {Cerfontaine}\ \emph
  {et~al.}(2020{\natexlab{a}})\citenamefont {Cerfontaine}, \citenamefont
  {Botzem}, \citenamefont {Ritzmann}, \citenamefont {Humpohl}, \citenamefont
  {Ludwig}, \citenamefont {Schuh}, \citenamefont {Bougeard}, \citenamefont
  {Wieck},\ and\ \citenamefont {Bluhm}}]{Cerfontaine2020}%
  \BibitemOpen
  \bibfield  {author} {\bibinfo {author} {\bibfnamefont {P.}~\bibnamefont
  {Cerfontaine}}, \bibinfo {author} {\bibfnamefont {T.}~\bibnamefont {Botzem}},
  \bibinfo {author} {\bibfnamefont {J.}~\bibnamefont {Ritzmann}}, \bibinfo
  {author} {\bibfnamefont {S.~S.}\ \bibnamefont {Humpohl}}, \bibinfo {author}
  {\bibfnamefont {A.}~\bibnamefont {Ludwig}}, \bibinfo {author} {\bibfnamefont
  {D.}~\bibnamefont {Schuh}}, \bibinfo {author} {\bibfnamefont
  {D.}~\bibnamefont {Bougeard}}, \bibinfo {author} {\bibfnamefont {A.~D.}\
  \bibnamefont {Wieck}},\ and\ \bibinfo {author} {\bibfnamefont
  {H.}~\bibnamefont {Bluhm}},\ }\bibfield  {title} {\bibinfo {title}
  {Closed-loop control of a gaas-based singlet-triplet spin qubit with 99.5{\%}
  gate fidelity and low leakage},\ }\href
  {https://doi.org/10.1038/s41467-020-17865-3} {\bibfield  {journal} {\bibinfo
  {journal} {Nature Communications}\ }\textbf {\bibinfo {volume} {11}},\
  \bibinfo {pages} {4144} (\bibinfo {year} {2020}{\natexlab{a}})}\BibitemShut
  {NoStop}%
\bibitem [{\citenamefont {Cerfontaine}\ \emph
  {et~al.}(2020{\natexlab{b}})\citenamefont {Cerfontaine}, \citenamefont
  {Otten}, \citenamefont {Wolfe}, \citenamefont {Bethke},\ and\ \citenamefont
  {Bluhm}}]{Cerfontaine2020b}%
  \BibitemOpen
  \bibfield  {author} {\bibinfo {author} {\bibfnamefont {P.}~\bibnamefont
  {Cerfontaine}}, \bibinfo {author} {\bibfnamefont {R.}~\bibnamefont {Otten}},
  \bibinfo {author} {\bibfnamefont {M.~A.}\ \bibnamefont {Wolfe}}, \bibinfo
  {author} {\bibfnamefont {P.}~\bibnamefont {Bethke}},\ and\ \bibinfo {author}
  {\bibfnamefont {H.}~\bibnamefont {Bluhm}},\ }\bibfield  {title} {\bibinfo
  {title} {High-fidelity gate set for exchange-coupled singlet-triplet
  qubits},\ }\href {https://doi.org/10.1103/PhysRevB.101.155311} {\bibfield
  {journal} {\bibinfo  {journal} {Phys. Rev. B}\ }\textbf {\bibinfo {volume}
  {101}},\ \bibinfo {pages} {155311} (\bibinfo {year}
  {2020}{\natexlab{b}})}\BibitemShut {NoStop}%
\bibitem [{\citenamefont {Fedele}\ \emph {et~al.}(2021)\citenamefont {Fedele},
  \citenamefont {Chatterjee}, \citenamefont {Fallahi}, \citenamefont {Gardner},
  \citenamefont {Manfra},\ and\ \citenamefont {Kuemmeth}}]{Fedele2021}%
  \BibitemOpen
  \bibfield  {author} {\bibinfo {author} {\bibfnamefont {F.}~\bibnamefont
  {Fedele}}, \bibinfo {author} {\bibfnamefont {A.}~\bibnamefont {Chatterjee}},
  \bibinfo {author} {\bibfnamefont {S.}~\bibnamefont {Fallahi}}, \bibinfo
  {author} {\bibfnamefont {G.~C.}\ \bibnamefont {Gardner}}, \bibinfo {author}
  {\bibfnamefont {M.~J.}\ \bibnamefont {Manfra}},\ and\ \bibinfo {author}
  {\bibfnamefont {F.}~\bibnamefont {Kuemmeth}},\ }\bibfield  {title} {\bibinfo
  {title} {Simultaneous operations in a two-dimensional array of
  singlet-triplet qubits},\ }\href
  {https://doi.org/10.1103/PRXQuantum.2.040306} {\bibfield  {journal} {\bibinfo
   {journal} {PRX Quantum}\ }\textbf {\bibinfo {volume} {2}},\ \bibinfo {pages}
  {040306} (\bibinfo {year} {2021})}\BibitemShut {NoStop}%
\bibitem [{\citenamefont {Jirovec}\ \emph {et~al.}(2021)\citenamefont
  {Jirovec}, \citenamefont {Hofmann}, \citenamefont {Ballabio}, \citenamefont
  {Mutter}, \citenamefont {Tavani}, \citenamefont {Botifoll}, \citenamefont
  {Crippa}, \citenamefont {Kukucka}, \citenamefont {Sagi}, \citenamefont
  {Martins}, \citenamefont {Saez-Mollejo}, \citenamefont {Prieto},
  \citenamefont {Borovkov}, \citenamefont {Arbiol}, \citenamefont {Chrastina},
  \citenamefont {Isella},\ and\ \citenamefont {Katsaros}}]{Jirovec2021}%
  \BibitemOpen
  \bibfield  {author} {\bibinfo {author} {\bibfnamefont {D.}~\bibnamefont
  {Jirovec}}, \bibinfo {author} {\bibfnamefont {A.}~\bibnamefont {Hofmann}},
  \bibinfo {author} {\bibfnamefont {A.}~\bibnamefont {Ballabio}}, \bibinfo
  {author} {\bibfnamefont {P.~M.}\ \bibnamefont {Mutter}}, \bibinfo {author}
  {\bibfnamefont {G.}~\bibnamefont {Tavani}}, \bibinfo {author} {\bibfnamefont
  {M.}~\bibnamefont {Botifoll}}, \bibinfo {author} {\bibfnamefont
  {A.}~\bibnamefont {Crippa}}, \bibinfo {author} {\bibfnamefont
  {J.}~\bibnamefont {Kukucka}}, \bibinfo {author} {\bibfnamefont
  {O.}~\bibnamefont {Sagi}}, \bibinfo {author} {\bibfnamefont {F.}~\bibnamefont
  {Martins}}, \bibinfo {author} {\bibfnamefont {J.}~\bibnamefont
  {Saez-Mollejo}}, \bibinfo {author} {\bibfnamefont {I.}~\bibnamefont
  {Prieto}}, \bibinfo {author} {\bibfnamefont {M.}~\bibnamefont {Borovkov}},
  \bibinfo {author} {\bibfnamefont {J.}~\bibnamefont {Arbiol}}, \bibinfo
  {author} {\bibfnamefont {D.}~\bibnamefont {Chrastina}}, \bibinfo {author}
  {\bibfnamefont {G.}~\bibnamefont {Isella}},\ and\ \bibinfo {author}
  {\bibfnamefont {G.}~\bibnamefont {Katsaros}},\ }\bibfield  {title} {\bibinfo
  {title} {A singlet-triplet hole spin qubit in planar ge},\ }\href
  {https://doi.org/10.1038/s41563-021-01022-2} {\bibfield  {journal} {\bibinfo
  {journal} {Nature Materials}\ }\textbf {\bibinfo {volume} {20}},\ \bibinfo
  {pages} {1106} (\bibinfo {year} {2021})}\BibitemShut {NoStop}%
\bibitem [{\citenamefont {Connors}\ \emph {et~al.}(2019)\citenamefont
  {Connors}, \citenamefont {Nelson}, \citenamefont {Qiao}, \citenamefont
  {Edge},\ and\ \citenamefont {Nichol}}]{Connors2019}%
  \BibitemOpen
  \bibfield  {author} {\bibinfo {author} {\bibfnamefont {E.~J.}\ \bibnamefont
  {Connors}}, \bibinfo {author} {\bibfnamefont {J.}~\bibnamefont {Nelson}},
  \bibinfo {author} {\bibfnamefont {H.}~\bibnamefont {Qiao}}, \bibinfo {author}
  {\bibfnamefont {L.~F.}\ \bibnamefont {Edge}},\ and\ \bibinfo {author}
  {\bibfnamefont {J.~M.}\ \bibnamefont {Nichol}},\ }\bibfield  {title}
  {\bibinfo {title} {Low-frequency charge noise in si/sige quantum dots},\
  }\href {https://doi.org/10.1103/PhysRevB.100.165305} {\bibfield  {journal}
  {\bibinfo  {journal} {Phys. Rev. B}\ }\textbf {\bibinfo {volume} {100}},\
  \bibinfo {pages} {165305} (\bibinfo {year} {2019})}\BibitemShut {NoStop}%
\bibitem [{\citenamefont {Struck}\ \emph {et~al.}(2020)\citenamefont {Struck},
  \citenamefont {Hollmann}, \citenamefont {Schauer}, \citenamefont {Fedorets},
  \citenamefont {Schmidbauer}, \citenamefont {Sawano}, \citenamefont {Riemann},
  \citenamefont {Abrosimov}, \citenamefont {Cywi{\'n}ski}, \citenamefont
  {Bougeard},\ and\ \citenamefont {Schreiber}}]{Struck2020}%
  \BibitemOpen
  \bibfield  {author} {\bibinfo {author} {\bibfnamefont {T.}~\bibnamefont
  {Struck}}, \bibinfo {author} {\bibfnamefont {A.}~\bibnamefont {Hollmann}},
  \bibinfo {author} {\bibfnamefont {F.}~\bibnamefont {Schauer}}, \bibinfo
  {author} {\bibfnamefont {O.}~\bibnamefont {Fedorets}}, \bibinfo {author}
  {\bibfnamefont {A.}~\bibnamefont {Schmidbauer}}, \bibinfo {author}
  {\bibfnamefont {K.}~\bibnamefont {Sawano}}, \bibinfo {author} {\bibfnamefont
  {H.}~\bibnamefont {Riemann}}, \bibinfo {author} {\bibfnamefont {N.~V.}\
  \bibnamefont {Abrosimov}}, \bibinfo {author} {\bibfnamefont
  {{\L}.}~\bibnamefont {Cywi{\'n}ski}}, \bibinfo {author} {\bibfnamefont
  {D.}~\bibnamefont {Bougeard}},\ and\ \bibinfo {author} {\bibfnamefont
  {L.~R.}\ \bibnamefont {Schreiber}},\ }\bibfield  {title} {\bibinfo {title}
  {Low-frequency spin qubit energy splitting noise in highly purified
  28si/sige},\ }\href {https://doi.org/10.1038/s41534-020-0276-2} {\bibfield
  {journal} {\bibinfo  {journal} {npj Quantum Information}\ }\textbf {\bibinfo
  {volume} {6}},\ \bibinfo {pages} {40} (\bibinfo {year} {2020})}\BibitemShut
  {NoStop}%
\bibitem [{\citenamefont {Connors}\ \emph {et~al.}(2022)\citenamefont
  {Connors}, \citenamefont {Nelson}, \citenamefont {Edge},\ and\ \citenamefont
  {Nichol}}]{Connors2022}%
  \BibitemOpen
  \bibfield  {author} {\bibinfo {author} {\bibfnamefont {E.~J.}\ \bibnamefont
  {Connors}}, \bibinfo {author} {\bibfnamefont {J.}~\bibnamefont {Nelson}},
  \bibinfo {author} {\bibfnamefont {L.~F.}\ \bibnamefont {Edge}},\ and\
  \bibinfo {author} {\bibfnamefont {J.~M.}\ \bibnamefont {Nichol}},\ }\bibfield
   {title} {\bibinfo {title} {Charge-noise spectroscopy of si/sige quantum dots
  via dynamically-decoupled exchange oscillations},\ }\href
  {https://doi.org/10.1038/s41467-022-28519-x} {\bibfield  {journal} {\bibinfo
  {journal} {Nature Communications}\ }\textbf {\bibinfo {volume} {13}},\
  \bibinfo {pages} {940} (\bibinfo {year} {2022})}\BibitemShut {NoStop}%
\bibitem [{\citenamefont {Yoneda}\ \emph {et~al.}(2023)\citenamefont {Yoneda},
  \citenamefont {Rojas-Arias}, \citenamefont {Stano}, \citenamefont {Takeda},
  \citenamefont {Noiri}, \citenamefont {Nakajima}, \citenamefont {Loss},\ and\
  \citenamefont {Tarucha}}]{Yoneda2023}%
  \BibitemOpen
  \bibfield  {author} {\bibinfo {author} {\bibfnamefont {J.}~\bibnamefont
  {Yoneda}}, \bibinfo {author} {\bibfnamefont {J.~S.}\ \bibnamefont
  {Rojas-Arias}}, \bibinfo {author} {\bibfnamefont {P.}~\bibnamefont {Stano}},
  \bibinfo {author} {\bibfnamefont {K.}~\bibnamefont {Takeda}}, \bibinfo
  {author} {\bibfnamefont {A.}~\bibnamefont {Noiri}}, \bibinfo {author}
  {\bibfnamefont {T.}~\bibnamefont {Nakajima}}, \bibinfo {author}
  {\bibfnamefont {D.}~\bibnamefont {Loss}},\ and\ \bibinfo {author}
  {\bibfnamefont {S.}~\bibnamefont {Tarucha}},\ }\bibfield  {title} {\bibinfo
  {title} {Noise-correlation spectrum for a pair of spin qubits in silicon},\
  }\href {https://doi.org/10.1038/s41567-023-02238-6} {\bibfield  {journal}
  {\bibinfo  {journal} {Nature Physics}\ }\textbf {\bibinfo {volume} {19}},\
  \bibinfo {pages} {1793} (\bibinfo {year} {2023})}\BibitemShut {NoStop}%
\bibitem [{\citenamefont {Dutta}\ and\ \citenamefont {Horn}(1981)}]{Dutta1981}%
  \BibitemOpen
  \bibfield  {author} {\bibinfo {author} {\bibfnamefont {P.}~\bibnamefont
  {Dutta}}\ and\ \bibinfo {author} {\bibfnamefont {P.~M.}\ \bibnamefont
  {Horn}},\ }\bibfield  {title} {\bibinfo {title} {Low-frequency fluctuations
  in solids: $\frac{1}{f}$ noise},\ }\href
  {https://doi.org/10.1103/RevModPhys.53.497} {\bibfield  {journal} {\bibinfo
  {journal} {Rev. Mod. Phys.}\ }\textbf {\bibinfo {volume} {53}},\ \bibinfo
  {pages} {497} (\bibinfo {year} {1981})}\BibitemShut {NoStop}%
\bibitem [{\citenamefont {Paladino}\ \emph {et~al.}(2014)\citenamefont
  {Paladino}, \citenamefont {Galperin}, \citenamefont {Falci},\ and\
  \citenamefont {Altshuler}}]{Paladino2014}%
  \BibitemOpen
  \bibfield  {author} {\bibinfo {author} {\bibfnamefont {E.}~\bibnamefont
  {Paladino}}, \bibinfo {author} {\bibfnamefont {Y.~M.}\ \bibnamefont
  {Galperin}}, \bibinfo {author} {\bibfnamefont {G.}~\bibnamefont {Falci}},\
  and\ \bibinfo {author} {\bibfnamefont {B.~L.}\ \bibnamefont {Altshuler}},\
  }\bibfield  {title} {\bibinfo {title} {{${1}/{f}$} noise: Implications for
  solid-state quantum information},\ }\href
  {https://doi.org/10.1103/RevModPhys.86.361} {\bibfield  {journal} {\bibinfo
  {journal} {Rev. Mod. Phys.}\ }\textbf {\bibinfo {volume} {86}},\ \bibinfo
  {pages} {361} (\bibinfo {year} {2014})}\BibitemShut {NoStop}%
\bibitem [{\citenamefont {Burkard}\ \emph {et~al.}(2023)\citenamefont
  {Burkard}, \citenamefont {Ladd}, \citenamefont {Pan}, \citenamefont
  {Nichol},\ and\ \citenamefont {Petta}}]{Burkard2023}%
  \BibitemOpen
  \bibfield  {author} {\bibinfo {author} {\bibfnamefont {G.}~\bibnamefont
  {Burkard}}, \bibinfo {author} {\bibfnamefont {T.~D.}\ \bibnamefont {Ladd}},
  \bibinfo {author} {\bibfnamefont {A.}~\bibnamefont {Pan}}, \bibinfo {author}
  {\bibfnamefont {J.~M.}\ \bibnamefont {Nichol}},\ and\ \bibinfo {author}
  {\bibfnamefont {J.~R.}\ \bibnamefont {Petta}},\ }\bibfield  {title} {\bibinfo
  {title} {Semiconductor spin qubits},\ }\href
  {https://doi.org/10.1103/RevModPhys.95.025003} {\bibfield  {journal}
  {\bibinfo  {journal} {Rev. Mod. Phys.}\ }\textbf {\bibinfo {volume} {95}},\
  \bibinfo {pages} {025003} (\bibinfo {year} {2023})}\BibitemShut {NoStop}%
\bibitem [{\citenamefont {Reed}\ \emph {et~al.}(2016)\citenamefont {Reed},
  \citenamefont {Maune}, \citenamefont {Andrews}, \citenamefont {Borselli},
  \citenamefont {Eng}, \citenamefont {Jura}, \citenamefont {Kiselev},
  \citenamefont {Ladd}, \citenamefont {Merkel}, \citenamefont {Milosavljevic},
  \citenamefont {Pritchett}, \citenamefont {Rakher}, \citenamefont {Ross},
  \citenamefont {Schmitz}, \citenamefont {Smith}, \citenamefont {Wright},
  \citenamefont {Gyure},\ and\ \citenamefont {Hunter}}]{Reed2016}%
  \BibitemOpen
  \bibfield  {author} {\bibinfo {author} {\bibfnamefont {M.~D.}\ \bibnamefont
  {Reed}}, \bibinfo {author} {\bibfnamefont {B.~M.}\ \bibnamefont {Maune}},
  \bibinfo {author} {\bibfnamefont {R.~W.}\ \bibnamefont {Andrews}}, \bibinfo
  {author} {\bibfnamefont {M.~G.}\ \bibnamefont {Borselli}}, \bibinfo {author}
  {\bibfnamefont {K.}~\bibnamefont {Eng}}, \bibinfo {author} {\bibfnamefont
  {M.~P.}\ \bibnamefont {Jura}}, \bibinfo {author} {\bibfnamefont {A.~A.}\
  \bibnamefont {Kiselev}}, \bibinfo {author} {\bibfnamefont {T.~D.}\
  \bibnamefont {Ladd}}, \bibinfo {author} {\bibfnamefont {S.~T.}\ \bibnamefont
  {Merkel}}, \bibinfo {author} {\bibfnamefont {I.}~\bibnamefont
  {Milosavljevic}}, \bibinfo {author} {\bibfnamefont {E.~J.}\ \bibnamefont
  {Pritchett}}, \bibinfo {author} {\bibfnamefont {M.~T.}\ \bibnamefont
  {Rakher}}, \bibinfo {author} {\bibfnamefont {R.~S.}\ \bibnamefont {Ross}},
  \bibinfo {author} {\bibfnamefont {A.~E.}\ \bibnamefont {Schmitz}}, \bibinfo
  {author} {\bibfnamefont {A.}~\bibnamefont {Smith}}, \bibinfo {author}
  {\bibfnamefont {J.~A.}\ \bibnamefont {Wright}}, \bibinfo {author}
  {\bibfnamefont {M.~F.}\ \bibnamefont {Gyure}},\ and\ \bibinfo {author}
  {\bibfnamefont {A.~T.}\ \bibnamefont {Hunter}},\ }\bibfield  {title}
  {\bibinfo {title} {Reduced sensitivity to charge noise in semiconductor spin
  qubits via symmetric operation},\ }\href
  {https://doi.org/10.1103/PhysRevLett.116.110402} {\bibfield  {journal}
  {\bibinfo  {journal} {Phys. Rev. Lett.}\ }\textbf {\bibinfo {volume} {116}},\
  \bibinfo {pages} {110402} (\bibinfo {year} {2016})}\BibitemShut {NoStop}%
\bibitem [{\citenamefont {Martins}\ \emph {et~al.}(2016)\citenamefont
  {Martins}, \citenamefont {Malinowski}, \citenamefont {Nissen}, \citenamefont
  {Barnes}, \citenamefont {Fallahi}, \citenamefont {Gardner}, \citenamefont
  {Manfra}, \citenamefont {Marcus},\ and\ \citenamefont
  {Kuemmeth}}]{Martins2016}%
  \BibitemOpen
  \bibfield  {author} {\bibinfo {author} {\bibfnamefont {F.}~\bibnamefont
  {Martins}}, \bibinfo {author} {\bibfnamefont {F.~K.}\ \bibnamefont
  {Malinowski}}, \bibinfo {author} {\bibfnamefont {P.~D.}\ \bibnamefont
  {Nissen}}, \bibinfo {author} {\bibfnamefont {E.}~\bibnamefont {Barnes}},
  \bibinfo {author} {\bibfnamefont {S.}~\bibnamefont {Fallahi}}, \bibinfo
  {author} {\bibfnamefont {G.~C.}\ \bibnamefont {Gardner}}, \bibinfo {author}
  {\bibfnamefont {M.~J.}\ \bibnamefont {Manfra}}, \bibinfo {author}
  {\bibfnamefont {C.~M.}\ \bibnamefont {Marcus}},\ and\ \bibinfo {author}
  {\bibfnamefont {F.}~\bibnamefont {Kuemmeth}},\ }\bibfield  {title} {\bibinfo
  {title} {Noise suppression using symmetric exchange gates in spin qubits},\
  }\href {https://doi.org/10.1103/PhysRevLett.116.116801} {\bibfield  {journal}
  {\bibinfo  {journal} {Phys. Rev. Lett.}\ }\textbf {\bibinfo {volume} {116}},\
  \bibinfo {pages} {116801} (\bibinfo {year} {2016})}\BibitemShut {NoStop}%
\bibitem [{\citenamefont {Xue}\ \emph {et~al.}(2019)\citenamefont {Xue},
  \citenamefont {Watson}, \citenamefont {Helsen}, \citenamefont {Ward},
  \citenamefont {Savage}, \citenamefont {Lagally}, \citenamefont {Coppersmith},
  \citenamefont {Eriksson}, \citenamefont {Wehner},\ and\ \citenamefont
  {Vandersypen}}]{Xue2019}%
  \BibitemOpen
  \bibfield  {author} {\bibinfo {author} {\bibfnamefont {X.}~\bibnamefont
  {Xue}}, \bibinfo {author} {\bibfnamefont {T.~F.}\ \bibnamefont {Watson}},
  \bibinfo {author} {\bibfnamefont {J.}~\bibnamefont {Helsen}}, \bibinfo
  {author} {\bibfnamefont {D.~R.}\ \bibnamefont {Ward}}, \bibinfo {author}
  {\bibfnamefont {D.~E.}\ \bibnamefont {Savage}}, \bibinfo {author}
  {\bibfnamefont {M.~G.}\ \bibnamefont {Lagally}}, \bibinfo {author}
  {\bibfnamefont {S.~N.}\ \bibnamefont {Coppersmith}}, \bibinfo {author}
  {\bibfnamefont {M.~A.}\ \bibnamefont {Eriksson}}, \bibinfo {author}
  {\bibfnamefont {S.}~\bibnamefont {Wehner}},\ and\ \bibinfo {author}
  {\bibfnamefont {L.~M.~K.}\ \bibnamefont {Vandersypen}},\ }\bibfield  {title}
  {\bibinfo {title} {Benchmarking gate fidelities in a si/sige two-qubit
  device},\ }\href {https://doi.org/10.1103/PhysRevX.9.021011} {\bibfield
  {journal} {\bibinfo  {journal} {Phys. Rev. X}\ }\textbf {\bibinfo {volume}
  {9}},\ \bibinfo {pages} {021011} (\bibinfo {year} {2019})}\BibitemShut
  {NoStop}%
\bibitem [{\citenamefont {Kirton}\ and\ \citenamefont
  {Uren}(1989)}]{Kirton1989}%
  \BibitemOpen
  \bibfield  {author} {\bibinfo {author} {\bibfnamefont {M.}~\bibnamefont
  {Kirton}}\ and\ \bibinfo {author} {\bibfnamefont {M.}~\bibnamefont {Uren}},\
  }\bibfield  {title} {\bibinfo {title} {Noise in solid-state microstructures:
  A new perspective on individual defects, interface states and low-frequency
  (1/ƒ) noise},\ }\href {https://doi.org/10.1080/00018738900101122} {\bibfield
   {journal} {\bibinfo  {journal} {Advances in Physics}\ }\textbf {\bibinfo
  {volume} {38}},\ \bibinfo {pages} {367} (\bibinfo {year} {1989})}\BibitemShut
  {NoStop}%
\bibitem [{\citenamefont {M\"uller}\ \emph {et~al.}(2015)\citenamefont
  {M\"uller}, \citenamefont {Lisenfeld}, \citenamefont {Shnirman},\ and\
  \citenamefont {Poletto}}]{Muller2015}%
  \BibitemOpen
  \bibfield  {author} {\bibinfo {author} {\bibfnamefont {C.}~\bibnamefont
  {M\"uller}}, \bibinfo {author} {\bibfnamefont {J.}~\bibnamefont {Lisenfeld}},
  \bibinfo {author} {\bibfnamefont {A.}~\bibnamefont {Shnirman}},\ and\
  \bibinfo {author} {\bibfnamefont {S.}~\bibnamefont {Poletto}},\ }\bibfield
  {title} {\bibinfo {title} {Interacting two-level defects as sources of
  fluctuating high-frequency noise in superconducting circuits},\ }\href
  {https://doi.org/10.1103/PhysRevB.92.035442} {\bibfield  {journal} {\bibinfo
  {journal} {Phys. Rev. B}\ }\textbf {\bibinfo {volume} {92}},\ \bibinfo
  {pages} {035442} (\bibinfo {year} {2015})}\BibitemShut {NoStop}%
\bibitem [{\citenamefont {M{\"u}ller}\ \emph {et~al.}(2019)\citenamefont
  {M{\"u}ller}, \citenamefont {Cole},\ and\ \citenamefont
  {Lisenfeld}}]{Muller2019}%
  \BibitemOpen
  \bibfield  {author} {\bibinfo {author} {\bibfnamefont {C.}~\bibnamefont
  {M{\"u}ller}}, \bibinfo {author} {\bibfnamefont {J.~H.}\ \bibnamefont
  {Cole}},\ and\ \bibinfo {author} {\bibfnamefont {J.}~\bibnamefont
  {Lisenfeld}},\ }\bibfield  {title} {\bibinfo {title} {Towards understanding
  two-level-systems in amorphous solids: insights from quantum circuits},\
  }\href {https://doi.org/10.1088/1361-6633/ab3a7e} {\bibfield  {journal}
  {\bibinfo  {journal} {Reports on Progress in Physics}\ }\textbf {\bibinfo
  {volume} {82}},\ \bibinfo {pages} {124501} (\bibinfo {year}
  {2019})}\BibitemShut {NoStop}%
\bibitem [{\citenamefont {Ball}\ \emph {et~al.}(2016)\citenamefont {Ball},
  \citenamefont {Stace}, \citenamefont {Flammia},\ and\ \citenamefont
  {Biercuk}}]{Ball2016}%
  \BibitemOpen
  \bibfield  {author} {\bibinfo {author} {\bibfnamefont {H.}~\bibnamefont
  {Ball}}, \bibinfo {author} {\bibfnamefont {T.~M.}\ \bibnamefont {Stace}},
  \bibinfo {author} {\bibfnamefont {S.~T.}\ \bibnamefont {Flammia}},\ and\
  \bibinfo {author} {\bibfnamefont {M.~J.}\ \bibnamefont {Biercuk}},\
  }\bibfield  {title} {\bibinfo {title} {Effect of noise correlations on
  randomized benchmarking},\ }\href
  {https://doi.org/10.1103/PhysRevA.93.022303} {\bibfield  {journal} {\bibinfo
  {journal} {Phys. Rev. A}\ }\textbf {\bibinfo {volume} {93}},\ \bibinfo
  {pages} {022303} (\bibinfo {year} {2016})}\BibitemShut {NoStop}%
\bibitem [{\citenamefont {Qi}\ and\ \citenamefont {Ng}(2021)}]{Qi2021}%
  \BibitemOpen
  \bibfield  {author} {\bibinfo {author} {\bibfnamefont {J.}~\bibnamefont
  {Qi}}\ and\ \bibinfo {author} {\bibfnamefont {H.~K.}\ \bibnamefont {Ng}},\
  }\bibfield  {title} {\bibinfo {title} {Randomized benchmarking in the
  presence of time-correlated dephasing noise},\ }\href
  {https://doi.org/10.1103/PhysRevA.103.022607} {\bibfield  {journal} {\bibinfo
   {journal} {Phys. Rev. A}\ }\textbf {\bibinfo {volume} {103}},\ \bibinfo
  {pages} {022607} (\bibinfo {year} {2021})}\BibitemShut {NoStop}%
\bibitem [{\citenamefont {Bravyi}\ and\ \citenamefont
  {Maslov}(2021)}]{Bravyi2021}%
  \BibitemOpen
  \bibfield  {author} {\bibinfo {author} {\bibfnamefont {S.}~\bibnamefont
  {Bravyi}}\ and\ \bibinfo {author} {\bibfnamefont {D.}~\bibnamefont
  {Maslov}},\ }\bibfield  {title} {\bibinfo {title} {Hadamard-free circuits
  expose the structure of the clifford group},\ }\href
  {https://doi.org/10.1109/TIT.2021.3081415} {\bibfield  {journal} {\bibinfo
  {journal} {IEEE Transactions on Information Theory}\ }\textbf {\bibinfo
  {volume} {67}},\ \bibinfo {pages} {4546} (\bibinfo {year}
  {2021})}\BibitemShut {NoStop}%
\bibitem [{\citenamefont {McKay}\ \emph {et~al.}(2019)\citenamefont {McKay},
  \citenamefont {Sheldon}, \citenamefont {Smolin}, \citenamefont {Chow},\ and\
  \citenamefont {Gambetta}}]{McKay2019}%
  \BibitemOpen
  \bibfield  {author} {\bibinfo {author} {\bibfnamefont {D.~C.}\ \bibnamefont
  {McKay}}, \bibinfo {author} {\bibfnamefont {S.}~\bibnamefont {Sheldon}},
  \bibinfo {author} {\bibfnamefont {J.~A.}\ \bibnamefont {Smolin}}, \bibinfo
  {author} {\bibfnamefont {J.~M.}\ \bibnamefont {Chow}},\ and\ \bibinfo
  {author} {\bibfnamefont {J.~M.}\ \bibnamefont {Gambetta}},\ }\bibfield
  {title} {\bibinfo {title} {Three-qubit randomized benchmarking},\ }\href
  {https://doi.org/10.1103/PhysRevLett.122.200502} {\bibfield  {journal}
  {\bibinfo  {journal} {Phys. Rev. Lett.}\ }\textbf {\bibinfo {volume} {122}},\
  \bibinfo {pages} {200502} (\bibinfo {year} {2019})}\BibitemShut {NoStop}%
\bibitem [{\citenamefont {Proctor}\ \emph {et~al.}(2019)\citenamefont
  {Proctor}, \citenamefont {Carignan-Dugas}, \citenamefont {Rudinger},
  \citenamefont {Nielsen}, \citenamefont {Blume-Kohout},\ and\ \citenamefont
  {Young}}]{Proctor2019}%
  \BibitemOpen
  \bibfield  {author} {\bibinfo {author} {\bibfnamefont {T.~J.}\ \bibnamefont
  {Proctor}}, \bibinfo {author} {\bibfnamefont {A.}~\bibnamefont
  {Carignan-Dugas}}, \bibinfo {author} {\bibfnamefont {K.}~\bibnamefont
  {Rudinger}}, \bibinfo {author} {\bibfnamefont {E.}~\bibnamefont {Nielsen}},
  \bibinfo {author} {\bibfnamefont {R.}~\bibnamefont {Blume-Kohout}},\ and\
  \bibinfo {author} {\bibfnamefont {K.}~\bibnamefont {Young}},\ }\bibfield
  {title} {\bibinfo {title} {Direct randomized benchmarking for multiqubit
  devices},\ }\href {https://doi.org/10.1103/PhysRevLett.123.030503} {\bibfield
   {journal} {\bibinfo  {journal} {Phys. Rev. Lett.}\ }\textbf {\bibinfo
  {volume} {123}},\ \bibinfo {pages} {030503} (\bibinfo {year}
  {2019})}\BibitemShut {NoStop}%
\bibitem [{\citenamefont {Hines}\ \emph {et~al.}(2023)\citenamefont {Hines},
  \citenamefont {Lu}, \citenamefont {Naik}, \citenamefont {Hashim},
  \citenamefont {Ville}, \citenamefont {Mitchell}, \citenamefont {Kriekebaum},
  \citenamefont {Santiago}, \citenamefont {Seritan}, \citenamefont {Nielsen},
  \citenamefont {Blume-Kohout}, \citenamefont {Young}, \citenamefont {Siddiqi},
  \citenamefont {Whaley},\ and\ \citenamefont {Proctor}}]{Hines2023}%
  \BibitemOpen
  \bibfield  {author} {\bibinfo {author} {\bibfnamefont {J.}~\bibnamefont
  {Hines}}, \bibinfo {author} {\bibfnamefont {M.}~\bibnamefont {Lu}}, \bibinfo
  {author} {\bibfnamefont {R.~K.}\ \bibnamefont {Naik}}, \bibinfo {author}
  {\bibfnamefont {A.}~\bibnamefont {Hashim}}, \bibinfo {author} {\bibfnamefont
  {J.-L.}\ \bibnamefont {Ville}}, \bibinfo {author} {\bibfnamefont
  {B.}~\bibnamefont {Mitchell}}, \bibinfo {author} {\bibfnamefont {J.~M.}\
  \bibnamefont {Kriekebaum}}, \bibinfo {author} {\bibfnamefont {D.~I.}\
  \bibnamefont {Santiago}}, \bibinfo {author} {\bibfnamefont {S.}~\bibnamefont
  {Seritan}}, \bibinfo {author} {\bibfnamefont {E.}~\bibnamefont {Nielsen}},
  \bibinfo {author} {\bibfnamefont {R.}~\bibnamefont {Blume-Kohout}}, \bibinfo
  {author} {\bibfnamefont {K.}~\bibnamefont {Young}}, \bibinfo {author}
  {\bibfnamefont {I.}~\bibnamefont {Siddiqi}}, \bibinfo {author} {\bibfnamefont
  {B.}~\bibnamefont {Whaley}},\ and\ \bibinfo {author} {\bibfnamefont
  {T.}~\bibnamefont {Proctor}},\ }\bibfield  {title} {\bibinfo {title}
  {Demonstrating scalable randomized benchmarking of universal gate sets},\
  }\href {https://doi.org/10.1103/PhysRevX.13.041030} {\bibfield  {journal}
  {\bibinfo  {journal} {Phys. Rev. X}\ }\textbf {\bibinfo {volume} {13}},\
  \bibinfo {pages} {041030} (\bibinfo {year} {2023})}\BibitemShut {NoStop}%
\bibitem [{\citenamefont {Viola}\ and\ \citenamefont
  {Lloyd}(1998)}]{Viola1998}%
  \BibitemOpen
  \bibfield  {author} {\bibinfo {author} {\bibfnamefont {L.}~\bibnamefont
  {Viola}}\ and\ \bibinfo {author} {\bibfnamefont {S.}~\bibnamefont {Lloyd}},\
  }\bibfield  {title} {\bibinfo {title} {Dynamical suppression of decoherence
  in two-state quantum systems},\ }\href
  {https://doi.org/10.1103/PhysRevA.58.2733} {\bibfield  {journal} {\bibinfo
  {journal} {Phys. Rev. A}\ }\textbf {\bibinfo {volume} {58}},\ \bibinfo
  {pages} {2733} (\bibinfo {year} {1998})}\BibitemShut {NoStop}%
\bibitem [{\citenamefont {Viola}\ \emph {et~al.}(1999)\citenamefont {Viola},
  \citenamefont {Knill},\ and\ \citenamefont {Lloyd}}]{Viola1999}%
  \BibitemOpen
  \bibfield  {author} {\bibinfo {author} {\bibfnamefont {L.}~\bibnamefont
  {Viola}}, \bibinfo {author} {\bibfnamefont {E.}~\bibnamefont {Knill}},\ and\
  \bibinfo {author} {\bibfnamefont {S.}~\bibnamefont {Lloyd}},\ }\bibfield
  {title} {\bibinfo {title} {Dynamical decoupling of open quantum systems},\
  }\href {https://doi.org/10.1103/PhysRevLett.82.2417} {\bibfield  {journal}
  {\bibinfo  {journal} {Phys. Rev. Lett.}\ }\textbf {\bibinfo {volume} {82}},\
  \bibinfo {pages} {2417} (\bibinfo {year} {1999})}\BibitemShut {NoStop}%
\bibitem [{\citenamefont {Grenander}(1950)}]{Grenander1950}%
  \BibitemOpen
  \bibfield  {author} {\bibinfo {author} {\bibfnamefont {U.}~\bibnamefont
  {Grenander}},\ }\bibfield  {title} {\bibinfo {title} {{Stochastic processes
  and statistical inference}},\ }\href {https://doi.org/10.1007/BF02590638}
  {\bibfield  {journal} {\bibinfo  {journal} {Arkiv för Matematik}\ }\textbf
  {\bibinfo {volume} {1}},\ \bibinfo {pages} {195 } (\bibinfo {year}
  {1950})}\BibitemShut {NoStop}%
\bibitem [{\citenamefont {Fleischer}\ and\ \citenamefont
  {Kooharian}(1958)}]{Fleischer1958}%
  \BibitemOpen
  \bibfield  {author} {\bibinfo {author} {\bibfnamefont {I.}~\bibnamefont
  {Fleischer}}\ and\ \bibinfo {author} {\bibfnamefont {A.}~\bibnamefont
  {Kooharian}},\ }\bibfield  {title} {\bibinfo {title} {{On the Statistical
  Treatment of Stochastic Processes}},\ }\href
  {https://doi.org/10.1214/aoms/1177706629} {\bibfield  {journal} {\bibinfo
  {journal} {The Annals of Mathematical Statistics}\ }\textbf {\bibinfo
  {volume} {29}},\ \bibinfo {pages} {544 } (\bibinfo {year}
  {1958})}\BibitemShut {NoStop}%
\bibitem [{\citenamefont {Diggle}\ and\ \citenamefont
  {Fisher}(1991)}]{Diggle1991}%
  \BibitemOpen
  \bibfield  {author} {\bibinfo {author} {\bibfnamefont {P.~J.}\ \bibnamefont
  {Diggle}}\ and\ \bibinfo {author} {\bibfnamefont {N.~I.}\ \bibnamefont
  {Fisher}},\ }\bibfield  {title} {\bibinfo {title} {Nonparametric comparison
  of cumulative periodograms},\ }\href
  {https://rss.onlinelibrary.wiley.com/doi/abs/10.2307/2347522} {\bibfield
  {journal} {\bibinfo  {journal} {Journal of the Royal Statistical Society:
  Series C (Applied Statistics)}\ }\textbf {\bibinfo {volume} {40}},\ \bibinfo
  {pages} {423} (\bibinfo {year} {1991})}\BibitemShut {NoStop}%
\bibitem [{\citenamefont {Kim}\ and\ \citenamefont {Kish}(2006)}]{Kim2006}%
  \BibitemOpen
  \bibfield  {author} {\bibinfo {author} {\bibfnamefont {J.~U.}\ \bibnamefont
  {Kim}}\ and\ \bibinfo {author} {\bibfnamefont {L.~B.}\ \bibnamefont {Kish}},\
  }\bibfield  {title} {\bibinfo {title} {Recognizing different types of
  stochastic processes},\ }\href {https://doi.org/10.1142/S0219477506003082}
  {\bibfield  {journal} {\bibinfo  {journal} {Fluctuation and Noise Letters}\
  }\textbf {\bibinfo {volume} {06}},\ \bibinfo {pages} {L1} (\bibinfo {year}
  {2006})}\BibitemShut {NoStop}%
\bibitem [{\citenamefont {Maraj-Zygm{\c a}t}\ \emph {et~al.}(2023)\citenamefont
  {Maraj-Zygm{\c a}t}, \citenamefont {Sikora}, \citenamefont {Pitera},\ and\
  \citenamefont {Wy{\l}oma{\'n}ska}}]{MarajZygmat2023}%
  \BibitemOpen
  \bibfield  {author} {\bibinfo {author} {\bibfnamefont {K.}~\bibnamefont
  {Maraj-Zygm{\c a}t}}, \bibinfo {author} {\bibfnamefont {G.}~\bibnamefont
  {Sikora}}, \bibinfo {author} {\bibfnamefont {M.}~\bibnamefont {Pitera}},\
  and\ \bibinfo {author} {\bibfnamefont {A.}~\bibnamefont
  {Wy{\l}oma{\'n}ska}},\ }\bibfield  {title} {\bibinfo {title}
  {{Goodness-of-fit test for stochastic processes using even empirical moments
  statistic}},\ }\href {https://doi.org/10.1063/5.0111505} {\bibfield
  {journal} {\bibinfo  {journal} {Chaos: An Interdisciplinary Journal of
  Nonlinear Science}\ }\textbf {\bibinfo {volume} {33}},\ \bibinfo {pages}
  {013128} (\bibinfo {year} {2023})}\BibitemShut {NoStop}%
\bibitem [{\citenamefont {Khodjasteh}\ and\ \citenamefont
  {Viola}(2009)}]{Khodjasteh2009}%
  \BibitemOpen
  \bibfield  {author} {\bibinfo {author} {\bibfnamefont {K.}~\bibnamefont
  {Khodjasteh}}\ and\ \bibinfo {author} {\bibfnamefont {L.}~\bibnamefont
  {Viola}},\ }\bibfield  {title} {\bibinfo {title} {Dynamically error-corrected
  gates for universal quantum computation},\ }\href
  {https://doi.org/10.1103/PhysRevLett.102.080501} {\bibfield  {journal}
  {\bibinfo  {journal} {Phys. Rev. Lett.}\ }\textbf {\bibinfo {volume} {102}},\
  \bibinfo {pages} {080501} (\bibinfo {year} {2009})}\BibitemShut {NoStop}%
\bibitem [{\citenamefont {Wang}\ \emph
  {et~al.}(2014{\natexlab{a}})\citenamefont {Wang}, \citenamefont {Bishop},
  \citenamefont {Barnes}, \citenamefont {Kestner},\ and\ \citenamefont
  {Sarma}}]{Wang2014}%
  \BibitemOpen
  \bibfield  {author} {\bibinfo {author} {\bibfnamefont {X.}~\bibnamefont
  {Wang}}, \bibinfo {author} {\bibfnamefont {L.~S.}\ \bibnamefont {Bishop}},
  \bibinfo {author} {\bibfnamefont {E.}~\bibnamefont {Barnes}}, \bibinfo
  {author} {\bibfnamefont {J.~P.}\ \bibnamefont {Kestner}},\ and\ \bibinfo
  {author} {\bibfnamefont {S.~D.}\ \bibnamefont {Sarma}},\ }\bibfield  {title}
  {\bibinfo {title} {Robust quantum gates for singlet-triplet spin qubits using
  composite pulses},\ }\href {https://doi.org/10.1103/PhysRevA.89.022310}
  {\bibfield  {journal} {\bibinfo  {journal} {Phys. Rev. A}\ }\textbf {\bibinfo
  {volume} {89}},\ \bibinfo {pages} {022310} (\bibinfo {year}
  {2014}{\natexlab{a}})}\BibitemShut {NoStop}%
\bibitem [{\citenamefont {Wang}\ \emph
  {et~al.}(2014{\natexlab{b}})\citenamefont {Wang}, \citenamefont
  {Calderon-Vargas}, \citenamefont {Rana}, \citenamefont {Kestner},
  \citenamefont {Barnes},\ and\ \citenamefont {Das~Sarma}}]{Wang2014b}%
  \BibitemOpen
  \bibfield  {author} {\bibinfo {author} {\bibfnamefont {X.}~\bibnamefont
  {Wang}}, \bibinfo {author} {\bibfnamefont {F.~A.}\ \bibnamefont
  {Calderon-Vargas}}, \bibinfo {author} {\bibfnamefont {M.~S.}\ \bibnamefont
  {Rana}}, \bibinfo {author} {\bibfnamefont {J.~P.}\ \bibnamefont {Kestner}},
  \bibinfo {author} {\bibfnamefont {E.}~\bibnamefont {Barnes}},\ and\ \bibinfo
  {author} {\bibfnamefont {S.}~\bibnamefont {Das~Sarma}},\ }\bibfield  {title}
  {\bibinfo {title} {Noise-compensating pulses for electrostatically controlled
  silicon spin qubits},\ }\href {https://doi.org/10.1103/PhysRevB.90.155306}
  {\bibfield  {journal} {\bibinfo  {journal} {Phys. Rev. B}\ }\textbf {\bibinfo
  {volume} {90}},\ \bibinfo {pages} {155306} (\bibinfo {year}
  {2014}{\natexlab{b}})}\BibitemShut {NoStop}%
\bibitem [{\citenamefont {Kaulakys}\ \emph {et~al.}(2005)\citenamefont
  {Kaulakys}, \citenamefont {Gontis},\ and\ \citenamefont
  {Alaburda}}]{Kaulakys2005}%
  \BibitemOpen
  \bibfield  {author} {\bibinfo {author} {\bibfnamefont {B.}~\bibnamefont
  {Kaulakys}}, \bibinfo {author} {\bibfnamefont {V.}~\bibnamefont {Gontis}},\
  and\ \bibinfo {author} {\bibfnamefont {M.}~\bibnamefont {Alaburda}},\
  }\bibfield  {title} {\bibinfo {title} {Point process model of {$1/f$} noise
  vs a sum of lorentzians},\ }\href
  {https://doi.org/10.1103/PhysRevE.71.051105} {\bibfield  {journal} {\bibinfo
  {journal} {Phys. Rev. E}\ }\textbf {\bibinfo {volume} {71}},\ \bibinfo
  {pages} {051105} (\bibinfo {year} {2005})}\BibitemShut {NoStop}%
\bibitem [{\citenamefont {Virtanen}\ \emph {et~al.}(2020)\citenamefont
  {Virtanen}, \citenamefont {Gommers}, \citenamefont {Oliphant}, \citenamefont
  {Haberland}, \citenamefont {Reddy}, \citenamefont {Cournapeau}, \citenamefont
  {Burovski}, \citenamefont {Peterson}, \citenamefont {Weckesser},
  \citenamefont {Bright}, \citenamefont {{van der Walt}}, \citenamefont
  {Brett}, \citenamefont {Wilson}, \citenamefont {Millman}, \citenamefont
  {Mayorov}, \citenamefont {Nelson}, \citenamefont {Jones}, \citenamefont
  {Kern}, \citenamefont {Larson}, \citenamefont {Carey}, \citenamefont {Polat},
  \citenamefont {Feng}, \citenamefont {Moore}, \citenamefont {{VanderPlas}},
  \citenamefont {Laxalde}, \citenamefont {Perktold}, \citenamefont {Cimrman},
  \citenamefont {Henriksen}, \citenamefont {Quintero}, \citenamefont {Harris},
  \citenamefont {Archibald}, \citenamefont {Ribeiro}, \citenamefont
  {Pedregosa}, \citenamefont {{van Mulbregt}},\ and\ \citenamefont {{SciPy 1.0
  Contributors}}}]{2020SciPy-NMeth}%
  \BibitemOpen
  \bibfield  {author} {\bibinfo {author} {\bibfnamefont {P.}~\bibnamefont
  {Virtanen}}, \bibinfo {author} {\bibfnamefont {R.}~\bibnamefont {Gommers}},
  \bibinfo {author} {\bibfnamefont {T.~E.}\ \bibnamefont {Oliphant}}, \bibinfo
  {author} {\bibfnamefont {M.}~\bibnamefont {Haberland}}, \bibinfo {author}
  {\bibfnamefont {T.}~\bibnamefont {Reddy}}, \bibinfo {author} {\bibfnamefont
  {D.}~\bibnamefont {Cournapeau}}, \bibinfo {author} {\bibfnamefont
  {E.}~\bibnamefont {Burovski}}, \bibinfo {author} {\bibfnamefont
  {P.}~\bibnamefont {Peterson}}, \bibinfo {author} {\bibfnamefont
  {W.}~\bibnamefont {Weckesser}}, \bibinfo {author} {\bibfnamefont
  {J.}~\bibnamefont {Bright}}, \bibinfo {author} {\bibfnamefont {S.~J.}\
  \bibnamefont {{van der Walt}}}, \bibinfo {author} {\bibfnamefont
  {M.}~\bibnamefont {Brett}}, \bibinfo {author} {\bibfnamefont
  {J.}~\bibnamefont {Wilson}}, \bibinfo {author} {\bibfnamefont {K.~J.}\
  \bibnamefont {Millman}}, \bibinfo {author} {\bibfnamefont {N.}~\bibnamefont
  {Mayorov}}, \bibinfo {author} {\bibfnamefont {A.~R.~J.}\ \bibnamefont
  {Nelson}}, \bibinfo {author} {\bibfnamefont {E.}~\bibnamefont {Jones}},
  \bibinfo {author} {\bibfnamefont {R.}~\bibnamefont {Kern}}, \bibinfo {author}
  {\bibfnamefont {E.}~\bibnamefont {Larson}}, \bibinfo {author} {\bibfnamefont
  {C.~J.}\ \bibnamefont {Carey}}, \bibinfo {author} {\bibfnamefont
  {{\.I}.}~\bibnamefont {Polat}}, \bibinfo {author} {\bibfnamefont
  {Y.}~\bibnamefont {Feng}}, \bibinfo {author} {\bibfnamefont {E.~W.}\
  \bibnamefont {Moore}}, \bibinfo {author} {\bibfnamefont {J.}~\bibnamefont
  {{VanderPlas}}}, \bibinfo {author} {\bibfnamefont {D.}~\bibnamefont
  {Laxalde}}, \bibinfo {author} {\bibfnamefont {J.}~\bibnamefont {Perktold}},
  \bibinfo {author} {\bibfnamefont {R.}~\bibnamefont {Cimrman}}, \bibinfo
  {author} {\bibfnamefont {I.}~\bibnamefont {Henriksen}}, \bibinfo {author}
  {\bibfnamefont {E.~A.}\ \bibnamefont {Quintero}}, \bibinfo {author}
  {\bibfnamefont {C.~R.}\ \bibnamefont {Harris}}, \bibinfo {author}
  {\bibfnamefont {A.~M.}\ \bibnamefont {Archibald}}, \bibinfo {author}
  {\bibfnamefont {A.~H.}\ \bibnamefont {Ribeiro}}, \bibinfo {author}
  {\bibfnamefont {F.}~\bibnamefont {Pedregosa}}, \bibinfo {author}
  {\bibfnamefont {P.}~\bibnamefont {{van Mulbregt}}},\ and\ \bibinfo {author}
  {\bibnamefont {{SciPy 1.0 Contributors}}},\ }\bibfield  {title} {\bibinfo
  {title} {{{SciPy} 1.0: Fundamental Algorithms for Scientific Computing in
  Python}},\ }\href {https://doi.org/10.1038/s41592-019-0686-2} {\bibfield
  {journal} {\bibinfo  {journal} {Nature Methods}\ }\textbf {\bibinfo {volume}
  {17}},\ \bibinfo {pages} {261} (\bibinfo {year} {2020})}\BibitemShut
  {NoStop}%
\bibitem [{\citenamefont {Kasdin}(1995)}]{Kasdin1995}%
  \BibitemOpen
  \bibfield  {author} {\bibinfo {author} {\bibfnamefont {N.}~\bibnamefont
  {Kasdin}},\ }\bibfield  {title} {\bibinfo {title} {Discrete simulation of
  colored noise and stochastic processes and $1/f^\alpha$ power law noise
  generation},\ }\href {https://doi.org/10.1109/5.381848} {\bibfield  {journal}
  {\bibinfo  {journal} {Proceedings of the IEEE}\ }\textbf {\bibinfo {volume}
  {83}},\ \bibinfo {pages} {802} (\bibinfo {year} {1995})}\BibitemShut
  {NoStop}%
\end{thebibliography}
%

\appendix
\section{Simulation Parameters} \label{app:SimParameters}

In our simulations, we choose the following parameters for our singlet-triplet Hamiltonian: a static magnetic field gradient of $\Delta b_z/h=10\MHz$, a residual exchange of $J_0/h=0.075\MHz$, and an insensitivity of $\mathcal{I}=18 \mathrm{mV}$. Our assumed value of $\mathcal{I}$ falls in the range reported in Ref.~\cite{Reed2016}. 
For our applied voltages $V_0(t)$, we target a $\pi$ rotation about the $z$-axis to be around $100\mV$ for a $30\ns$ voltage pulse, with full baseband control of the qubit in the $70\mV-120\mV$ range. These operating ranges of the singlet-triplet qubit are in line with other work~\cite{Reed2016}. 

\section{Simulation framework}\label{sec:simulation_framework}
In this section, we outline how we simulate state evolution under the singlet-triplet Hamiltonian $H(t,J(t),\Delta b_z(t))$ described by
\begin{equation}
    H(t,J(t),\Delta b_z(t))=\frac{J(t)}2\sigma^z+\frac{\Delta b_z(t)}2\sigma^x \ ,
\end{equation}
where $J(t)$ is an exchange pulse sequence controlled electrically, and $\Delta b_z(t)$ is assumed to be approximately constant in time. To implement arbitrary single qubit operations, we optimize the exchange pulse sequence to implement said operation.
\subsection{State propagation}\label{sec:state_propagation}
Given some time-dependent Hamiltonian $H(t,J(t),\Delta b_z(t))$, our simulator approximates unitaries using a first order Magnus expansion:
\begin{eqnarray}
    U(t,J(t),\Delta b_z(t)) &\approx& \nonumber \\
    && \hspace{-2cm} \prod_{k=1}^{t/\Delta t}\exp\br{-\frac{i\Delta t}{\hbar}H(k\Delta t,J(t),\Delta b_z(t))} \ .
    \label{eq:approximate_unitary}
\end{eqnarray}
In Eq.~\eqref{eq:approximate_unitary}, $\Delta t=1/f_s$ is the sampling time (chosen to be on the order of nanoseconds; see Appendix~\ref{sec:discrete_sampling_effects} for justification of this choice). 
In our simulations, we utilize the assumption that relaxation times are long relative to the coherence lifetime of the qubit; if this assumption cannot be confidently made, the propagation method should be updated to account for relaxation effects.

\subsection{Operator compilation}\label{sec:operator_compilation}
In order to implement a target unitary operation $U_\mathrm{tgt}$, we specify $J(t)$ and an elapsed time $T$ such that $U(T,J(t),\Delta b_z(t))$ implements $U_\mathrm{tgt}$. We find the exchange pulse sequence $J(t)$ via optimization, which is the topic of the following sections. Pulse optimization techniques have been the subject of extensive prior research and can be chosen to achieve qualities such as noise robustness~\cite{Khodjasteh2009,Wang2014,Wang2014b} or to account for additional hardware-related considerations~\cite{Cerfontaine2020b}. In our work, we perform straightforward pulse-shaping and optimization as described in the following sections.
\subsubsection{Pulse sequence parameterization}
There are a number of considerations to be made when constructing a pulse sequence $J(t)$. The first is the specification of the control time series ``shape'' parameters: number of pulses, pulse duration(s), rise/peak/decay times, and elapsed time. These parameters should be chosen to reflect pulse sequences created by realistic AWGs. In our simulations, we target sequences with 1-5 discrete pulses, a peak pulse duration of $30-50\ns$, and rise/decay times of approximately $4-6\ns$. To create rounded pulses from square pulses, we utilize Gaussian convolution with standard deviation determining the rise/decay time.
\subsubsection{Objective function}\label{sec:objective_function}
\begin{figure}[h]
    \centering
    \includegraphics[width=\linewidth]{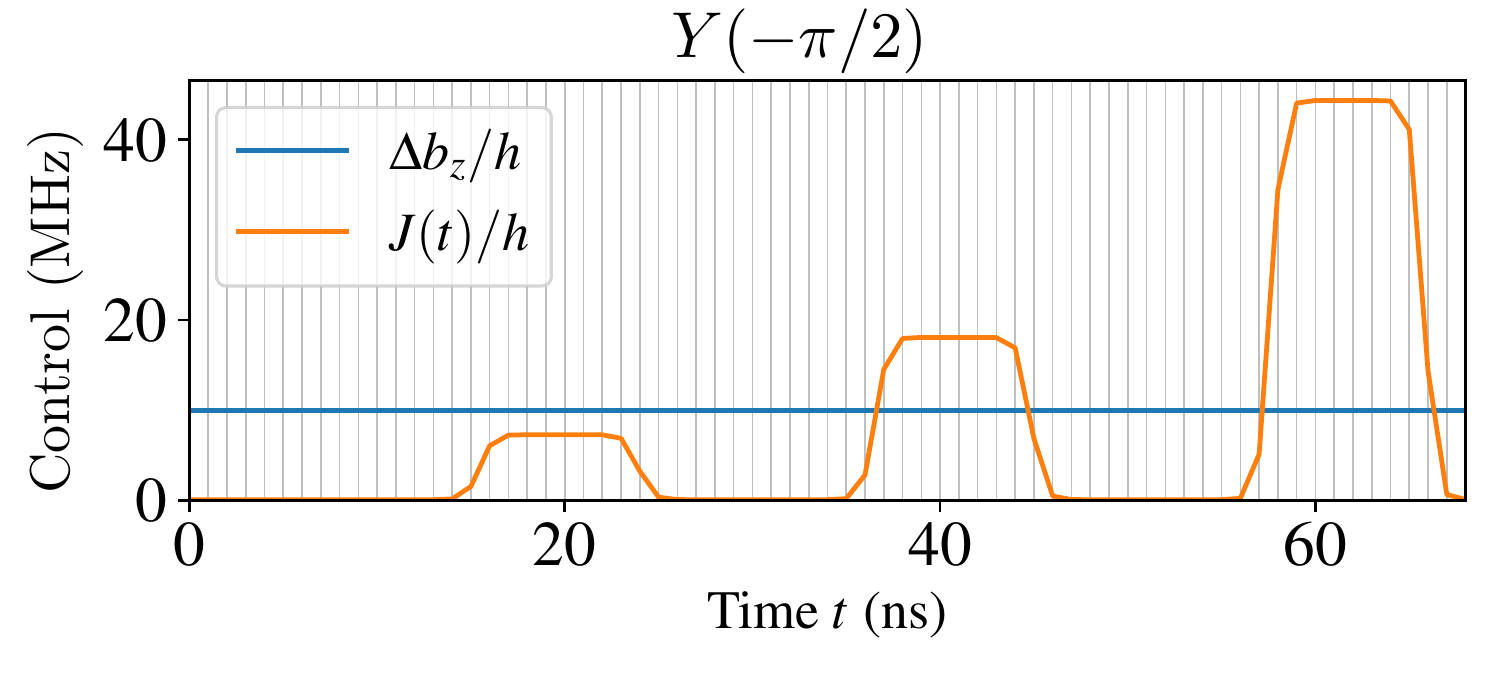}
    \caption{\textbf{Example Clifford pulse schedule.} A sample exchange pulse sequence ($J(t)$, orange) for a single Clifford operator $Y(-\pi/2)$ (a rotation about the $y$-axis by an angle $-\pi/2$), under a constant field gradient $\Delta b_z/h=10\MHz$ (blue). The infidelity associated with this operator in the absence of voltage and magnetic noise is on the order of $10^{-7}$. The operator above uses a sampling time of $\Delta t=1\ns$ (vertical gray) with a residual exchange of $J_0/h=75\kHz\ll\Delta b_z$.}
    \label{fig:sample_pulse_sequence}
\end{figure}

In our optimizations, we neglect charge and magnetic noise such that the effective field gradient is a constant $\Delta b_z(t)=\Delta b_z$. To construct an exchange pulse sequence that implements some arbitrary operator $U_\mathrm{tgt}$, we use the unitary fidelity as the objective function in our optimizations:
\begin{equation}
	J(t)=\argmax_{J_c(t)}\frac14\absq{\Tr{U_\mathrm{tgt}^\dag U(T,J_c(t),\Delta b_z)}} \ .
	\label{eq:objective_function}
\end{equation}
In Eq.~\eqref{eq:objective_function}, $U(T,J_c(t),\Delta b_z)$ is the unitary described in Eq.~\eqref{eq:approximate_unitary} with elapsed time $T$, candidate exchange pulse sequence $J_c(t)$, and field gradient $\Delta b_z$. It is generated by $H(t)=J_c(t)\sigma^z/2+\Delta b_z\sigma^x/2$ with $t\in[0,T]$. Generally speaking, there are multiple pulse sequences $J(t)$ solving the maximization problem in Eq.~\eqref{eq:objective_function}. A ``preferred'' exchange pulse sequence may be chosen from this set, as determined by factors such as timing constraints or noise sensitivities. 
We show a sample exchange pulse sequence $J(t)$ for a single Clifford operation in Fig~\ref{fig:sample_pulse_sequence}.

\subsubsection{Operation sequences}\label{sec:operation_sequences}
Let $J_i(t)$ be the exchange pulse sequence implementing an operator $U_{\mathrm{tgt},i}$, with the corresponding approximate unitary. A sequence of operations $U_{\mathrm{tgt},1}, U_{\mathrm{tgt},2}, ...$ is described by concatenating pulse sequences $J_1(t),J_2(t), ...$, defined on $[0,T_1), [0,T_2),...$. That is,
\begin{equation}
\begin{split}
    \prod_iU_{\mathrm{tgt},i}&\approx U\pr{\sum_iT_i+\tilde T_i,J'(t)+\delta J(t),\Delta b_z(t)}\\
    \textrm{with }J'(t)&\equiv\sum_iJ_i\pr{t-t_i}\mathbf1_{[t_i,t_i+T_i)}(t)\\&\quad\quad\phantom{ii}+J_0\mathbf1_{[t_i+T_i,t_i+T_i+\tilde T_i)}(t)\\
    t_i&\equiv \sum_{j<i}T_j+\tilde T_j
\end{split}
    \label{eq:operation_sequences}
\end{equation}
In Eq.~\eqref{eq:operation_sequences}, $T_i$ is the elapsed time of the $i$-th sequence, $\tilde T_i$ is an associated idling time after it, and $t_i,t_i+T_i$ are the sequence's start and finish times. $\Delta b_z(t)=\Delta b_z+\delta b_z(t)$ is the noisy field gradient term defined over the entire sequence of operations, $[0, \sum_iT_i+\tilde T_i)$. $\delta J(t)$ is a time-dependent exchange noise term, defined over this same interval. $J'(t)$ is the concatenation of all pulse sequences $J_i(t)$, such that the $i$-th exchange pulse sequence is ``active'' during the interval $\left[t_i,t_i+T_i\right)$. The residual exchange $J_0$ is active during the idling intervals $[t_i+T_i,t_i+T_i+\tilde T_i)$, although it may be ignored given $J_0\ll\Delta b_z$.

\subsubsection{Single-qubit Clifford gate set}
To perform randomized benchmarking, we first compile the single-qubit Clifford gates in terms of our exchange pulse sequences. Let $C_i$ be a single-qubit Clifford gate, and $J_i(t)$ be a pulse sequence implementing $C_i$ (i.e., $J_i(t)$ solves Eq~\eqref{eq:objective_function} with $U_\mathrm{tgt}=C_i$). $J_i(t)$ may be found in two ways: by directly optimizing the objective function in Eq.~\eqref{eq:objective_function}, or by representing $C_i$ in terms of generators and concatenating optimal pulse sequences implementing these generators. The former is more computationally expensive, but can result in more efficient operations. The latter is less expensive (especially as the number of qubits increases) and can be closer to native gate sets on hardware of interest but may result in much longer operations.

Let $\mathcal C_1$ be the generators of the single-qubit Clifford group $\mathbf C_1$, so each Clifford $C_i\in \mathbf C_1$ may be written as a product of $\mathcal C_1$'s elements. We may choose $\mathcal C_1=
 \sr{\identity,X(\pm\pi/2), Z(\pm\pi/2),X(\pi),Z(\pm\pi)}$ to generate $\mathbf C_1$~\cite{Xue2019}.
\begin{equation}
    C_i=\prod_{j}G_{k_j}\quad (G_{k_j}\in\mathcal C_1) \ .
    \label{eq:clifford_generators}
\end{equation}
The pulse sequence $j_k(t)$ implementing a generator $G_k\in\mathcal C_1$ solves Eq.~\eqref{eq:objective_function} with $U_\mathrm{tgt}=G_k$. Using Eq.~\eqref{eq:operation_sequences} and the generator representation of Cliffords, $C_i$ may be simulated as
\begin{equation}
    C_i\approx U\pr{\sum_{j}T_{k_j}, J_i(t)+\delta J(t),\Delta b_z(t)} \ ,
    \label{eq:clifford_generators2}
\end{equation}
with $J_i(t)=\sum_{j}j_{k_j}(t-t_{k_j})\mathbf1_{[t_{k_j},t_{k_j}+T_{k_j})}(t)$ and $t_{k_j}=\sum_{l<j}T_{k_l}$.
In Eq.~\eqref{eq:clifford_generators2}, $T_{k_j}$ is the elapsed time of the pulse sequence $j_{k_j}(t)$ which implements generator $G_{k_j}$. There is no idling between generators. In our simulations, we use a tabulation of Cliffords in terms of the generators $\identity$, $X(\pm\pi/2)$ and $Z(\pm\pi/2)$, as given in Ref.~\cite{Xue2019}.

\section{Noise propagation}\label{sec:noise_propagation}
A computationally efficient method for generating $1/f^\alpha$ noise is of high importance for simulating the effects of correlated noise on qubit fidelity. In the context of ``wall-clock'' simulations, it is especially useful to avoid generating noise during long idling windows such as during state preparation and measurement; this is the primary reason for our need of a stateful noise model that can generate correlated noise. One convenient method for generating stateful correlated noise is by adding independent Ornstein-Uhlenbeck (OU) processes, which has been previously shown to be able to mimic $1/f^\alpha$ noise~\cite{Kaulakys2005}. 

Let $\eta_i(t)$ be independent OU processes, each characterized by power $p_i/2$ and frequency $f_i$ as described in Sec.~\ref{sec:noise_model_and_calibration}. The correlation function of the individual OU process is given by~\cite{Kaulakys2005} 
\begin{equation}
    \expect{\eta_i(t)\eta_j(t+\tau)}=\frac{p_i}{2}e^{-2\pi f_i\abs\tau}\delta_{ij} \ .
\end{equation}
The noise signal $\eta(t)$ is modeled as a sum of these processes, $\eta(t) = \sum_i\eta_i(t)$. The autocorrelation function of the noise signal is therefore given by
\begin{equation}
    \expect{\eta(t)\eta(t+\tau)} = \sum_i\frac{p_i}{2}e^{-2\pi f_i\abs\tau} \ .
    \label{eq:correlation_function}
\end{equation}
The power spectral density $S(f)$ of $\eta(t)$ is derived from the Fourier transform of the autocorrelation function:
\begin{equation}
    S(f) = \sum_i\frac{p_if_i}{\pi(f^2+f_i^2)} \ .
\end{equation}
The distribution associated with the process $\eta(t)$ is a Gaussian, with mean $\sum_i \langle \eta_i(t) \rangle$ and variance $\sum_ip_i/2$~\cite{Kaulakys2005}. In our simulations, we take $\expect{\eta_i(0)}=0$ for all $i$. By the stationarity of OU processes, we have $\expect{\eta_i(t)}=0$, consequently giving $\expect{\eta(t)}=0$. The implication of this constraint is that our noise processes $\eta(t)$ are zero-mean Gaussians.

\section{Analytical results for free induction decay}\label{sec:analytical_results_for_free_induction_decay}
\subsection{Charge $T_2^*$}\label{sec:charge_T2}
Consider the first version of FID introduced in Sec.~\ref{sec:noise_model_and_calibration}. Given the field gradient term is turned off ($\Delta b_z(t) = 0$), a singlet-triplet qubit will evolve according to the Hamiltonian $H_{ST_0}(t) = J(t)\sigma^z/2 = J_0e^{(V_{\rm FID}+\delta V(t))/\mathcal{I}}\sigma^z/2 = (J_{\rm FID}+\delta J(t))\sigma^z/2$, where $V_{\rm FID}$ is a noiseless applied voltage, $\delta V(t)$ is the charge noise term, $J_{\rm FID}=J_0e^{V_{\rm FID}/\mathcal{I}}$ is the exchange driven at $V_{\rm FID}$, and $\delta J(t) = J(e^{\delta V(t) / \mathcal{I}}-1)$ is the exchange noise. Given the singlet-triplet qubit is prepared in the $\ket{+} = (\ket{0}+\ket{1})/\sqrt2$ state, the mean probability of returning to this state at time $t\geq0$ is
\begin{equation}
    \expect{P(+ | +,(V(t),0),t)}=\expect{\cos^2\br{\frac{1}{2\hbar}\int_0^tJ(t')dt'}} \ .
\end{equation}
For small charge noise, we can approximate $\delta J(t)$ with $1+J_{\rm FID}\delta V(t)/\mathcal{I}$, and we then have
\begin{eqnarray}
\expect{P(+|+,(V(t),0),t)}&\approx&\frac12+\frac12\expect{\cos\br{\delta\theta_V(t)}}\cos\br{\frac{J_{\rm FID}t}{h}} \nonumber \\
&& \hspace{-1cm} -\frac12\expect{\sin\br{\delta\theta_V(t)}}\sin\br{\frac{J_{\rm FID}t}{h}} \ ,
\end{eqnarray}
where we have defined $\delta\theta_V(t)\equiv\frac{ J}{\mathcal{I} \hbar}\int_0^t\delta V(t')dt'$.
Since $\delta V(t)$ is a zero-mean Gaussian, $\delta\theta_V(t)$ is as well. Let $\sigma_{\delta\theta_V}^2(t)$ denote the variance of $\delta\theta_V(t)$. We have $\expect{\delta\theta_V(t)^{2k+1}}=0$ and $\expect{\delta\theta_V(t)^{2k}}=\frac{(2k)!}{2^kk!}\sigma_{\delta\theta_V}^{2k}(t)$ for integers $k$. Consequently, we have
\begin{eqnarray}
\expect{\cos\br{\delta\theta_V(t)}}&=&\frac12\sum_{n=0}^\infty\frac1{n!}\expect{(i\delta\theta_V(t))^n}(1+(-1)^n)  \nonumber \\
&=&e^{-\sigma_{\delta\theta_V}^{2}(t)/2} \ ,\nonumber \\
\expect{\sin\br{\delta\theta_V(t)}}&=&\frac1{2i}\sum_{n=0}^\infty\frac1{n!}\expect{(i\delta\theta_V(t))^n}(1-(-1)^n) \nonumber \\
&=&0 \ .
\end{eqnarray}
Substituting this result in our expression for the return probability, we have
\begin{eqnarray}
    \expect{P(+|+,(V(t),0),t)}&\approx& \nonumber \\
    && \hspace{-2cm} \frac12\pr{1+e^{-\sigma_{\delta\theta_V}^{2}(t)/2}\cos\br{\frac{J_{\rm FID}t}{h}}} \ .
\end{eqnarray}
The variance $\sigma_{\delta\theta_V(t)}^{2}$ of $\delta\theta_V(t)$ is computed directly from its definition, as
\begin{eqnarray}
\sigma_{\delta\theta_V}^{2}(t)&=&\pr{\frac{J_{\rm FID}}{\mathcal{I} \hbar}}^2\expect{\pr{\int_0^t\delta V(t')dt'}^2} \nonumber \\
&=&\pr{\frac{ J_{\rm FID}}{\mathcal{I} \hbar}}^2\iint_0^t\expect{\delta V(t_1)\delta V(t_2)}dt_1dt_2 \ .
\end{eqnarray}
Since $\delta V(t)$ is a sum of independent OU processes $\delta V_i(t)$ associated with powers $p_i/2$ and characteristic frequencies $f_i$, the above can be rewritten in terms of the autocorrelation functions of these independent processes (see Eq.~\eqref{eq:correlation_function}):
\begin{eqnarray}
\sigma_{\delta\theta_V}^{2}(t)&=&\pr{\frac{ J_{\rm FID}}{\mathcal{I} \hbar}}^2\sum_i\iint_0^t\expect{\delta V_i(t_1)\delta V_i(t_2)}dt_1dt_2 \nonumber \\
&=&\pr{\frac{ J_{\rm FID}}{\mathcal{I}  \sqrt2\hbar}}^2\sum_i p_i\iint_0^te^{-2\pi f_i\abs{t_1-t_2}}dt_1dt_2 \nonumber \\
&=&\pr{\frac{ J_{\rm FID}}{\mathcal{I}  h}}^2\sum_i \frac{p_i}{f_i^2}\pr{e^{-2\pi f_it}+2\pi f_it-1} \ . \nonumber \\
\end{eqnarray}
The quantity $\sigma_{\delta\theta_V}^{2}(t)$ corresponds to the decay time of Ramsey oscillations, and its approximate quadratic scaling with $t$ in the $f_it\ll1$ limit shows that the envelope will be close to Gaussian. The charge $T_2^*$ should then solve $\sigma_{\delta\theta_V}^{2}(T_2^*)/2=1$.

\subsection{Magnetic $T_2^*$}\label{sec:magnetic_T2}
The derivation for mean Ramsey oscillations under magnetic noise with exchange off is similar to but simpler than the charge noise case. We now have the singlet state $\ket 0$ as the initial state evolving under the Hamiltonian $H_{ST_0}(t) = \Delta b_z(t)\sigma^x/2 = (\Delta b_z + \delta b_z(t))\sigma^x/2$, with $\Delta b_z$ the noiseless field gradient and $\delta b_z(t)$ the magnetic noise. We follow the same steps as in the previous section, but with $\ket{+}\to\ket 0$, $J_{\rm FID}\to\Delta b_z$, $\delta J(t)\to \delta b_z(t)$; in contrast to the charge noise case, there is no need to make a linear approximation of the magnetic noise, so there is no magnetic noise analog of the additional factor $( J /\mathcal{I})^2$ in $\sigma_{\delta\theta_{b_z}(t)}^2$ from the previous section. The analytical results for free induction decay under magnetic noise are thus written as
\begin{eqnarray}
    \expect{P(0|0,(0,\Delta b_z(t)),t)}&=& \nonumber \\
    && \hspace{-1cm} \frac12\pr{1+e^{-\sigma_{\delta\theta_{b_z}(t)}^{2}/2}\cos\br{\frac{\Delta b_zt}{h}}} \ , \nonumber \\
    \sigma_{\delta\theta_{b_z}}^{2}(t)&=& \nonumber \\
    && \hspace{-1cm} \frac1{h^2}\sum_i \frac{p_i}{f_i^2}\pr{e^{-2\pi f_it}+2\pi f_it-1}  \ , \nonumber \\
    \sigma_{\delta\theta_{b_z}}^{2}(T_2^*)/2&=&1 \ .
\end{eqnarray}
\section{RB fit quality}\label{sec:rb_fit_quality}
\begin{figure}[h]
    \centering
    \includegraphics[width=\linewidth]{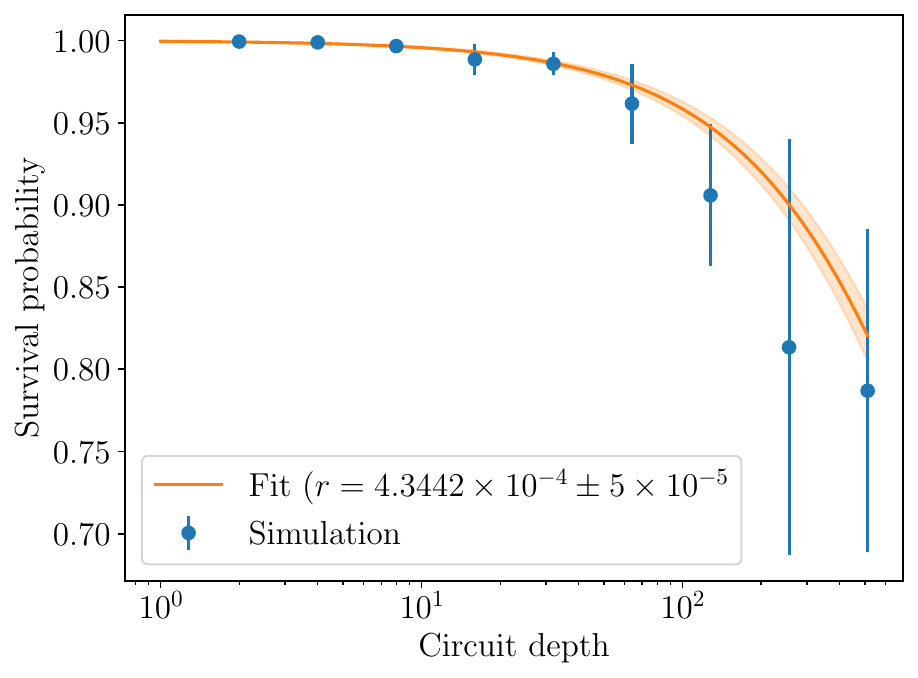}
    \caption{Survival probability of a single RB pass using Model 7. Each simulation data point is the mean probability over 10 circuits at a given depth, and the error bars are the $2\sigma$ confidence interval calculated using a bootstrap over the 10 circuits performed for each depth.  We identify the fit parameter $r$ in Eq.~\eqref{eqt:RBfit} using the SciPy \emph{curve\_fit} function \cite{2020SciPy-NMeth}, and the orange shaded region corresponds to the $2 \sigma$ confidence interval generated by the fit.}
    \label{fig:rb_decay_fit}
\end{figure}

\begin{figure}[h]
    \centering
    \includegraphics[width=\linewidth]{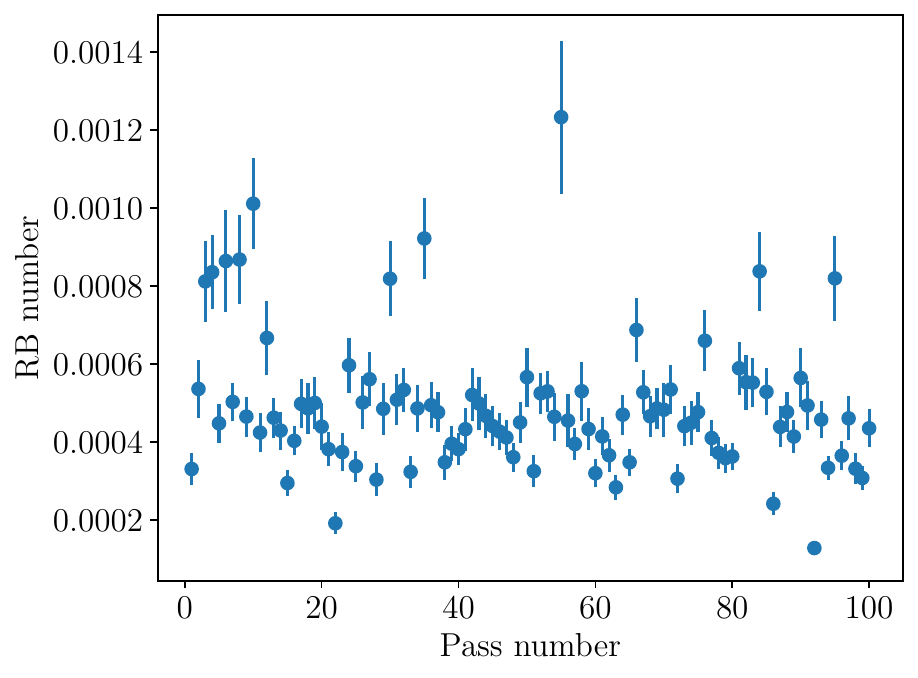}
    \caption{Pass-to-pass RB number using Model 7 for a single noise realization. We identify the fit parameter $r$ in Eq.~\eqref{eqt:RBfit} using the SciPy \emph{curve\_fit} function \cite{2020SciPy-NMeth}, and the error bars corresponds to the $2 \sigma$ confidence interval of fit.}
    \label{fig:rb_fit_passtopass}
\end{figure}

%
In Fig.~\ref{fig:rb_decay_fit}, we give an example of the results of a single RB pass for Model 7.  We see that at the largest depth, the variation in survival probabilities between the 10 different circuits can be large.  We show in Fig.~\ref{fig:rb_fit_passtopass} how the RB number varies from pass-to-pass for one noise realization.

\section{KS-Test} \label{app:KS}
%

\subsubsection{Empirical distributions}
In order to define the K-S statistic, we must first compute the empirical cumulative distribution function (eCDF) of the metric that we use in our tests. If $X_i$ is one out of $n$ samples of this metric $X$, then its eCDF is canonically defined as
\begin{equation}
F_n(x) = \frac1n\sum_{i=1}^n\indicator{(-\infty,x]}{X_i} \ .
\label{eq:eCDF}
\end{equation}
In the context of this work, $X_i$ is associated with one out of many datasets, so our notation for eCDFs will read as $F_{X, j, s}(x)$ with $j$ being a label for the dataset generated under a given noise model, and $s$ labeling the random seed used in generating the noise trajectory in simulations. We include this labeling of the seed in order to distinguish between data generated under a fixed noise model with different noise time traces.

\subsubsection{K-S statistics}
After the eCDF $F_{X, j, s}(x)$ for each $j$ is computed, the two-sample K-S statistic testing data generated under $j$ against that generated under $j'$ is defined as

\begin{equation}
D_{X, jj', ss'}\equiv\sup_x\abs{F_{X, j, s}(x)-F_{X, j', s'}(x)} \ .
\label{eq:k_s_statistic}
\end{equation}
%
\subsubsection{Threshold values for type I and type II error rates}
The threshold values for the ten models for the results in Fig.~\ref{fig:ks_test_error_rates} are: $\lbrace 0.57, 0.62, 0.41, 0.35, 0.62, 0.39, 0.36, 0.25, 0.20, 0.18 \rbrace$.
\section{Single noise realization error attribution}\label{sec:single_noise_error_attribution}

\begin{figure}[h!]
    \centering
    \subfigure[]{\includegraphics[width=\linewidth]{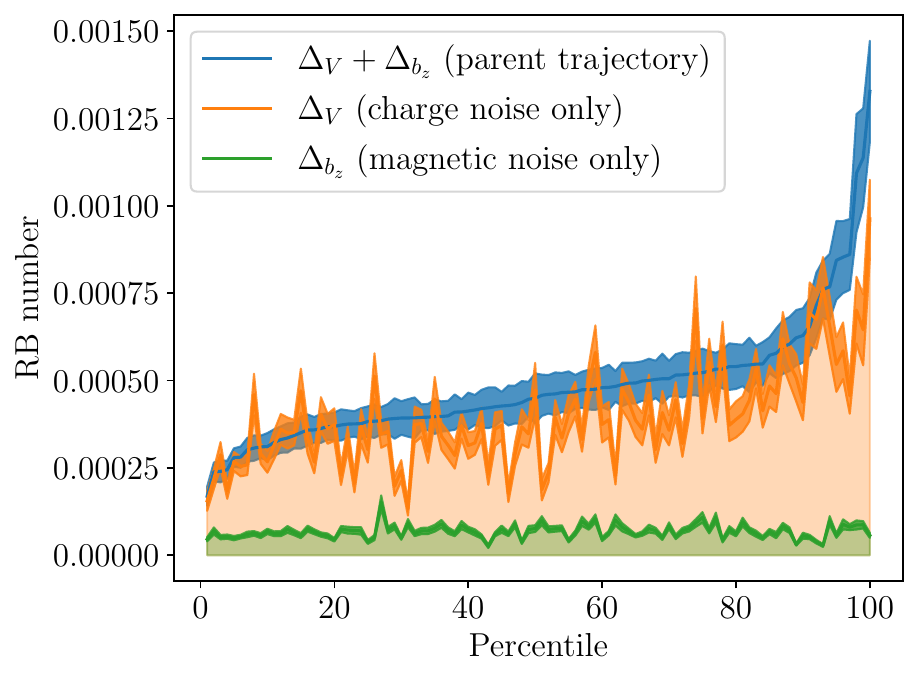}}
    \caption{\textbf{Single noise RB error attribution.} Similar to Fig.~\ref{fig:rb_number_error_attribution} in the main text except the results are for only a single noise realization.  In (a) the dark shaded regions are the 95\% confidence intervals of the RB number computed using the SciPy \emph{curve\_fit} function \cite{2020SciPy-NMeth}.}
    \label{fig:rb_number_error_attribution_single}
\end{figure}

In Fig.~\ref{fig:rb_number_error_attribution_single} we give an example of error attribution for single noise realizations. Unlike Fig.~\ref{fig:rb_number_error_attribution} in the main text, we see that additivity of the noise does not quite hold when studying a single noise realization. 
If we dial down the strength of the noise, we observe the restoration of additivity (not shown here). This suggests that for the noise levels considered here, nonlinear effects are relevant when considering individual noise trajectories. Additionally, when looking at the `parent trajectory', when both noise sources (along orthogonal directions) are on, it appears that the individual noise sources combine in a nontrivial manner -- they partially echo each other out. However, as shown in the main text, when studied as an ensemble of trajectories, the violation of additivity seen here is effectively averaged out.

\section{Different frequency partition} \label{app:differentpartition}

\begin{figure}[h!]
    \centering
    \subfigure[]{\includegraphics[width=\linewidth]{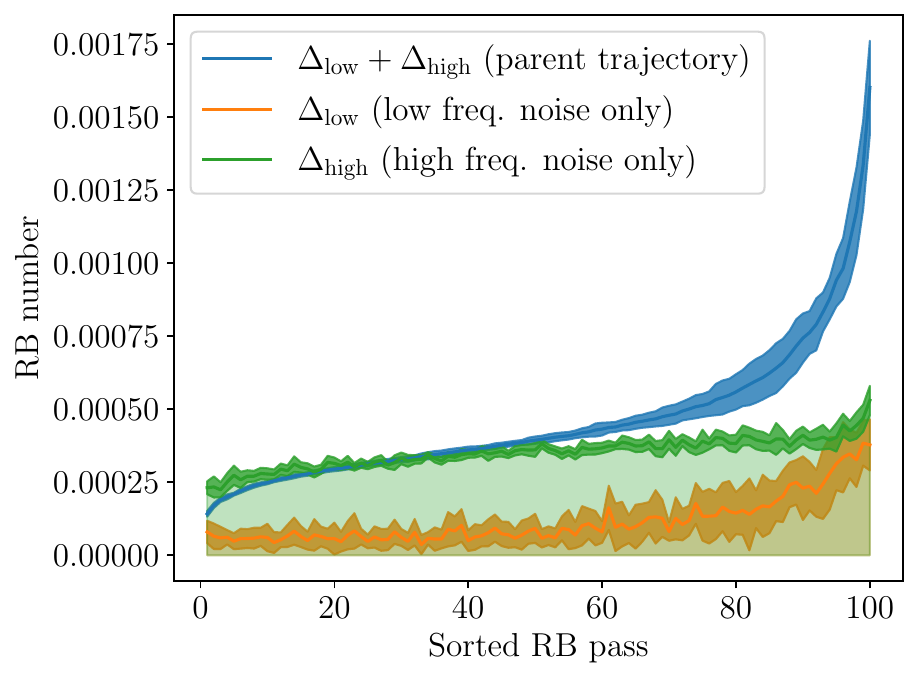}}
    \caption{\textbf{RB error attribution for Model 7 using a different partition.} For $10^2$ independent noise realization, 100 RB passes are performed.  For each noise realization, the RB numbers from the 100 passes are ordered according to the parent trajectory. This same ordering is used for the split trajectories, where we consider only charge noise and partition it into low and high-frequency components according to Eq.~\eqref{eq:partition2}. The solid lines are the mean of the median of these sorted RB numbers, and the dark shaded regions are the 95\% confidence intervals.  Both are computed using a bootstrap over the 100 noise realizations.} \label{fig:rb_number_error_attribution2}
\end{figure}

Here we show what happens if we consider a different partitioning for the frequency components. Since a single RB pass takes approximately $1s$, we can consider the partitioning:
\begin{eqnarray}  \label{eq:partition2}
    \mathcal{P}_{\mathrm{low}}&=&\sr{\eta_{10^{-3}\Hz}(t),\eta_{10^{-2}\Hz}(t),...,\eta_{10^0\Hz}(t)}  \ , \nonumber \\
    \mathcal{P}_{\mathrm{high}}&=&\sr{\eta_{10^1\Hz}(t),\eta_{10^2\Hz}(t),...,\eta_{10^7\Hz}(t)} \ .
\end{eqnarray}
We show results for this choice in Fig.~\ref{fig:rb_number_error_attribution2}. Unlike in Fig.~\ref{fig:rb_number_error_attribution}, additivity at the largest error rates is not well satisfied.

\section{Per-circuit error attribution}\label{sec:per_circuit_error_attribution}
\begin{figure}[h!]
    \centering
   \subfigure[]{\includegraphics[width=\linewidth]{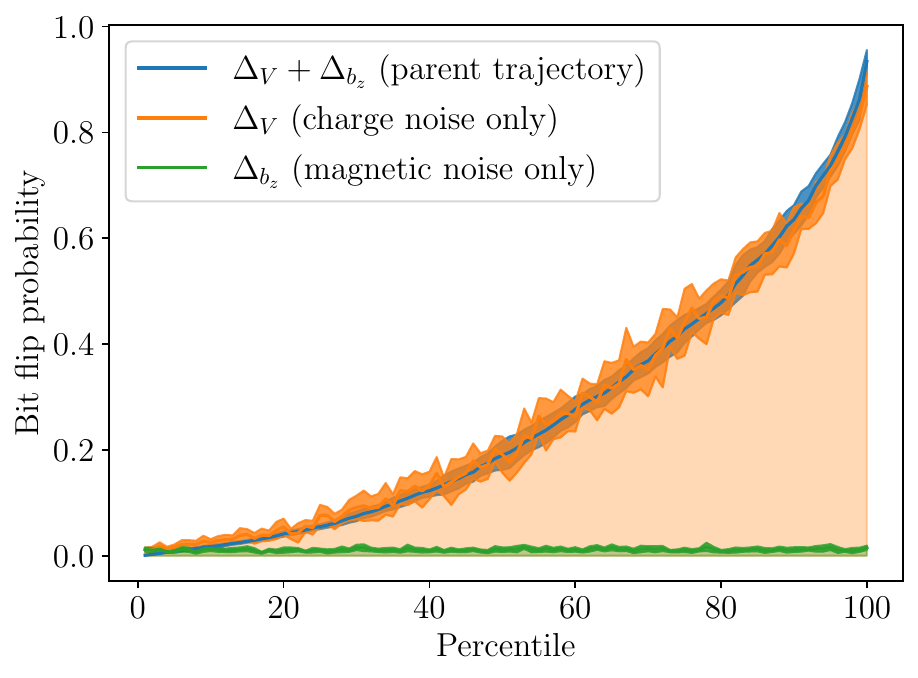}}
    \subfigure[]{\includegraphics[width=\linewidth]{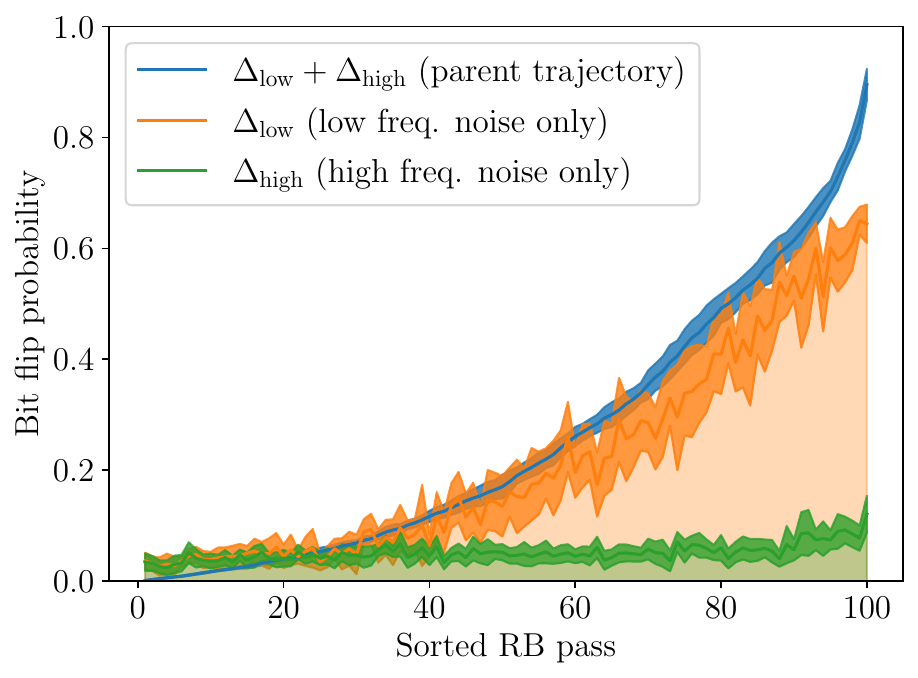}}
    \label{fig:per_circuit_error_attribution_components}
    \caption{\textbf{Per-circuit error attribution for Model 7.} The per-circuit error attribution for the second RB circuit of depth 128, $U_{128,2}$. The results displayed here are similar to those in Fig.~\ref{fig:rb_number_error_attribution}, but with $X=P(1|0,\Delta,U_{128,2})$ (the bitflip probability under the noise trajectory $\Delta$) instead of $X=r$.For 100 independent noise realization, 100 RB passes are performed. For each noise realization, the RB numbers from the 100 passes are ordered according to the parent trajectory. This same ordering is used for the split trajectories, where in (a) we split the noise into its different axes and in (b) we consider only charge noise and partition it into low and high-frequency components according to Eqn.~\eqref{eq:partition}. The solid lines are the mean of the median of these sorted bit flip probabilities, and the dark shaded regions are the 95\% confidence intervals.  Both are computed using a bootstrap over the 100 noise realizations.}
    \label{fig:per_circuit_error_attribution}
\end{figure}

For information about a circuit $U_i$'s sensitivity to various noise sources or frequencies, we may redo the tasks outlined in sections~\ref{sec:error_attribution_by_axis} and \ref{sec:error_attribution_by_OU_component} with $X=P(1|0,\Delta,U_i)$ (the bitflip probability for an RB circuit $U_i$) instead of $X=r$. Figure~\ref{fig:per_circuit_error_attribution} shows per-circuit error attribution results with the same noise trajectories $\Delta$ used to generate the results for Secs.~\ref{sec:error_attribution_by_axis} and \ref{sec:error_attribution_by_OU_component}. The data shows that bitflip errors for this circuit are primarily caused by charge noise and specifically the low frequency components according to our partition in Eq.~\eqref{eq:partition}.

\begin{figure}[b]
    \centering
    \includegraphics[width=\linewidth]{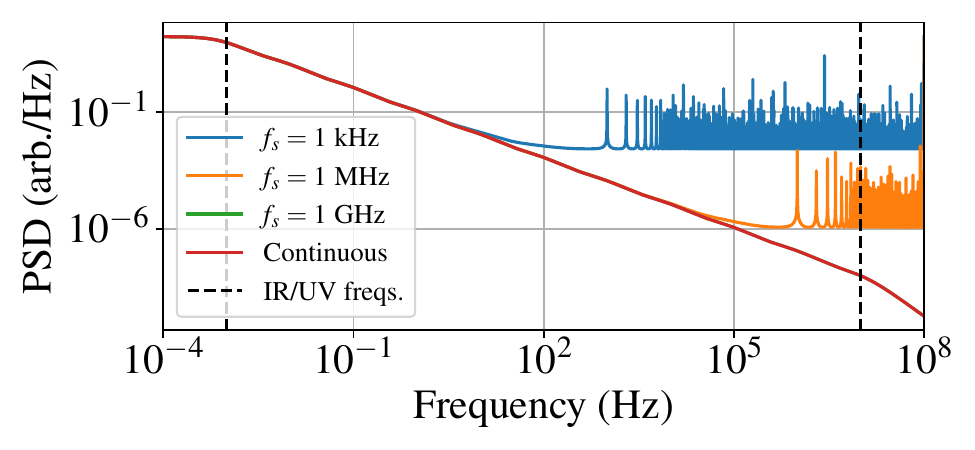}
    \caption{\textbf{Aliasing effects due to discrete time sampling.} The discretely-sampled PSD $S_d(f)$ under various sampling rates $f_s$ plotted with the continuously-sampled PSD $S(f)$. The set of OU process powers and frequencies are fixed. Aliasing effects are negligible for $f_s=1\GHz$, making this an appropriate choice for a sampling rate in our simulations. The power spectral density chosen in this example is that of one OU process per decade for $f_{i} \in \lbrace 10^{-3},..., 10^{7} \rbrace$ Hz and unit amplitude $p_{i}=1$.}
    \label{fig:aliasing_effects}
\end{figure}
%
\section{Discrete sampling effects on power spectrum}\label{sec:discrete_sampling_effects}

In the case of a finite sampling rate $f_s$, aliasing effects will change the PSD shown in Eq.~\eqref{eq:psd}. These may be incorporated by deriving the discrete-time PSD as is outlined in Ref.~\cite{Kasdin1995}. Let $S_d(f)$ denote this discrete-time PSD:
\begin{eqnarray}
S_d(f) &=& \sum_{n=-\infty}^\infty S(f+nf_s) \nonumber \\
&=&\sum_{j=1}^m\frac{p_j\coth\br{\pi f_j/f_s}\pr{\cot^2\br{\pi f/f_s}+1}}{f_s\pr{\cot^2\br{\pi f/f_s}+\coth^2\br{\pi f_j/f_s}}} \ .
\label{eq:Kasdin1995}
\end{eqnarray}
Note that in Eq.~\eqref{eq:Kasdin1995}, we recover the continuous-time PSD $S(f)$ in the $f_s\to\infty$ limit. The true simulated PSD is given by $S_d(f)$. We wish to choose a sampling rate large enough such that $S_d(f)$ approximates $S(f)$ well. In Fig~\ref{fig:aliasing_effects}, we depict how the choice of sampling rate affects the degree of aliasing artifacts in our simulated power spectrum. As can be seen in the figure, $f_s\sim1\GHz$ is large enough such that aliasing effects are negligible, and our simulated PSD $S_d(f)$ approximates the continuous-time PSD $S(f)$. Our corresponding sampling time is $\Delta t=1/f_s\sim1\ns$. Larger sampling rates are computationally expensive, and thus we keep our sampling rates and times in these ranges.
%

\section{Test error tradeoffs}
\label{sec:test_error_tradeoffs}
In defining the model validation test error rates $\alpha_{X,j}$ and $\beta_{X,jj'}$ in Sec.~\ref{sec:statistical_methods}, we observe that there is a tradeoff between type I and II error rates. This may be shown using the definition of $\beta_{X,jj'}$, assuming a fixed dataset:
\begin{eqnarray}
\beta_{X,jj'}&=&P(D_{X,jj',ss'}\leq d_{X,j}|j\not\sim j')\nonumber \\
&& \hspace{-1.5cm} =P(D_{X,jj',ss'}\leq \mathcal{P}_{100(1-\alpha_{X,j})}\pr{\sr{D_{X,jj,ss'}}_{s<s'}}|j\not\sim j') \ , \nonumber \\
\label{eq:expanded_type_II_error_definition}
\end{eqnarray}
where $\mathcal{P}_i(S)$ refers to the $i$-th percentile of the set $S$. Let $\alpha_{X,j}\leq\tilde\alpha_{X,j}$ be two chosen type I error rates; we associate $\tilde d_{X,j}$ and $\tilde\beta_{X,jj'}$ with $\tilde\alpha_{X,j}$, and similarly for $d_{X,j}$ and $\beta_{X,jj'}$ with $\alpha_{X,j}$. This gives $100(1-\tilde\alpha_{X,j})\leq100(1-\alpha_{X,j})$, and consequently 
\begin{eqnarray} 
\mathcal{P}_{100(1-\tilde\alpha_{X,j})}\pr{\sr{D_{X,jj,ss'}}_{s<s'}} &\leq& \nonumber  \\
&& \hspace{-3cm} \mathcal{P}_{100(1-\alpha_{X,j})}\pr{\sr{D_{X,jj,ss'}}_{s<s'}} \ . \label{eqt:percentileinequality}
\end{eqnarray}
Since $d_{X,j}=\mathcal{P}_{100(1-\alpha_{X,j})}\pr{\sr{D_{X,jj,ss'}}_{s<s'}}$ is the rejection threshold associated with $\alpha_{X,j}$, we can express Eq.~\eqref{eqt:percentileinequality} as $\tilde d_{X,j}\leq d_{X,j}$. This fact in conjunction with the monotonicity of $P(D_{X,jj',ss'}\leq p|j\not\sim j')$ gives $P(D_{X,jj',ss'}\leq \tilde d_{X,j}|j\not\sim j')\leq P(D_{X,jj',ss'}\leq d_{X,j}|j\not\sim j')$. Recall that these are the definitions of type II errors $\beta_{X,jj'}'$ and $\beta_{X,jj'}$ respectively. Put concisely,
\begin{equation}
    \alpha_{X,j}\leq\tilde\alpha_{X,j}\implies \tilde\beta_{X,jj'}\leq \beta_{X,jj'} \ .
\end{equation}
That is, choosing to decrease (increase) type I error rates may only increase (decrease) type II error rates given a fixed dataset. This implies that in order to decrease one type of error without increasing the other, the scheduling of the experiment or the amount of data collected should be modified, as is discussed in Sec.~\ref{sec:model_validation_results}.

\end{document}